\documentclass[twoside,11pt]{article}

\usepackage{jmlr2e}
\usepackage{enumerate}
\usepackage{url} % not crucial - just used below for the URL 
\usepackage{xr}
\usepackage[utf8]{inputenc} % allow utf-8 input
\usepackage{booktabs}       % professional-quality tables
\usepackage{amsfonts}       % blackboard math symbols
\usepackage{nicefrac}       % compact symbols for 1/2, etc.
\usepackage{latexsym}
\usepackage{float}
\usepackage{indentfirst}
\usepackage{makecell,diagbox}
\usepackage{bbm,mathrsfs,bm}
\usepackage{amsmath}
\usepackage{graphics}
\usepackage{psfrag,pst-node,subfigure,rotating,picture,   upgreek}
\usepackage{multirow}
\usepackage{threeparttable}

\newtheorem{Cor}{ Corollary}

\def\0{\mathbf 0}

\def\A{\mathbf A}

\def\Y{\mathbf Y}
\def\X{\mathbf X}
\def\Z{\mathbf Z}

\def\W{\mathbf W}
\def\I{\mathbf I}

\def\E{\mathrm E}

\def\W{\mathbf W}
\def\G{\mathbf G}

\def\T{\mathrm T}
\newcommand*{\dif}{\,\mathrm{d}}
\def\bfalpha{\boldsymbol \alpha}

\newcommand{\bfGamma}{\boldsymbol {\Gamma}}
\newcommand{\bfmu}{\boldsymbol {\mu}}

\newcommand{\VAR}{\operatorname {Var}}
\newcommand{\DPI}{\operatorname {DPI}}
\newcommand{\MRP}{\operatorname {\texttt{MRP}}}
\usepackage{lastpage}
\ShortHeadings{Two Sample Testing for High-dimensional Functional Data}{Wang,  Cao, Liu,  You and  Liu} 
\firstpageno{1}

\begin{document}

\title{Two Sample Testing for High-dimensional Functional Data:
 A Multi-resolution Projection  Method}

\author{\name Shouxia Wang \email wangshouxia20@sufe.edu.cn \\
       \addr School of Statistics and Data Science\\
     Shanghai University of Finance and Economics\\
          \AND
       \name Jiguo Cao \email jiguo\_cao@sfu.ca \\
       \addr Department of Statistics and Actuarial Science\\
  Simon Fraser University\\
         \AND
     \name   Hua Liu  \email liuhua\_22@xjtu.edu.cn  \\
   \addr School of Economics and Finance\\
  Xi'an Jiaotong University\\
      \AND
\name Jinhong You \email johnyou07@163.com \\	
		\addr School of Statistics and Data Science\\
	Shanghai University of Finance and Economics\\
     \AND
     \name  Jicai Liu  \email liujicai1234@126.com  \\
   \addr School of Statistics and Mathematics\\
 Shanghai Lixin University of Accounting and Finance\\
}

\editor{}

\maketitle

\begin{abstract}%  
It is of great interest to test the equality of the means in two samples of functional data. Past research has predominantly concentrated on low-dimensional functional data, a focus that may not hold up in high-dimensional scenarios.  In this article, we propose a novel two-sample test for the mean functions of high-dimensional functional data, employing a multi-resolution projection (MRP)  method. We establish the asymptotic normality of the proposed MRP test statistic and investigate its power performance when the dimension of the functional variables is high.  In practice, functional data are observed only at discrete and usually asynchronous points. We further explore the influence of function reconstruction on our test statistic theoretically. Finally, we assess the finite-sample performance of our test through extensive simulation studies and demonstrate its practicality via two real data applications. Specifically, our analysis of global climate data uncovers significant differences in the functional means of climate variables in the years 2020-2069 when comparing intermediate greenhouse gas emission pathways (e.g., RCP4.5) to high greenhouse gas emission pathways (e.g., RCP8.5). 
% Testing the equality of the means in two samples of functional data is a fundamental problem in statistics. Existing research has primarily focused on low-dimensional functional data, which may be invalid when confronted with high-dimensional scenarios.
% In this study, we propose a novel two-sample test for mean functions of high-dimensional functional data, employing a multi-resolution projection (MRP)  technique. We establish the asymptotic normality of the proposed MRP 
% % mr-PD  MPD  mRPD 
% test statistic and investigate its power performance when the dimension of the functional variables is high.  In practice,  functional data are only observed on discrete points. We further explore the influence of function reconstruction on our test statistic theoretically. Finally, we assess the finite-sample performance of our test through extensive simulation studies and demonstrate its applicability via two real data applications. Specifically, the analysis of global climate data reveals notable differences in climate characteristics such as daily maximum air temperature, daily minimum air temperature, and daily precipitation when comparing intermediate greenhouse gas emission pathways (e.g., RCP4.5) to high greenhouse gas emission pathways (e.g., RCP8.5) over mid to long-term periods, with more significant differences in the second 25 years. However, these differences may not be significant when examining the first 25 years.
\end{abstract}


\section{Introduction}\label{Introduction}
% \section{Introduction}\label{Introduction}
% \setcounter{page}{1}
Over the past decade, functional data analysis (FDA) has gained significant importance in the field of statistics, primarily due to the growing demand in various disciplines to extract valuable information from complex datasets that are represented by discretized curves or surfaces. These curves or surfaces, known as functional data, are considered as random samples of functions defined over continuous domains, such as time or spatial location
\citep{2005Functional, ferraty2006nonparametric, Horvp2012, hsing2015theoretical,wang2016functional,tangClusteringForecastingMultiple2022a}. Along with advances in measuring technology, 
high-resolution observations are increasingly available, which motivates high-dimensional functional data.
Scientific applications are witnessing a growing amount of high-dimensional functional data, such as chemometrics, neuroscience, and climate science.

 For example, in climate change research, the NASA Center for Climate Simulation (NCCS) provides the projected global climate data based on two Representative Concentration Pathways (RCPs), RCP4.5 and RCP8.5. They are two scenarios that represent future greenhouse gas emission pathways 
% riahi2011rcp,
\citep{Moss2010TheNG,thomson2011rcp4,meinshausen2011rcp,chou2014assessment}, where the numerical values 4.5 and 8.5 refer to the radiative forcing produced by CO$_2$ emissions in 2100. RCP4.5 corresponds to an intermediate level of greenhouse gas emissions, whereas RCP8.5 signifies high emissions. 
% its daily estimators during one year at the different grid points  can be viewed  as high-dimensional functional data. Thus,  the problem can be treated as a two-sample test of the means for high-dimensional functional data.
Many scientists think that current emissions are tracking close to the high greenhouse gas emission, RCP8.5 pathway. At present, lots of countries and organizations are implementing various measures to reduce emissions in order to achieve the RCP 4.5 pathway, a medium-level scenario for future greenhouse gas concentration and climate change mitigation. Thus,
it is of scientific interest to test whether the climate around the whole world is different between the RCP4.5 and RCP8.5 pathways .

\iffalse
However, this test problem involves high-dimensional functional data owing to the fact that  
the global climate data provided by the NCCS includes three climate variables, daily minimum/maximum air temperature and daily precipitation from 2006 to 2100, measuring across \textcolor{red}{$1,028,032$} geographical coordinates worldwide.
\fi

Each climate variable in the global climate dataset is measured across 1,028,032 geographical coordinates worldwide from 2006 to 2100. Intuitively, each climate variable measured at a specific spatial grid point in a given year can be viewed as a univariate functional curve. Consequently, when considering the climate variable measurements at all spatial grid points for each year, we are dealing with an extremely high-dimensional functional data, which has 1,028,032 dimensions. 
% \textcolor{red}{(Hua: I think it's better to use the real data as an example of high-dimensional functional data not only functional data. Otherwise, it is strange that the next sentence is followed by the previous one (only mentioning the functional data)).}
% \textcolor{red}{(Hua: Maybe we can rewrite the previous sentence: Intuitively, each climate variable at all spatial grid points measured in each year can be viewed as high-dimensional functional data.
% It is of scientific interest to test whether each climate variable around the whole world in the RCP4.5 and RCP8.5 pathways is different. )}
Motivated by this application, this article aims to develop a two-sample test for high-dimensional functional data to determine if the population means of two high-dimensional functional data are equal.

% This test problem involves high-dimensional functional data which is challenging. This is because each climate variable in the global climate dataset is measured across 1,028,032 geographical coordinates worldwide from 2006 to 2100.
% % , making it
% % high-dimensional in nature.
% Intuitively, each climate variable measured at a specific spatial grid point in a given year can be viewed as a univariate functional curve. Consequently, when considering the climate variable measurements at all spatial grid points for each year, we are dealing with an extremely high-dimensional functional data set of 1,028,032 dimensions. 
% % \textcolor{red}{(Hua: I think it's better to use the real data as an example of high-dimensional functional data not only functional data. Otherwise, it is strange that the next sentence is followed by the previous one (only mentioning the functional data)).}
% % \textcolor{red}{(Hua: Maybe we can rewrite the previous sentence: Intuitively, each climate variable at all spatial grid points measured in each year can be viewed as high-dimensional functional data.
% % It is of scientific interest to test whether each climate variable around the whole world in the RCP4.5 and RCP8.5 pathways is different. )}
% Motivated by this application, this article aims to develop a two-sample test for high-dimensional functional data to determine if the population means of two high-dimensional functional data are equal.

During the last decade, significant research efforts have focused on conducting sample testing of mean functions for functional data.  For example,  \citet{cuevas2004anova} proposed a $ L_2$ norm-based test for functional Analysis of Variance (ANOVA) by calculating the norms of the difference between two sample means.  \citet{2005Functional} extended the classical $F$-test to the context of functional data by proposing a pointwise $F$-type test. \citet{zhang2014one}  developed a global $F$-type test by integrating the  pointwise   $F$-test over a specified time interval.
Additionally, \citet{Horvp2012} proposed a test statistic based on the functional principal component analysis. \citet{horvath2013estimation}  introduced two test statistics based on orthogonal projections of the difference between the mean functions. \cite{zhang2013analysis} provided a comprehensive review of hypothesis testing methodologies for functional data.

All of the aforementioned tests are designed for univariate functional data. Little literature has explored two-sample tests of means for multivariate functional data. 
To the best of our knowledge, there is only one existing study that has investigated two-sample tests for multivariate functional data, in which the authors introduced two Hotelling $T^2$ type test statistics \citep{qiuTwosampleTestsMultivariate2021}. {Nevertheless, it is important to note that these two Hotelling $T^2$ type test statistics utilizing the inverse of sample covariance function are invalid for high-dimensional functional data due to the fact that 
 the sample covariance function is singular and inconsistent under the high-dimensional case.}  
% $\boldsymbol\Gamma(t, t)^{-1}$ and $\boldsymbol\Gamma(s, t)^{-1}$, which is difficult to be estimated consistently unless there are some additional assumptions such as sparsity. 
As far as our understanding goes, 
no method has been developed for testing the equality of the means for high-dimensional functional data, although extensive studies exist on testing high-dimensional mean for vector-valued data, such as \citet{bai1996effect}, \citet{escanciano2006consistent}, \citet{chenTwosampleTestHighdimensional2010},  \citet{tony2014two}, \citet{kim2020robust}, \citet{liu2022multiple} and \citet{Jiang2022jasa}.

% \citet{Baringhaus2004On}, 
Two sample testing for high-dimensional
functional data is different from that of high-dimensional vector-valued data.  
Due to the infinite-dimensional characteristic of functional data, the existing methods for high-dimensional vector-valued data are only suitable for each scalar component,  and cannot be applied to high-dimensional functional data, where each functional variable is infinite-dimensional. In addition, for functional data or longitudinal data with asynchronous observations for different curves, the vector-based methods are not applicable. The reason is that different samples have different dimensions if treated as vectors and thus are not comparable. To tackle the problem,  we develop a  multi-resolution projection (MRP) technique for the first time and propose a novel MRP-based test statistic. We project the high-dimensional functional process into a one-dimensional process and then project the process into a scalar random variable. As the projection is on two different spaces, the $p$-dimensional vector space and the Hilbert space of square-integrable functions, we call the proposed method the multi-resolution projection. 
% This is the first time to develop a multi-resolution projection
{Specifically,  we first define the distance between two high-dimensional random functions based on the multi-resolution projection. Next, we derive a computationally tractable form for the MRP distance and formulate the MRP test statistic. We then establish the asymptotic normality of this test statistic, which enables us to conduct hypothesis testing fast. }
% \sout{Additionally, we investigate the power of the proposed MRP test, providing insights into its performance in practice.}

{This article has four main contributions. Firstly, to the best of our knowledge, the proposed MRP test is the first attempt to test whether the two sample means of high-dimensional functional data are different. Secondly, the asymptotic variance of the MRP test is easily and consistently estimated, so we do not require a random permutation or bootstrap to approximate the asymptotic null distributions. Thus, the MRP test is computationally efficient. Thirdly, we explore the influence of function reconstruction on the proposed test as we could only get the discretized observations which are usually asynchronous instead of the continuous functions in practice. It is shown that the proposed MRP test with reconstructed functional data has the same asymptotic distribution as that with the true  functions. 
Finally, when applied to the global climate data, we find that there are notable differences in climate characteristics between  RCP4.5 and  RCP8.5 over mid to long-term periods, especially in the years 2045-2069, while no significant differences are found in the years 2020-2044.

}

% An important contribution of this article is 
% that we derive the asymptotic normality of the MRP-bast test statistic under the high-dimensional null hypotheses \textcolor{red}{(Hua: high-dimensional null hypotheses?)}, without any assumption on the relative growth rate between
% data dimension and sample size \textcolor{red}{(Shouxia and Hua: maybe we need to delete this sentence)}. Because the asymptotic variance is easily estimated,  we do not require a random permutation or bootstrap to approximate the asymptotic null distributions. Thus, the MRP-bast test is computationally efficient.  Furthermore,  we also derive the asymptotic power function of the proposed test.  To the best of our knowledge, these asymptotic results have not been developed for high-dimensional functional data.

 %{\color{red}The theoretical results can be viewed as %an extension of those of %\citet{chenTwosampleTestHighdimensional2010} to some %extent. However, it should be pointed out that the %extension is non-trivial due to the infinite %dimensionality of functional data.} {\color{blue}  %wsx: I think it also makes the contribution of our %work weak. }
% It should be noted that, similar to \citet{chenTwosampleTestHighdimensional2010}, 

The rest of the paper is organized as follows.   In Section \ref{MRPtest},
we introduce our test statistic through the multi-resolution projection technique.
In Section \ref{Asymptotic}, we 
study the asymptotic behaviors of our test statistic under the null and alternative hypotheses, as well as its power analysis.  In Section \ref{Reconstruction},
we explore theoretically the influence of function reconstruction on the proposed test. Numerical studies and real data applications are presented in Section \ref{Simulation} and Section \ref{Realdata}, respectively.
The conclusion is given in Section \ref{Discussion}. All technical details are postponed
to the Supplementary.

\section{Methodology and Theory}\label{Method}
Let $\mathbf{X}=\{\mathbf{X}(t),t\in[0,1]\}=\{(X_{1}(t), \ldots, X_{p}(t) )^{\T}, t\in[0,1]\}$ and $\mathbf{Y}=\{\mathbf{Y}(t),t\in[0,1]\}=\{(Y_{1}(t), \ldots, Y_{p}(t) )^{\T}, t\in[0,1]\}$ be two $p$-dimensional random functions, where the dimension $p$ is high, and each $X_{k}(t)$ and $Y_{k}(t)$ are defined on $L_2([0, 1])$, a Hilbert space of square-integrable functions on $[0, 1]$ with the inner product $\langle f, g\rangle=\int_{0}^{1} f(t) g(t) \dif t$ for any $f, g\in L_{2}([0, 1])$.
The mean functions of $\X(t)$ and $\Y(t)$ are defined by  $\bfmu_{1}(t)=\E\{\mathbf{X}(t)\}$  and $\bfmu_{2}(t)=\E\{\mathbf{Y}(t)\}$,  for $t\in[0, 1]$.
Let $\G_1(s, t)=\operatorname{Cov}\{\mathbf{X}(s),\mathbf{X}(t)\}$ and $\G_2(s, t)=\operatorname{Cov}\{\mathbf{Y}(s),\mathbf{Y}(t)\}$,  for $s,t\in[0, 1]$, be the covariance functions of
$\mathbf{X}(t)$ and $\mathbf{Y}(t)$. We aim  to test
 \begin{equation}\label{fun-twotest}
H_{0}: \bfmu_{1}(t)=\bfmu_{2}(t) \text{ for all } t \in[0, 1] \, \text { versus } \, H_{1}: \bfmu_{1}(t)\neq \bfmu_{2}(t) \text{ for some } t \in[0, 1].
 \end{equation}

\subsection{MRP Distance Based Test }
\label{MRPtest}
In this section, we propose a new projection-based approach to test the problem \eqref{fun-twotest} for high-dimensional functional data The proposed method depends on the following multi-resolution projection technique.
\begin{lemma}\label{lemma1}
Let $\X \in  L_{2}^{p}([0, 1]) $ and $\Y \in L_{2}^{p}([0, 1])$  be two $p$-dimensional random functions,  where $L_{2}^{p}([0, 1])=L_{2}([0, 1])\times\cdots\times L_{2}([0, 1])$. Then, we have that
 \begin{equation*}\label{equivalent}
 \E\{\X\}=\E\{\Y\}  \Longleftrightarrow   \E\{\langle\bfalpha^{\T}  \X, \gamma\rangle\}=\E\{ \langle \bfalpha^{\T}\Y, \gamma \rangle\}, \text{ for any }  \bfalpha \in\mathbb{R}^{p}, \gamma \in L_{2}([0, 1]),
\end{equation*}
 where   $``\Longleftrightarrow"$ stands for  `` equivalent to".
 \end{lemma}

 Lemma \ref{lemma1} suggests that the test problem \eqref{fun-twotest} is equivalent to testing the equality of expectations of the two univariate random variables $\langle\bfalpha^{\T}  \mathbf{X}, \gamma\rangle$ and $\langle\bfalpha^T  \mathbf{Y}, \gamma\rangle$, for any $\bfalpha \in\mathbb{R}^{p}$    and $\gamma \in L_{2}([0, 1])$.  Note that  $\langle\bfalpha^{\T}  \mathbf{X}, \gamma\rangle$ and $\langle\bfalpha^T  \mathbf{Y}, \gamma\rangle$ are obtained through  projecting  a $p$-dimensional random function  into  a random variable by the two directions   $\bfalpha$ and $\gamma$. Thus, we call it the multi-resolution projection approach.
 By Lemma \ref{lemma1}, we   define the following distance to measure the difference between $\E\{\X\}$ and $\E\{\Y\}$.
%  \\
% \textcolor{red}{Jiguo: MRP or MPD??}\\
\begin{definition}
Let $\X \in L_{2}^{p}([0, 1])$ and $\Y \in  L_{2}^{p}([0, 1])$  be two $p$-dimensional random functions. The MRP distance   of $\mathbf{X}$ and  $\mathbf{Y}$ is defined by
 \begin{equation}\label{equvlent-dpi}
\MRP(\X,\Y)   = \int_{L_{2} }  \int_{\mathbb{R}^{p } } \left(\E\{\langle\bfalpha^{\T}  (\X-\Y), \gamma\rangle\}\right)^2 \dif G(\bfalpha)\nu(\mathrm{d}\gamma),
\end{equation}
 where  $G(\cdot)$  is the CDF of $\bfalpha$,    $G(\bfalpha)=\Pi_{k=1}^pG_k(\alpha_k),$ $G_k(\alpha_k)$ is a  CDF  of $\alpha_k$ with mean 0 and variance 1, $k=1,\ldots,p$,  $\nu(\cdot)$ is the CDF of a centered process $\gamma$ in  $L_{2}([0, 1])$ with a positive definite  covariance function   $v(s, t)$,  and $\bfalpha$ and $\gamma$ are independent of  $\X$ and $\Y$. 
\end{definition}
% where  $G(\bfalpha)$  is the CDF  of $N_{p}(0, \mathbf{I}_{p})$, and  $\nu(\gamma)$ is a centered Gaussian measure in  $L_{2}([0, 1])$ with a specified covariance function $v(s, t)$.
%$G(\bfalpha)$  is the cumulative  distribution function (CDF)  
%of $\bfalpha$ satisfying 
%$G(\bfalpha)=\Pi_{k=1}^pG_0(\alpha_k),$ where  $G_0(\cdot)$ is a  CDF  %with a mean $0$ and variance $1$
% $N_{p}(0, \mathbf{I}_{p})$, 
% $\bfalpha=(\alpha_1,\ldots,\alpha_p)^{\T}$, where $\alpha_k %\stackrel{i.i.d.}\sim (0,1)$
%and  $\nu(\gamma)$ is a centered
% Gaussian
%process in  $L_{2}([0, 1])$ with a specified covariance function $v(s, t)$. Here, $\bfalpha$ and $\gamma$ are independent of  $\X$ and $\Y$.
 The multi-resolution projection integral in \eqref{equvlent-dpi} can be viewed as the averages over all the two-directions projections with respective to the weights $G(\bfalpha)$ and $\nu(\gamma)$. 
 %{\color{red}The integral is an extension of the one-%direction projection-averaging method for random %vectors, see  %\citet{escanciano2006consistent,Zhu2017Projection,kim2%020robust,Jiang2022jasa}. }
 Here the choice of $\bfalpha$ is very flexible. For example, we could choose $\bfalpha \sim N_{p}(0, \mathbf{I}_{p})$.
 Additionally, note that $\MRP(\X,\Y)$ depends on the function $v(s, t)$. In fact, by the Kolmogorov existence theorem,  for any positive semi-definite function
$v(s, t)$,  there exists a centered Gaussian process with the covariances given by $v(s, t)$.
Thus, any positive semi-definite function $v(s, t)$ can be used for $\MRP(\X,\Y)$.  In practice,  we can choose different covariance functions $v(s, t)$ based on some commonly used Gaussian processes. For example, we can consider two special choices:
 (i) If  $\gamma$ follows from a Wiener process, then $v(s, t)\propto \min \{s, t\}$. (ii) If $\gamma$ is a stationary Ornstein-Uhlenbeck process, we have $v(s, t) \propto a^{-1} \exp (-a|s-t|)$, for some $a>0$.
%  \begin{description}
%  \item[(i)] If  $\gamma$ follows from a Wiener process, we have that $v(s, t)\propto \min \{s, t\}$.
%   \item[(ii)] If $\gamma$ is a stationary Ornstein-Uhlenbeck process, we have $v(s, t) \propto a^{-1} \exp (-a|s-t|)$, for some $a>0$.
% \end{description}

%In the definition of $\MRP(\X,\Y)$,  the weight function $\nu(\gamma)$ %is chosen to be a centered process and  $G(\bfalpha)$  is the CDF of  %$\bfalpha$ satisfying 
%$G(\bfalpha)=\Pi_{k=1}^pG_0(\alpha_k),$ where  $G_0(\cdot)$ is a  CDF  %with a mean $0$ and variance $1$.
% and  as the weight functions $G(\bfalpha)$  and   $\nu(\gamma)$. 

Note that the multi-resolution projection integral in \eqref{equvlent-dpi}  may suffer from computational issues due to intractable integration. In the following theorem, we show that $\MRP(\mathbf{X},  \mathbf{Y})$  has a closed-form expression.
\begin{theorem} \label{Theorem11}
\begin{itemize}
 \item[(i)]  Suppose that $\mathbf{X}_{1}$ and $\mathbf{X}_{2}$ are  independent and
identically distributed (i.i.d.) copies of ${\mathbf{X}}$ and, independently, $\mathbf{Y}_{1}$ and $\mathbf{Y}_{2}$ are i.i.d. copies of ${\mathbf{Y}}$. Then, we have 
  \begin{equation}\label{closed-form-MRP}
 \MRP(\X,  \Y)
   = \int_{0}^{1}\int_{0}^{1}\left[\bfmu_{1}(s)-\bfmu_{2}(s)\right]^{\T}\left[\bfmu_{1}(t)-\bfmu_{2}(t)\right]v(s, t) \dif s \dif t.
  \end{equation}
  \item[(ii)] $\MRP(\X,  \Y) \geq 0$ and
    $\MRP(\X,  \Y)=0$ if and only if $\bfmu_{1}(t)=\bfmu_{2}(t)$,  for all $t \in[0, 1]$.
\end{itemize}
\end{theorem}

 Theorem \ref{Theorem11}(i) indicates that $\MRP(\mathbf{X},  \mathbf{Y})$ has a closed form and therefore is computationally tractable.
Theorem \ref{Theorem11}(ii) implies that $ \MRP(\mathbf{X},  \mathbf{Y})$ is a reasonable measurement for the difference between the means of two random functions.

We propose the following MRP test statistic based on an unbiased empirical estimate of $\texttt{MRP}(\mathbf{X},  \mathbf{Y})$:
\begin{equation*}
\begin{aligned}
\widehat{\texttt{MRP}} (\mathbf{X},\mathbf{Y}) =&\quad\frac{1}{ n(n-1) }\sum_{i, j=1 \atop i \neq j}^{n}\int_{0}^{1}\int_{0}^{1}\X_i(s)^{\T}\X_j(t)v(s, t) \dif s \dif t\\
 &+  \frac{1}{ m(m-1) }\sum_{i, j=1 \atop i \neq j}^{m}\int_{0}^{1}\int_{0}^{1}\Y_i(s)^{\T}\Y_j(t)v(s, t) \dif s \dif t\\
   &-\frac{2}{ nm }\sum_{i=1}^{n}\sum_{j=1}^{m}  \int_{0}^{1}\int_{0}^{1}\X_i(s)^{\T}\Y_j(t)v(s, t) \dif s \dif t,
   \end{aligned}
  \end{equation*}
where 
 $\{\X_{i}(t), t\in[0,1],i=1, \ldots, n\}$ and $\{\Y_{i}(t), t\in[0,1],i=1, \ldots, m\}$ are i.i.d. copies of $\X$ and $\Y$.  We would reject $H_0$ if $\widehat{\MRP} (\mathbf{X},\mathbf{Y})$ is large. To derive the critical value of the proposed MRP test, we need to establish the asymptotic normality of  $\widehat{\MRP}(\mathbf{X},\mathbf{Y})$. 
\subsection{Asymptotic Normality}\label{Asymptotic}
To facilitate the presentation, we define the following notation 
\begin{equation*}
 \begin{aligned}
\texttt{ITR}(\G_k, \G_j)&= \iiiint \operatorname{tr}\left\{\G_k(s,s_1)\G_j(t,t_1)\right\} v(s,t)v(s_1,t_1) \dif s \dif t \dif s_1 \dif t_1,\\
\texttt{IMD}(\G_k, \bfmu_1,\bfmu_2)&=\iiiint [\bfmu_{1}(t)-\bfmu_{2}(t)]^{\T}\G_k(s,s_1)[\bfmu_{1}(t_1)-\bfmu_{2}(t_1)]
	v(s,t)v(s_1,t_1) \dif s \dif t \dif s_1 \dif t_1
\end{aligned}
\end{equation*}
 where $\G_k(s,t)$ and $\G_j(s,t)$ are  $p\times p$ covariance matrices for  $s,t\in [0,1]$ for $k,j=1,2$. 
We then introduce the conditions needed to derive the asymptotic distribution. 
\begin{itemize}
\item[(C1).]  Assume that $\X$ and $\Y$ satisfy the following general multivariate functional model:
\begin{equation}\label{factor}
\begin{aligned}
& \mathbf{X}_j(t)=\boldsymbol{\mu}_1(t)+\boldsymbol{\Gamma}_1(t) \mathbf{Z}_{1 j}(t),  t\in[0,1], \text {  for  } j=1, \cdots, n, \\
& \mathbf{Y}_j(t)=\boldsymbol{\mu}_2(t)+\boldsymbol{\Gamma}_2(t) \mathbf{Z}_{2 j}(t), t\in[0,1],   \text {  for  } j=1, \cdots, m,
\end{aligned}
\end{equation}
where $\bfGamma_{1}(t)$ and $\bfGamma_{2}(t)$  are    $p\times d$-matrices  for any $t\in[0,1]$ with $d \geq p$, and
    $\mathbf{Z}_{i j}=\{\mathbf{Z}_{i j}(t)=( {Z}_{ij 1}(t), \cdots, {Z}_{ij d}(t))^{\T}, t\in[0,1]\}$ is a  $d$-dimensional random function  for $i=1, j=1,\cdots, n$ and $i=2, j=1,\cdots, m$, satisfying
\begin{itemize}
\item[(1)] $\left\{\mathbf{Z}_{1 j}(t) \right\}_{j=1}^{n}$ and $\left\{\mathbf{Z}_{2 j}(t)\right\}_{j=1}^{m}$ are i.i.d. with $\E \{\mathbf{Z}_{1 j}(t)\}=\mathbf{0}$, $\operatorname{Cov}\{\mathbf{Z}_{1 j}(s), \mathbf{Z}_{1 j}(t)\}= c_1(s,t)\mathbf{I}_{d}$, $\E\{\mathbf{Z}_{2 j}(t)\}=\mathbf{0}$, and $\operatorname{Cov}\{\mathbf{Z}_{2 j}(s), \mathbf{Z}_{2 j}(t)\}= c_2(s,t)\mathbf{I}_{d}$, where $\mathbf{I}_{d}$ is the $d\times d$ identity matrix, $c_1(s,t)$ and $c_2(s,t)$ are bounded for  $s,t \in [0,1]$.

\item[(2)]$c_1(s,t)\bfGamma_{1}(s) \bfGamma_{1}^{\T}(t)=\G_{1}(s,t)$ and $c_2(s,t)\bfGamma_{2}(s) \bfGamma_{2}^{\T}(t)=\G_{2}(s,t)$.

\item[(3)] $\sup _{i,j, k,t} \E\{ Z_{i j k}^{8}(t)\}<\infty$ and $\E\{ Z^{4}_{i j k}(t) \}=3+\Delta_{i}$
    for some constant $\Delta_{i}$, $i=1,2$.
     % where $C_1(t)$ and $C_2(t)$ are bounded for any $t \in [0,1]$.
    Also,
$
\E\{Z_{i j l_{1}}^{a_{1}}(t_1) \cdots Z_{i j l_{q}}^{a_{q}}(t_q) \}=E\{Z_{i j l_{1}}^{a_{1}}(t_1)  \} \cdots \E\{Z_{i j l_{q}}^{a_{q}}(t_q) \},
$
for any positive integers $q$ and $a_{l}$ such that $\sum_{l=1}^{q} a_{l} \leq 8$, and $l_{1}, l_{2}, \cdots, l_{q}$ are distinct indices.
\end{itemize}

\item[(C2).] As $\min \{m, n\} \rightarrow \infty,$ $n /(m+n) \rightarrow \tau \in(0,1)$.

\item[(C3).]  Assume that, for $k= 1,2,$
\begin{equation*}
	\texttt{IMD}(\G_{k}, \bfmu_1,\bfmu_2)
=o\Big({(n+m)}^{-1} \texttt{ITR}(\G_{12}, \G_{12})\Big), \,\text{where 
 } \G_{12}(s,t)=\G_{1}(s,t)+\G_{2}(s,t).
\end{equation*}
% \begin{align}\label{condition1}
% 	&\iiiint [\bfmu_{1}(t)-\bfmu_{2}(t)]^{\T}\G_{k}(s,s_1)[\bfmu_{1}(t_1)-\bfmu_{2}(t_1)]
% 	v(s,t)v(s_1,t_1) \dif s \dif t \dif s_1 \dif t_1
% 	 \nonumber\\ &
%   \quad=o\Big({(n+m)}^{-1} \texttt{ITR}(\G_{12}, \G_{12})\Big),
% \end{align}

\item[(C4).]  As $p\rightarrow \infty$, and for any $i, j, k, l \in\{1,2\}$,
\begin{align}
&\idotsint\operatorname{tr}\{\G_{i}(s,s_1) \G_{j}(t,t_1)\G_{k}(s_2,s_3) \G_{l}(t_2,t_3) \} v(s,t)v(s_1,t_1)v(s_2,t_2)\nonumber\\
&\quad\quad\quad \times v(s_3,t_3) \dif s \dif t \dif s_1 \dif t_1 \dif s_2 \dif t_2 \dif s_3 \dif t_3  =o\Big\{\Big(\texttt{ITR}(\G_{12}, \G_{12})\Big)^2\Big\}.
\end{align}
\end{itemize}

Condition  (C1)  gives a general multivariate model for high-dimensional functional data and \cite{chenTwosampleTestHighdimensional2010} used a similar multivariate model for high-dimensional vector-valued data.
% According to \cite{chenTwosampleTestHighdimensional2010},
The condition  $d \geq p$ means that the rank and eigenvalues of $\G_{1}(s,t)$ or $\G_{2}(s,t)$ are not affected by the transformation.
Condition  (C2) guarantees that $m$ and $n$ go to infinity proportionally, which is standard in two-sample test problems.

Condition (C3) is satisfied under $H_0$, and enables the variance
of  $\widehat{\MRP}(\mathbf{X},\mathbf{Y})$ to be asymptotically characterized by $\widehat{\sigma}_{nm}^{2}$ given in the following Theorem \ref{Theorem2}.
To better understand Condition (C3) for high-dimensional situations,  we consider a special case that
$\bfmu_{1}(t)-\bfmu_{2}(t)=\delta(t)\mathbf{1}_p$ and $\G_{1}(s,t)=\G_{2}(s,t)=c(s,t)\mathbf{I}_{p}$.
Then,  Condition  (C3)  implies that  the order of $\texttt{IMD}(\G_{k}, \bfmu_1,\bfmu_2)$
% at the left-hand side of \eqref{condition1} 
is $o((n+m)^{-1/2})$ which is  smaller  than $O((n+m)^{-1/2})$ for a fixed $p$. This means we can detect smaller differences for each component than that in the fixed-dimension situation. Thus, Condition (C3)  can be viewed as a version of local alternatives.

 If Condition (C3) does not satisfy, we can replace it with  the following
 condition:
 \begin{itemize}
\item[(C3').]  Assume that, for $k= 1,2,$
% \begin{align*}
%   {(n+m)}^{-1}\texttt{ITR}(\G_{12}, \G_{12})& =o\Big(\iiiint [\bfmu_{1}(t)-\bfmu_{2}(t)]^{\T}\G_{k}(s,s_1)[\bfmu_{1}(t_1)-\bfmu_{2}(t_1)]\\
% 	&\quad\quad\quad \times v(s,t)v(s_1,t_1) \dif s \dif t \dif s_1 \dif t_1\Big).
% \end{align*}
\begin{equation*}
  {(n+m)}^{-1}\texttt{ITR}(\G_{12}, \G_{12}) =o\Big(\texttt{IMD}(\G_{k}, \bfmu_1,\bfmu_2)\Big).
\end{equation*}
 \end{itemize}
 
% It is similar to Condition (3.5) in \cite{chenTwosampleTestHighdimensional2010}.
The condition (C3') means that the Mahanalobis distance between $\bfmu_{1}(t)$ and $\bfmu_{2}(t)$ is a larger order than that of the left side. Thus, it can be viewed as a version of fixed alternatives.
% \textcolor{red}{To facilitate the presentation, we mainly focus on  Condition  (C3) but we also discuss the results under Condition  (C3'). } 
Condition (C4) is typical to establish  the
asymptotic distribution by the martingale central limit theorem \citep{hallMartingaleLimitTheory1980}.  Similar conditions can be found in \cite{chenTwosampleTestHighdimensional2010}.  Additionally, we do not require explicit restrictions on the relative growth rate between
the dimension $p$ and sample size $n$ directly. The only condition on the dimension is   Condition (C4).    For example,   Condition (C4) holds if all the eigenvalues of $\G_{k}(s,t)$ are bounded away from zero and infinity.

\begin{theorem} \label{Theorem2}
Under Conditions (C1)-(C4), as $p, n, m  \rightarrow \infty$,   we have
 $$
 Q(\X, \Y) = \frac{\widehat\MRP(\X, \Y)-\MRP(\X, \Y)}{\sigma_{nm}(\X, \Y)} \stackrel{ {D}}{\longrightarrow} N(0,1),
 $$
 where
 \begin{equation}\label{sigma21-propulation}
\sigma_{nm}^{2}(\X, \Y)=\frac{2}{n\left(n-1\right)} 
\texttt{ITR}(\G_{1}, \G_{1})
+\frac{2}{m\left(m-1\right)}\texttt{ITR}(\G_{2}, \G_{2})
+\frac{4}{nm}\texttt{ITR}(\G_{1}, \G_{2}).
\end{equation}
\end{theorem}
\begin{remark}\label{theo1remark}
  If Condition (C3) is replaced by (C3'), then $\operatorname{Var}(\widehat\MRP(\X, \Y))=\sigma_{nm_2}^2$\\$=4n^{-1} \texttt{IMD}(\G_{1}, \bfmu_1,\bfmu_2)+4m^{-1} \texttt{IMD}(\G_{2}, \bfmu_1,\bfmu_2)
$. Theorem \ref{Theorem2} still holds with $\sigma_{nm}$ replaced by $\sigma_{nm_2}$.
 As under $H_0$, we have $\MRP(\X, \Y)=0$ and Condition (C3) holds, then 
$
{\sigma_{nm}(\X, \Y)}^{-1}\\
{\widehat\MRP(\X, \Y)}\stackrel{D}{\longrightarrow} N(0,1) \text { under }  H_0, \text { as } p,n,m  \rightarrow \infty.
$   
\end{remark}

As the variance $\sigma_{nm}^{2}(\X, \Y)$ defined in \eqref{sigma21-propulation} is unknown, we next establish the estimation of $\sigma_{nm}^{2}(\X, \Y)$. 
%defined in \eqref{sigma21-propulation}.
% It follows from \cite{chenTwosampleTestHighdimensional2010} that 
% $\mathcal{TR}_{11}(s,
 % s_1,t,t_1)=\operatorname{tr}\left\{\G_{1}(s,s_1)\G_{1}(t,t_1)\right\},$ $\mathcal{TR}_{22}(s,s_1,t,t_1)=\operatorname{tr}\left\{\G_{2}(s,s_1)\G_{2}(t,t_1)\right\}$ and  $\mathcal{TR}_{12}(s,s_1,t,t_1)=\operatorname{tr}\left\{\G_{1}(s,s_1)\G_{2}(t,t_1)\right\}$
Define  $\texttt{TR}_{kj}(s,
 s_1,t,t_1)=\operatorname{tr}\left\{\G_{k}(s,s_1)\G_{j}(t,t_1)\right\},k,j=1,2$ and 
we can estimate $\texttt{TR}_{kj}(s,
 s_1,t,t_1),k,j=1,2$
 as follows:
 \begin{equation*}
\begin{aligned}
	\widehat{\texttt{TR}}_{11}(s,s_1,t,t_1)	&=\frac{1}{n(n\!-\!1)}\! \operatorname{tr} \left\{\sum_{j \neq k}^{n}\left[\X_{j}(s)-\overline{\X}_{(j, k)}(s)\right] \X_{ j}^{\T}(s_1)\left[\X_{k}(t)-\overline{\X}_{(j, k)}(t)\right] \X_{k}^{\T}(t_1) \right\},\\
	\widehat{\texttt{TR}}_{22}(s,s_1,t,t_1)&=\frac{1}{m(m\!-\!1)} \!\operatorname{tr}\left\{\sum_{j \neq k}^{m}\big[\Y_{j}(s)-\overline{\Y}_{(j, k)}(s)\big] \Y_{ j}^{\T}(s_1)\big[\Y_{k}(t)-\overline{\Y}_{(j, k)}(t)\big] \Y_{k}^{\T}(t_1)\right\},\\
 \widehat{\texttt{TR}}_{12}(s,s_1,t,t_1)&=\frac{1}{nm} \!\operatorname{tr} \left\{\sum_{j=1}^{n} \sum_{k=1}^{m}\big[\X_{j}(s)-\overline{\X}_{(j)}(s)\big] \X_{ j}^{\T}(s_1)\big[\Y_{k}(t)-\overline{\Y}_{( k)}(t)\big]\Y_{k}^{\T}(t_1)\right \},
\end{aligned}
 \end{equation*}
where $\overline{\X}_{(j, k)}$ (or $\overline{\Y}_{(j, k)}$) is the sample mean after excluding $\X_{j}$  and $\X_{k}$ (or $\Y_{j}$ and $\Y_{k}$), and $\overline{\X}_{(j)}$(or $\overline{\Y}_{(k)}$) is the sample mean without $\X_{j}$ (or $\Y_{k}$). Then, we can obtain
an estimator of $\sigma_{nm}^{2}(\X, \Y)$,  given by
 \begin{equation}\label{sigma21}
\widehat{\sigma}_{nm}^{2}(\X, \Y)=\frac{2}{n\left(n-1\right)} 
\widehat{\texttt{ITR}}({\G}_{1},{\G}_{1})
+\frac{2}{m\left(m-1\right)}\widehat{\texttt{ITR}}({\G}_{2}, {\G}_{2})
+\frac{4}{nm}\widehat{\texttt{ITR}}({\G}_{1},{\G}_{2}),
\end{equation}
where 
$\widehat{\texttt{ITR}}({\G}_{k}, {\G}_{j})=  \iiiint\widehat{\mathcal{TR}}_{kj}(s,
 s_1,t,t_1) v(s,t)v(s_1,t_1) \dif s \dif t \dif s_1 \dif t_1,
 \text{ for } j,k=1,2.$

The next theorem establishes the consistency of the above estimators $\widehat{\texttt{ITR}}({\G}_{k}, {\G}_{j})$.
\begin{theorem} \label{Theorem3}
Under Conditions (C1)-(C4), as $p, n, m  \rightarrow \infty$,  we have
\begin{equation*}
 \frac{\widehat{\texttt{ITR}}({\G}_{k}, {\G}_{j})
}{\texttt{ITR}({\G}_{k},{\G}_{j})  } \stackrel{{P}}{\longrightarrow} 1, \text{   for    }    k,j=1, 2.
\end{equation*}
\end{theorem}

Theorem \ref{Theorem3} suggests  that $\widehat{\sigma}_{nm}^{2}(\X, \Y)$ is a consistent estimator of $\sigma_{nm}^{2}(\X, \Y)$.  Note that  $\MRP(\X, \Y)=0$ under $H_0:\bfmu_{1}(t)=\bfmu_{2}(t)$. Then, by Theorems \ref{Theorem2} and \ref{Theorem3}, we can obtain the following result.

\begin{Cor} \label{Cor-2}
Under Conditions (C1)-(C4) and $H_0:\bfmu_{1}(t)=\bfmu_{2}(t)$,   we have
 $$
 \widehat Q_{n}(\X, \Y)=\frac{\widehat\MRP(\X, \Y)}{\widehat{\sigma}_{nm}(\X, \Y)} \stackrel{{D}}{\longrightarrow} N(0,1),\,\text{as }p, n, m  \rightarrow \infty.
$$
\end{Cor}

According to Corollary \ref{Cor-2},  the proposed MRP-based  test
rejects $H_{0}$ at a significance level of  $\alpha$  if and only if
 $
\widehat Q_{n}(\X, \Y) > z_{\alpha},
$
where $z_{\alpha}$ is the upper $\alpha$ quantile of $N(0,1)$. %However, it is worth noting that $\widehat Q_{n}(\X, \Y)$ is infeasible in practice due to the unobservability of the true curves $\{\X_{i}(t),t\in[0,1]\}_{i=1}^n$ and $\{\Y_{i}(t),t\in[0,1]\}_{i=1}^m$. We consider the problem in the next subsection.

We then investigate the power of the proposed  MRP  test. Denote
 $$
\Delta_{nm}(\bfmu_{1}-\bfmu_{2}, \G_{1},\G_{2})=\frac{(n+m) \tau(1-\tau)\MRP(\X,\Y)}{\sqrt{2 \texttt{ITR}({\G}_{\tau},{\G}_{\tau})}},
$$
where $\G_\tau(s,t)=(1-\tau) \G_{1}(s,t)+\tau\G_{2}(s,t)$.

\begin{theorem} \label{theorem-power}
Under Conditions (C1)-(C4) and $H_1:\bfmu_{1}(t)\neq\bfmu_{2}(t)$,    we have
 \begin{align*}
 \lim_{p, n, m  \rightarrow \infty}\operatorname{P}\{ \widehat Q_{n}(\X, \Y)> z_{\alpha}\}
 =\lim_{p, n, m  \rightarrow \infty}\Phi\left(-z_{\alpha}+\Delta_{nm}(\bfmu_{1}-\bfmu_{2}, \G_{1},\G_{2})\right),
\end{align*}
where  $\Phi$ is the standard normal distribution function.
\end{theorem}
%  The power under Condition (C3')  is
% \begin{align*}
% g_2\left(\MRP(\X,\Y)\right)=\Phi\left(-\frac{\sigma_{nm}}{\sigma_{nm_2}} z_{\alpha}+\frac{\MRP(\X,\Y)}{\sigma_{nm_2}}\right)=\Phi\left(\frac{\DPI(\X,\Y)}{\sigma_{nm_2}}\right)
% \end{align*}
% as $\sigma_{nm} / \sigma_{nm_2} \rightarrow 0$ under condition Condition (C3'). Substitute the expression for $\sigma_{nm_2}$, and we have
% \begin{align}
% g_2\left(\MRP(\X,\Y)\right)=\Phi\left(\frac{\sqrt{(n+m) \tau(1-\tau)}\MRP(\X,\Y)}{2\sqrt{\texttt{IMD}(\G_{\tau}, \bfmu_1,\bfmu_2)}}\right)
% \label{power2}
% \end{align}
\begin{remark}
   The power under Condition (C3')  is
\begin{equation}
 \lim_{p, n, m  \rightarrow \infty}\operatorname{P}\{ \widehat Q_{n}(\X, \Y)> z_{\alpha}\}
 =\lim_{p, n, m  \rightarrow \infty}\Phi\left(
\Delta_{nm2}(\bfmu_{1}-\bfmu_{2}, \G_{1},\G_{2})\right),
\label{power2}
\end{equation}
where $\Delta_{nm2}(\bfmu_{1}-\bfmu_{2}, \G_{1},\G_{2})={\sqrt{(n+m) \tau(1-\tau)}\MRP(\X,\Y)}/{\sqrt{4\texttt{IMD}(\G_{\tau}, \bfmu_1,\bfmu_2)}}$.
This together with Theorem \ref{theorem-power}  indicate that $\widehat Q_{n}(\X, \Y)$  has non-trivial power under    the alternative  hypotheses as long as $\Delta_{nm}(\bfmu_{1}-\bfmu_{2}, \G_{1},\G_{2})$
does not vanish to 0 as  $p, n, m  \rightarrow \infty$. 
\end{remark}

 \subsection{Influence of Function Reconstruction}\label{Reconstruction}
In practice, the continuous functional data is only observed on a series of grid points. Therefore,  we need to reconstruct the true underlying function by these finite observations to compute the above test statistic.
% For notational convenience,  
% we here assume the number of the grid points are equal 
Denote the discrete observations for the  $k$th dimension of  the $i$th sample by  $ \mathcal{I} X_{ik} =\left(X_{ik}\left(t_{1}\right), \ldots, X_{ik}\left(t_{N_{ik}}\right)\right)^{\T}$, $k=1,\ldots,p,$ $i=1,\ldots,n,$ where $\mathcal{I}
% : {X}_{ik} \rightarrow \mathbb{R}^{N_{ik}\times 1}
$ 
is a discretization map.  That is,  
% the observed samples 
% are $\{\mathcal{I} \X_{i}\}_{i=1}^{n}=\{\left(\mathcal{I} X_{i,1}, \ldots, \mathcal{I} X_{i,p}\right)^{\T}\}_{i=1}^{n}$.
 $\{X_{ik}(t), t\in[0,1]\}_{k=1\, i=1}^{p \quad n}$,  each dimension of the random samples  $\{\X(t), t\in[0,1]\}$, can be observed at $\left\{t_{ik}, \ldots, t_{N_{ik}}\right\}$ 
 and  the observed samples 
are $\{\mathcal{I}X_{ik}\}_{k=1\, i=1}^{p \quad n}$.
 
Using these discrete observations $\{\mathcal{I} X_{ik}\}_{k=1\, i=1}^{p \quad n}$ and $\{\mathcal{I} Y_{ik}\}_{k=1\, i=1}^{p \quad m}$, we can reconstruct the random functions by some commonly used smoothing methods, such as kernel smoothing or spline smoothing and the reconstructed samples are $\{\widehat X_{ik}=\mathcal{R}(\mathcal{I} X_{ik})\}_{j=1\, i=1}^{p \quad n}$ and $\{\widehat Y_{ik}=\mathcal{R} (\mathcal{I} Y_{ik})\}_{k=1\, i=1}^{p \quad m}$, where $\mathcal{R}$ is the reconstruction map.
  % are  $\{\widehat \X_{i}=\mathcal{R}(\mathcal{I} \X_{i})\}_{i=1}^{n}$ and $\{\widehat \Y_i=\mathcal{R} (\mathcal{I} \Y_{i})\}_{i=1}^{m}.$

To ensure the effectiveness of the test, the reconstructed functional data need to satisfy the following condition:
\begin{itemize}
\item[(C5).]
	 As $p, n, m  \rightarrow \infty$,
	 \begin{align}
	&\int_{0}^{1}\Big \{\E|\widehat X_{k}(t)-X_{k}(t) |+\E |\widehat Y_{k}(t)-Y_{k}(t) |\Big\}\dif t \rightarrow 0,\label{as41}\\
& \sum_{k=1}^{p}\int_{0}^{1}\Big\{\E |\widehat X_{k}(t)-X_{k}(t) |+ \E |\widehat Y_{k}(t)-Y_{k}(t) |\Big\}\dif t =o(\sigma_{nm}(\X,\Y)).\label{as42}
\end{align}
\end{itemize}
In our numerical studies,  the B-spline method is used to reconstruct the random functions with the number of knots $K_{N_{ik}}=N_{ik}^{r},  k=1,\ldots,p, i=1,\ldots,n$ for some $0<r\leq 1$. 
% and $N=\min_{1\leq i \leq n}{N_i}$.
We provide some remarks on Condition (C5).
\begin{remark}
 If  each $X_{k}(t)$ and $Y_{k}(t)$, $k=1,\ldots,p$,  has twice-order  bounded continuous derivatives on $[0, 1]$, then $\sup _{t}\E\{|\widehat X_{k}(t)-X_{k}(t) |\}=O(K_{N_{k}}^{-2})$. Then provided that $\sup _{1\leq k\leq p}{N_{k}^{-2r}}\rightarrow 0$,    \eqref{as41} is satisfied.
 % \item[(ii)] Under the above assumption in (i),
(i) When $\G_1(s,t)=\G_2(s,t)=c(s,t)\I_p$,  where $c(s,t)$ is  uniformly bounded in $[0,1]^2$,  we have $\sigma_{nm}(\X,\Y)=O(p^{1/2}/n)$. Thus,  if 	$n\sum_{k=1}^{p}{N_{k}^{-2r}}p^{-1/2}\rightarrow 0$, 
% i.e., $N\gg {n}^{\frac{1}{2r}}p^{\frac{1}{4r}}$, 
\eqref{as42} holds. 
 % \item[(iii)] Under the above assumption in (i),  
(ii) When $\sup _{s,s_1,t,t_1}|\operatorname{tr}\left\{\G_{i}(s,s_1)\G_{j}(t,t_1)\right\}|=O(p^{1+a}),$  $i,j=1,2,$ for some $0\leq a <1$,   we have that $\sigma_{nm}(\X,\Y)=O(p^{(1+a)/2}/n)$. Thus, 
% if 	$N\gg {n}^{\frac{1}{2r}}p^{\frac{1-a}{4r}}$, 
if $n\sum_{k=1}^{p}{N_{k}^{-2r}}p^{-(1+a)/2}\rightarrow 0$, \eqref{as42} holds.
% \textcolor{red}{Note that the observations of the curves are allowed to be not that dense for finite dimensions among $p$.}
\end{remark}

Let  $\widehat\MRP(\widehat\X, \widehat\Y)$ and $\widehat{\sigma}_{nm}^{2}(\widehat\X, \widehat\Y)$
be the reconstructed versions of $\widehat\MRP(\X, \Y)$ and $\widehat{\sigma}_{nm}^{2}(\X, \Y)$ in \eqref{sigma21}, through replacing  the underlying curves  $\{\X_i\}_{i=1}^{n}$ and $\{ \Y_i\}_{i=1}^{m}$  by  the reconstructed samples $\{\widehat \X_i\}_{i=1}^{n}$ and $\{\widehat \Y_i\}_{i=1}^{m}$. The following Theorem \ref{Theorem4} establishes the asymptotic normality of $\widehat\MRP(\widehat\X, \widehat\Y)$.

\begin{theorem} \label{Theorem4}
Under Conditions (C1)-(C5), as $p, n, m  \rightarrow \infty$,   we have
 	$$	\frac{\widehat\MRP(\widehat\X, \widehat\Y)-\MRP(\X, \Y)}{\widehat\sigma_{nm}(\widehat\X, \widehat\Y)} \stackrel{ {D}}{\longrightarrow} N(0,1).
	$$
\end{theorem}
By Theorem \ref{Theorem4}, we can obtain the  asymptotic null distribution of $\widehat\MRP(\widehat\X, \widehat\Y)$ as follows.
\begin{Cor} \label{proposition2}
Assume that Conditions (C1)-(C5) hold.  Under  $H_0:\bfmu_{1}(t)=\bfmu_{2}(t)$,  we have
 \begin{equation}\label{Eqn:Q}
	\widehat Q_{n}(\widehat\X, \widehat\Y)=\frac{\widehat \MRP(\widehat\X, \widehat\Y) }{\widehat{\sigma}_{nm}(\widehat\X, \widehat\Y)} \stackrel{ {D}}{\longrightarrow} N(0,1), \text{ as }p, n, m  \rightarrow \infty.
 \end{equation}
\end{Cor}

By the asymptotic distribution in  Corollary \ref{proposition2}, we can obtain the critical value of $\widehat Q_{n}(\widehat\X, \widehat\Y)$ and test problem \eqref{fun-twotest}.  Specifically, the  proposed test with a nominal $\alpha$
level of significance rejects $H_{0}$ if $\widehat Q_{n}(\widehat\X, \widehat\Y)>z_{\alpha}$.

%\section{Power Analysis of MRP Test}\label{Power-Analysis}

By Theorem \ref{theorem-power} and Theorem \ref{Theorem4}, 
we can obtain the asymptotic power of the reconstructed test statistic $\widehat\MRP(\widehat\X, \widehat\Y)$ as follows. 
\begin{Cor} \label{proposition2-2}
Assume that Conditions (C1)-(C5) hold.   Under $H_1:\bfmu_{1}(t)\neq\bfmu_{2}(t)$,    we have
 \begin{equation*}
 \lim_{p, n, m  \rightarrow \infty}\operatorname{P}\{ \widehat Q_{n}(\widehat\X, \widehat\Y)> z_{\alpha}\}
 =\lim_{p, n, m  \rightarrow \infty}\Phi\left(-z_{\alpha}+\Delta_{nm}(\bfmu_{1}-\bfmu_{2}, \G_{1},\G_{2})\right).
\end{equation*}
\end{Cor}

 Corollary \ref{proposition2-2} suggests that the power of our final  test statistic $\widehat\MRP(\widehat\X, \widehat\Y)$  
%  is   non-trivial power under   the alternative  hypotheses as long as $\Delta_{nm}(\bfmu_{1}-\bfmu_{2}, \G_{1},\G_{2})$
% does not vanish to 0 as  $p, n, m  \rightarrow \infty$ and 
 can  converge  to 1  as long as $\Delta_{nm}(\bfmu_{1}-\bfmu_{2}, \G_{1},\G_{2})\rightarrow \infty$
 as  $p, n, m  \rightarrow \infty$.

\section{Monte Carlo Simulations}\label{Simulation}
In this section, we conduct Monte Carlo simulations to assess the finite sample performance of the proposed test based on  $\widehat Q_{n}(\widehat\X, \widehat\Y)$. For comparison, we also consider the test method proposed by \cite{qiuTwosampleTestsMultivariate2021} (QCZ test), which includes two global test statistics: $T_{n}$ by integrating a pointwise Hotelling $T^2$ type test statistic,  and $T_{n,max}$ by taking its maximum value over $[0,1]$.
Three simulation models for $\X$ and  $\Y$ are considered.  The first two models
are from functional moving average models,  which provide a general dependent structure; the third one has sparse alternative hypotheses to illustrate the performance of our
test under sparsity.  In the simulations, we repeat each experiment 400 times and report the empirical size and power for the tests.  The curves $\widehat\X, \widehat\Y$ are reconstructed by B-splines, which can be expressed by the basis functions and the corresponding coefficients. Then the integrals in the test statistic $\widehat Q_{n}(\widehat\X, \widehat\Y)$ can be 
calculated by integrating the basis functions together with the estimated coefficients.
 \subsection{Simulation I}\label{simulation1}
 In Simulation I, we generate data from  the following functional moving average model:
 \begin{equation}\label{simxynew}
\begin{aligned}
X_{i k}(t)&=\phi_1(t) Z_{1i (k-1)}(t) +
\phi_2( t) Z_{1i (k-2)}(t)+\cdots+\phi_p(t) Z_{1i (k-p)}(t) +\mu_{1 k}(t), t\in[0,1];\\
Y_{j k}(t)&=\phi_1(t) Z_{2j (k-1)}(t) +\phi_2( t) Z_{2j (k-2)}(t)
+\cdots+\phi_p(t) Z_{2j (k-p)}(t) +\mu_{2 k}(t), t\in[0,1],
\end{aligned}
\end{equation}
for $i=1,  \ldots, n$, $j=1,  \ldots, m$ and $k=1,  \ldots, p$,   where $\{Z_{1i k}(t), t\in[0,1]\}_{i=1}^n$ and $\{Z_{2j k}(t), t\in[0,1]\}_{j=1}^m$ are  independently generated from  a standard Brownian motion defined on $[0,1]$. The coefficient functions $\phi_k(t)$s are  defined as $\phi_k(t)= {c_{k}}\exp \left\{{-t^{2}} /{ 2}\right\}/\sqrt{0.7468}$, where  $c_{k}$s are scalars. In the above settings, it can be seen that  $ \int_{[0,1]}\phi_k^{2}(t) \mathrm{d} t =c_{k}^{2}$.
Without loss of generality,  we set  $\mu_{2 k}(t)= 0$.

We consider the following two kinds of dependence structures
for $X_{i k}(t)$ and $Y_{j k}(t)$.
% by the same way as
%  \citet{chenTwosampleTestHighdimensional2010}:
\begin{itemize}
\item[Case I:]   $c_{1}=0.5$, $c_{2}=0.3$,  and $c_{k}=0$ for $k\geq 3$.
\item[Case II:]  $c_{k}\sim U[0.1,0.6]$ for $k=1, \ldots,p$.
\end{itemize}
 Note that Case I  presents  a weaker dependence through a ``two-dependence" functional moving average structure, where $X_{i k_{1}}(t)$ and $X_{i k_{2}}(t)$ (also $Y_{j k_{1}}(t)$ and $Y_{j k_{2}}(t)$) are dependent only if $\left|k_{1}-k_{2}\right| \leq 2$. In  Case I,  the mean function  of $\X$  is defined  by
 \begin{equation}
	\mu_{1 k,\I}(t)=t\log(k/p+1)+\sin^2(2\pi t+k/p)+\cos(2\pi t+k/p), t\in[0,1],
		  \label{simmu1new}
\end{equation}
for $k=1, \ldots,p.$
Case II presents a strong dependence  through a ``full-dependence" functional moving average structure, where $X_{i k_{1}}(t)$ and $X_{i k_{2}}(t)$ (also $Y_{j k_{1}}(t)$ and $Y_{j k_{2}}(t)$) are dependent if $\left|k_{1}-k_{2}\right| \leq p$. In Case II,  we define  the mean function of $\X$   by   	
\begin{equation}
	\mu_{1 k,\I\I}(t)=\{t\log(k/p+1)+\sin^2(2\pi t+k/p)+\cos(2\pi t+k/p)\}\log(p^2/2), t\in[0,1],
	\label{simmu11}
\end{equation}
for $k=1, \ldots,p.$ In the following Figure \ref{mux}, we  display  the above mean functions
$\bfmu_{1,\I}(t)=(\mu_{1 1,\I}(t),\ldots,\mu_{1 p,\I}(t) )^T$  and  $\bfmu_{1,\I\I}(t)=(\mu_{1 1,\I\I}(t),\ldots,\mu_{1 p,\I\I}(t) )^T$ with  $p=100$.

\begin{figure}[!ht]
\subfigcapskip=-20pt
	\centering
	\subfigure[\scriptsize  $\mu_{1,\I}(t)$ for  Case I ]{
		\includegraphics[width=0.45\textwidth]{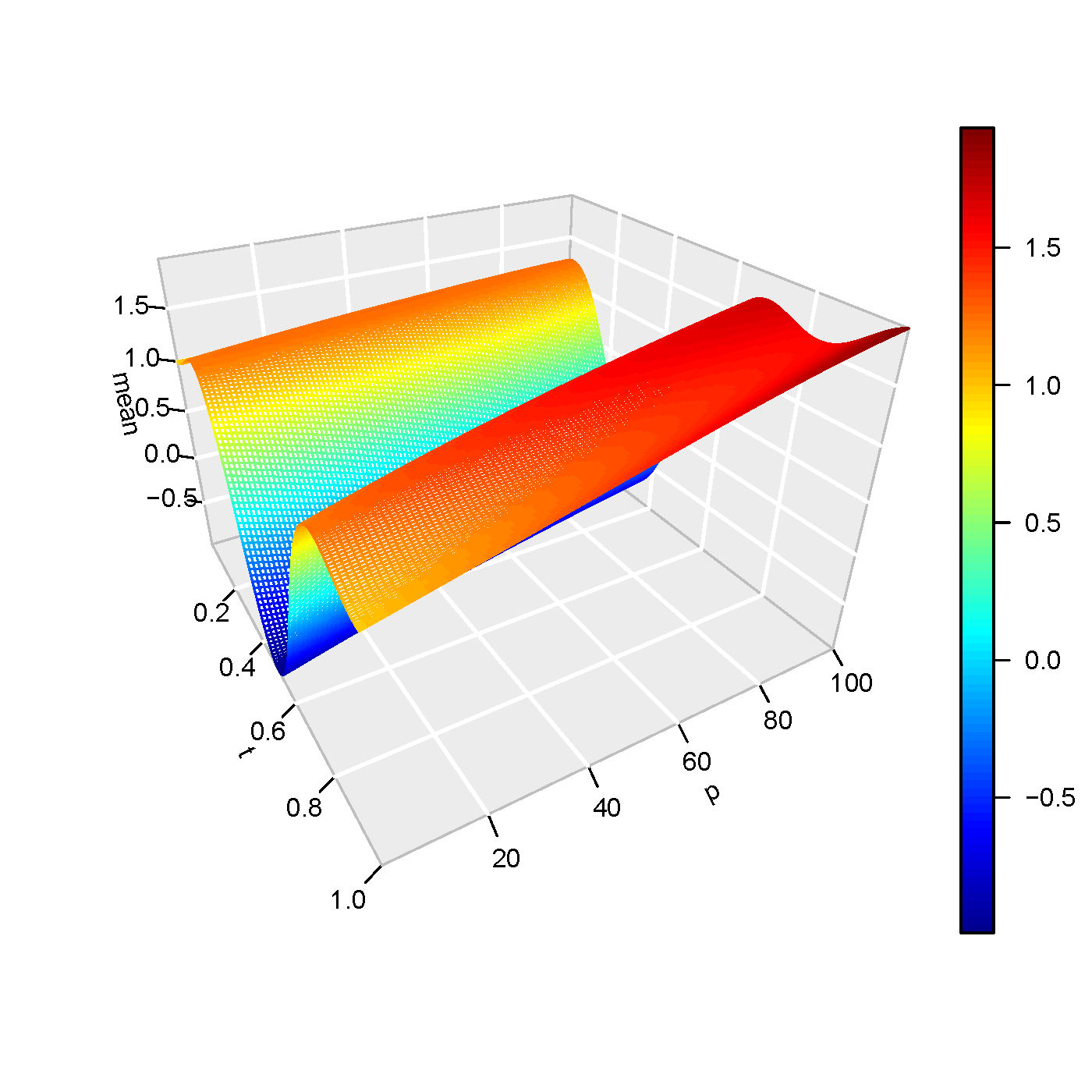}}
	\subfigure[\scriptsize   $\mu_{1,\I\I}(t)$ for  {Case II}]{
		\includegraphics[width=0.45\textwidth]{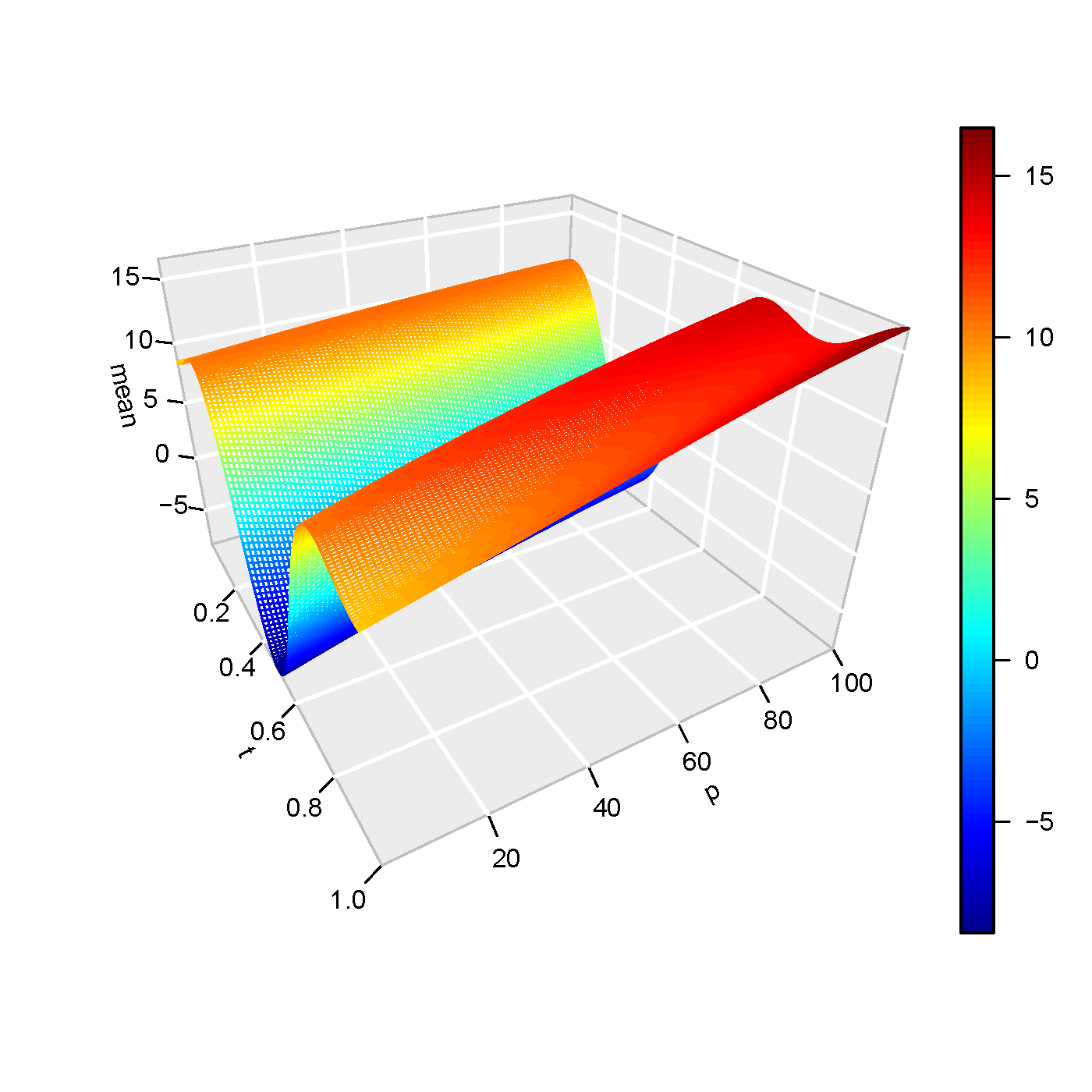}}
\caption{ The mean functions
$\bfmu_{1,\I}(t)$ in \eqref{simmu1new} and  $\bfmu_{1,\I\I}(t)$ in \eqref{simmu11} with the dimension of functional data $p=100$  for  model \eqref{simxynew} in Simulation I . }
 \label{mux}
\end{figure}

\begin{figure}[!htbp]
\subfigcapskip=-5pt
	\centering
	\subfigure[$n=m=25, p=20$]{
		\includegraphics[width=0.31\textwidth]{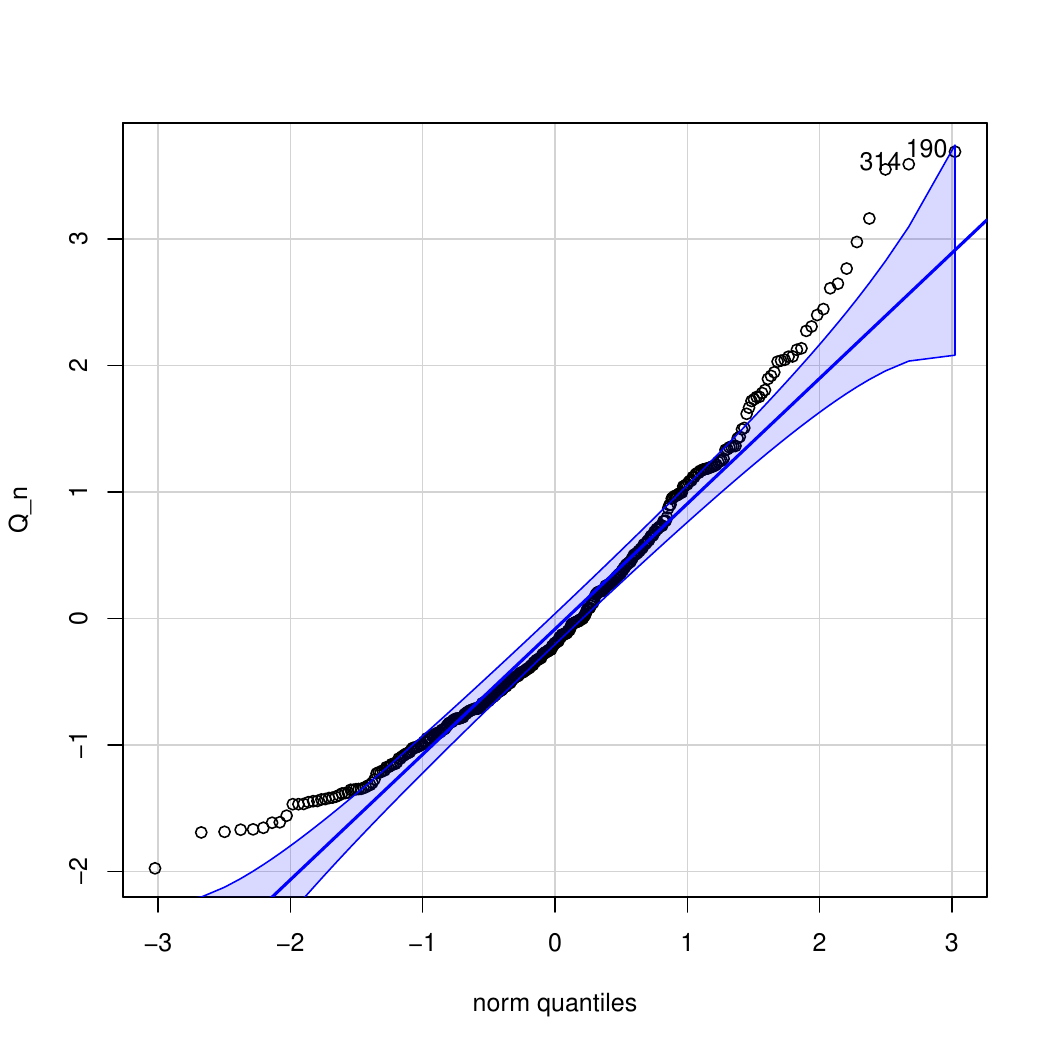}}
   \setcounter{subfigure}{4} 
  \subfigure[$n=m=40, p=20$]{
		\includegraphics[width=0.31\textwidth]{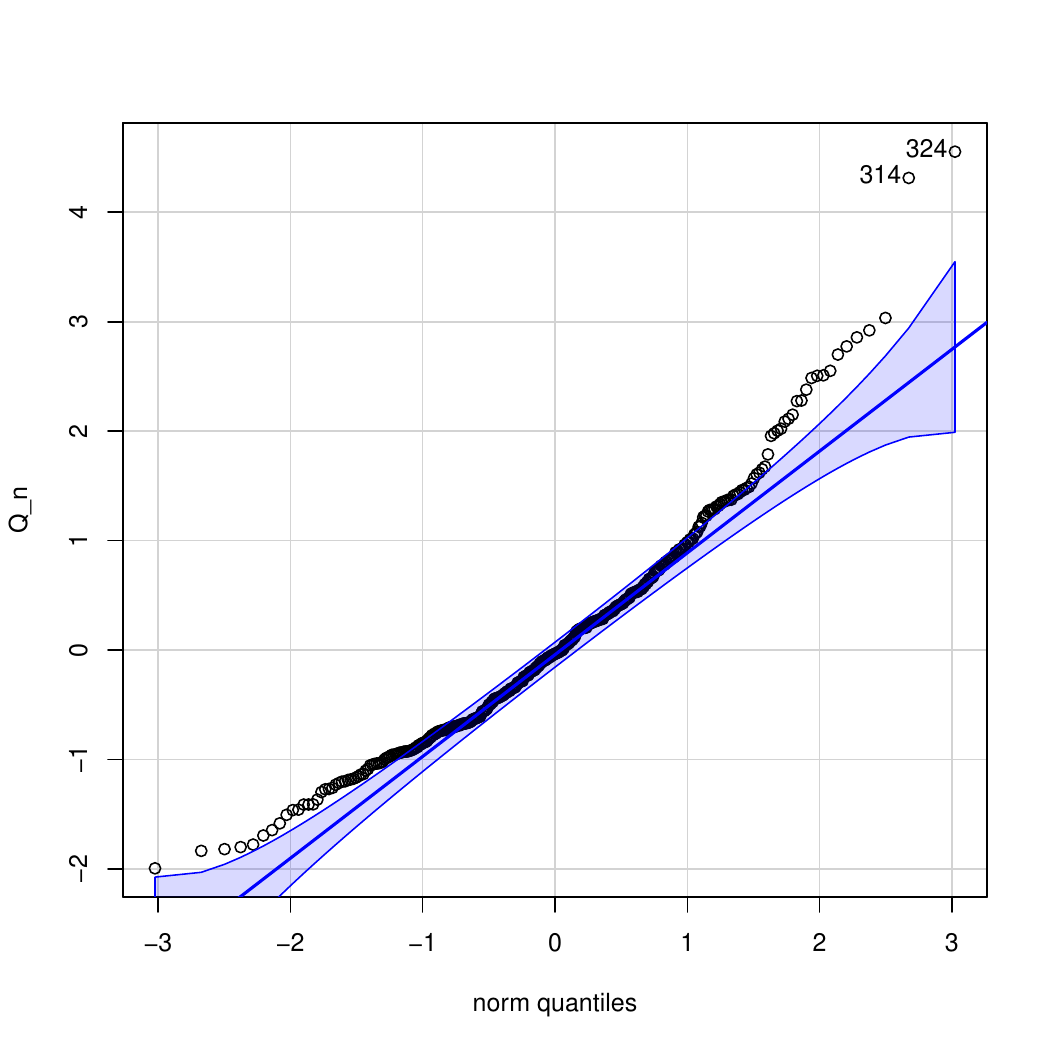}}\\
   \setcounter{subfigure}{1} 
	\subfigure[$n=m=25, p=50$]{
		\includegraphics[width=0.31\textwidth]{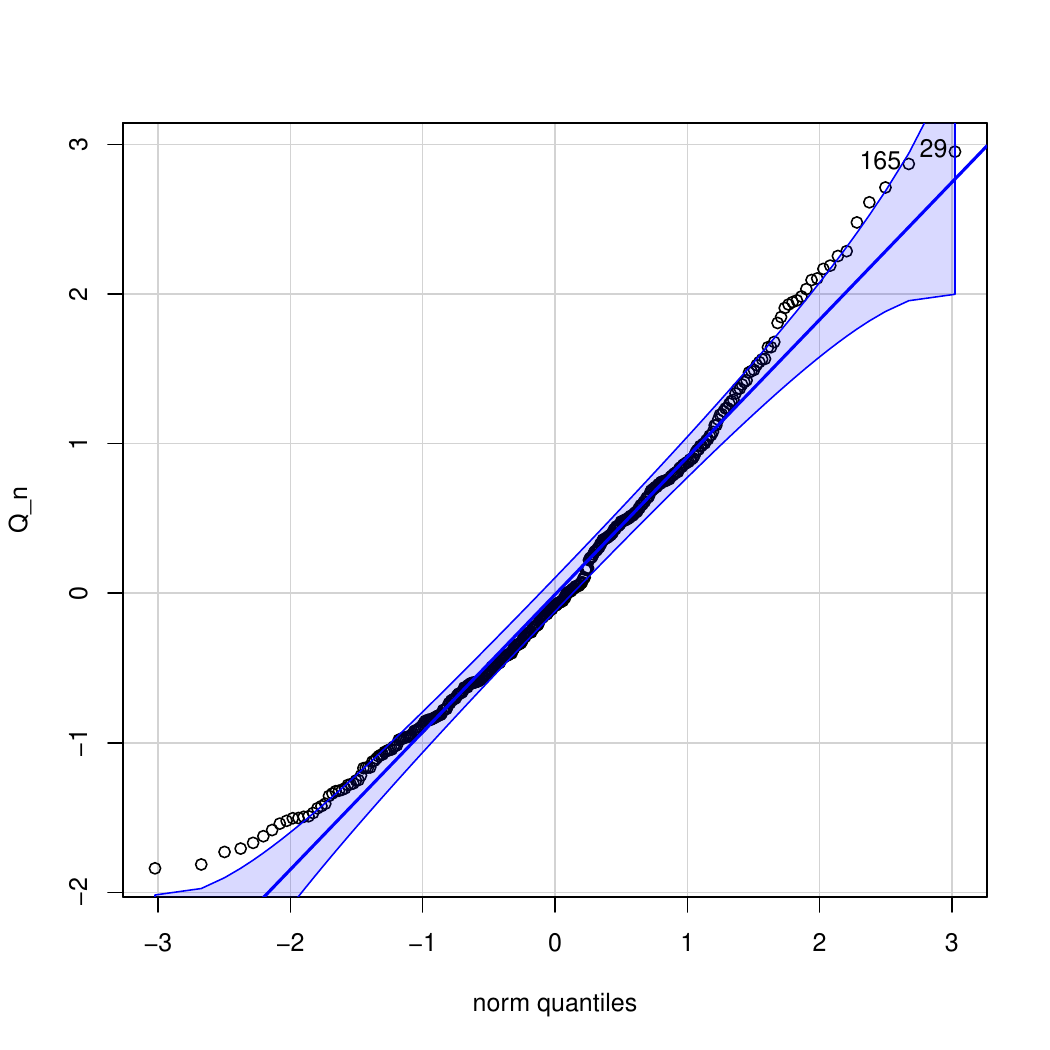}}
   \setcounter{subfigure}{5} 
  	\subfigure[$n=m=40, p=50$]{
		\includegraphics[width=0.31\textwidth]{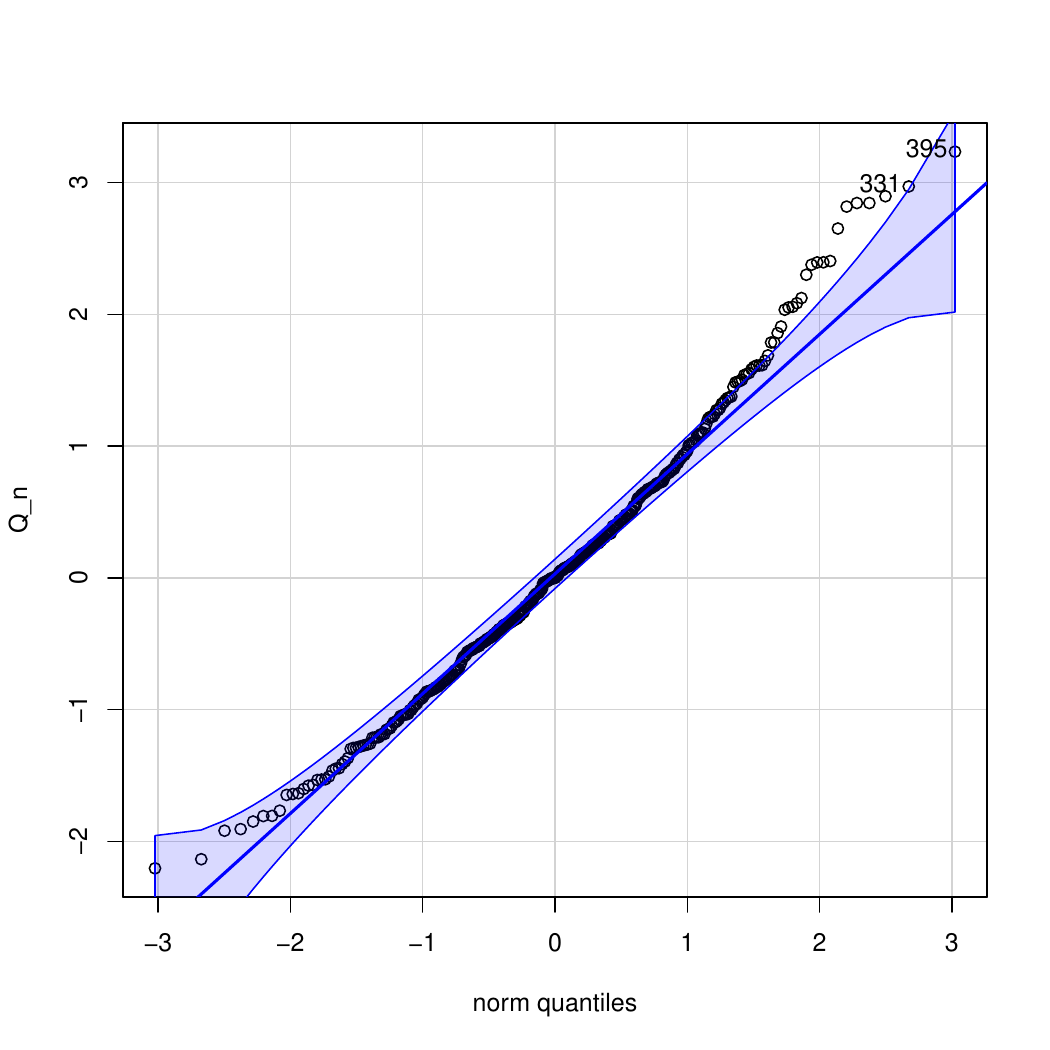}}\\
   \setcounter{subfigure}{2} 
  	\subfigure[$n=m=25, p=100$]{
		\includegraphics[width=0.31\textwidth]{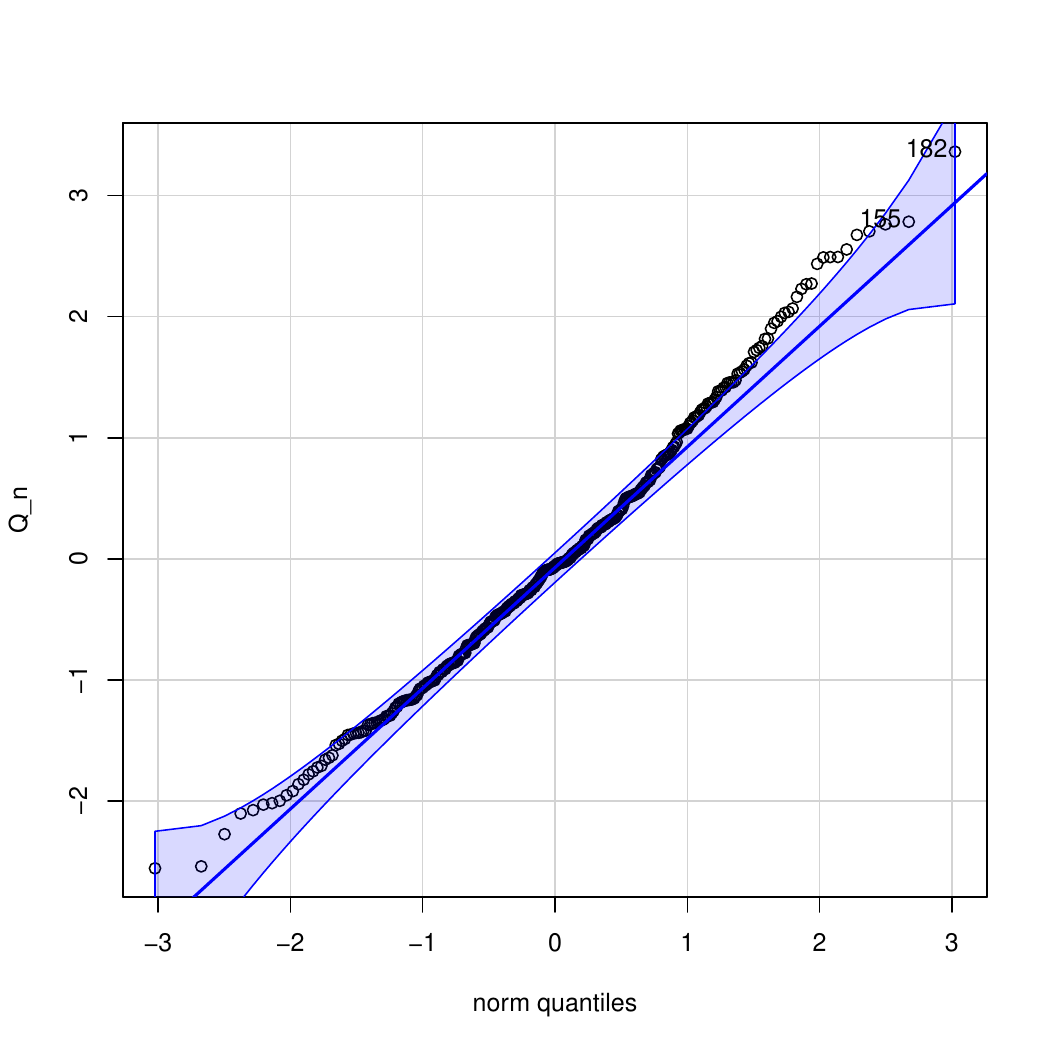}}
   \setcounter{subfigure}{6} 
    	\subfigure[$n=m=40, p=100$]{
		\includegraphics[width=0.31\textwidth]{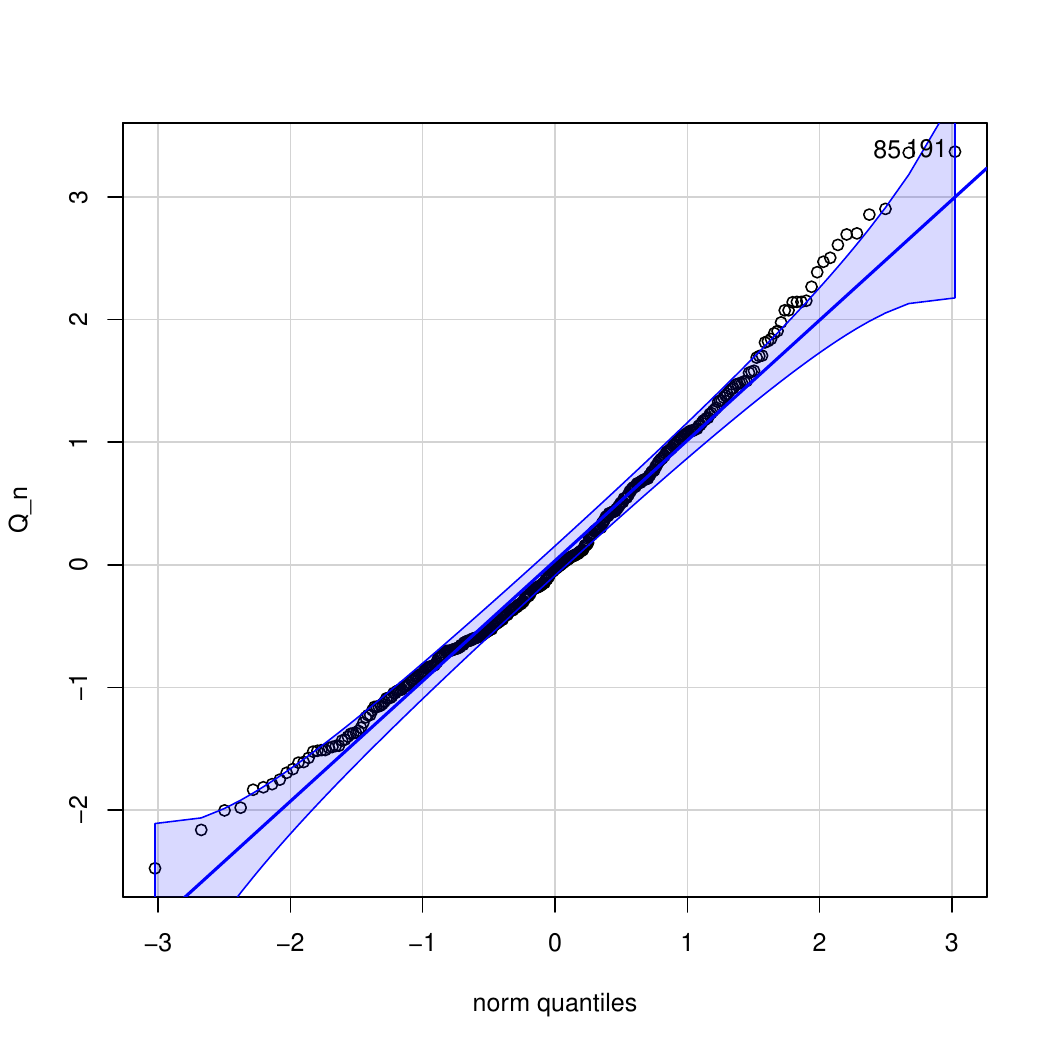}}\\
   \setcounter{subfigure}{3} 
	\subfigure[$n=m=25, p=200$]{
		\includegraphics[width=0.31\textwidth]{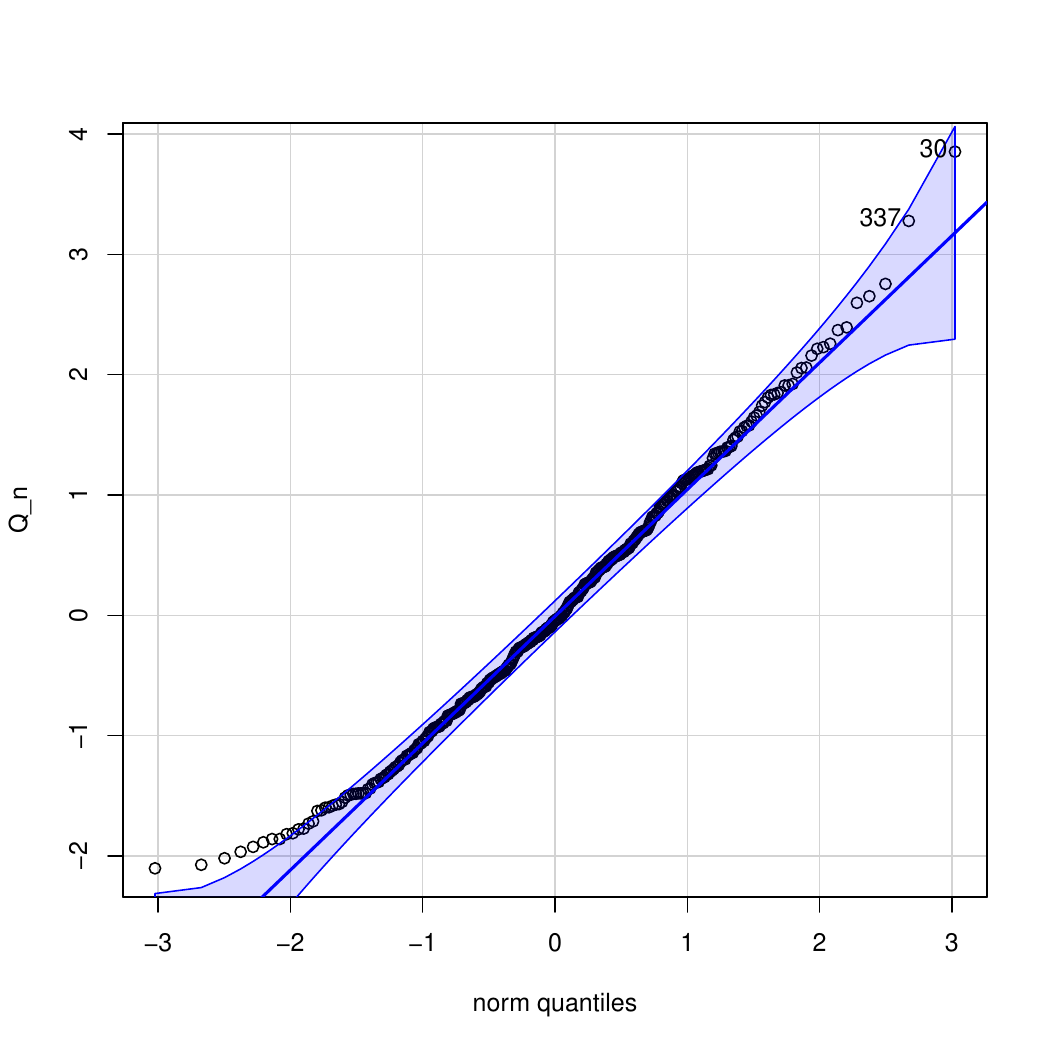}}
   \setcounter{subfigure}{7} 
	\subfigure[$n=m=40, p=200$]{
		\includegraphics[width=0.31\textwidth]{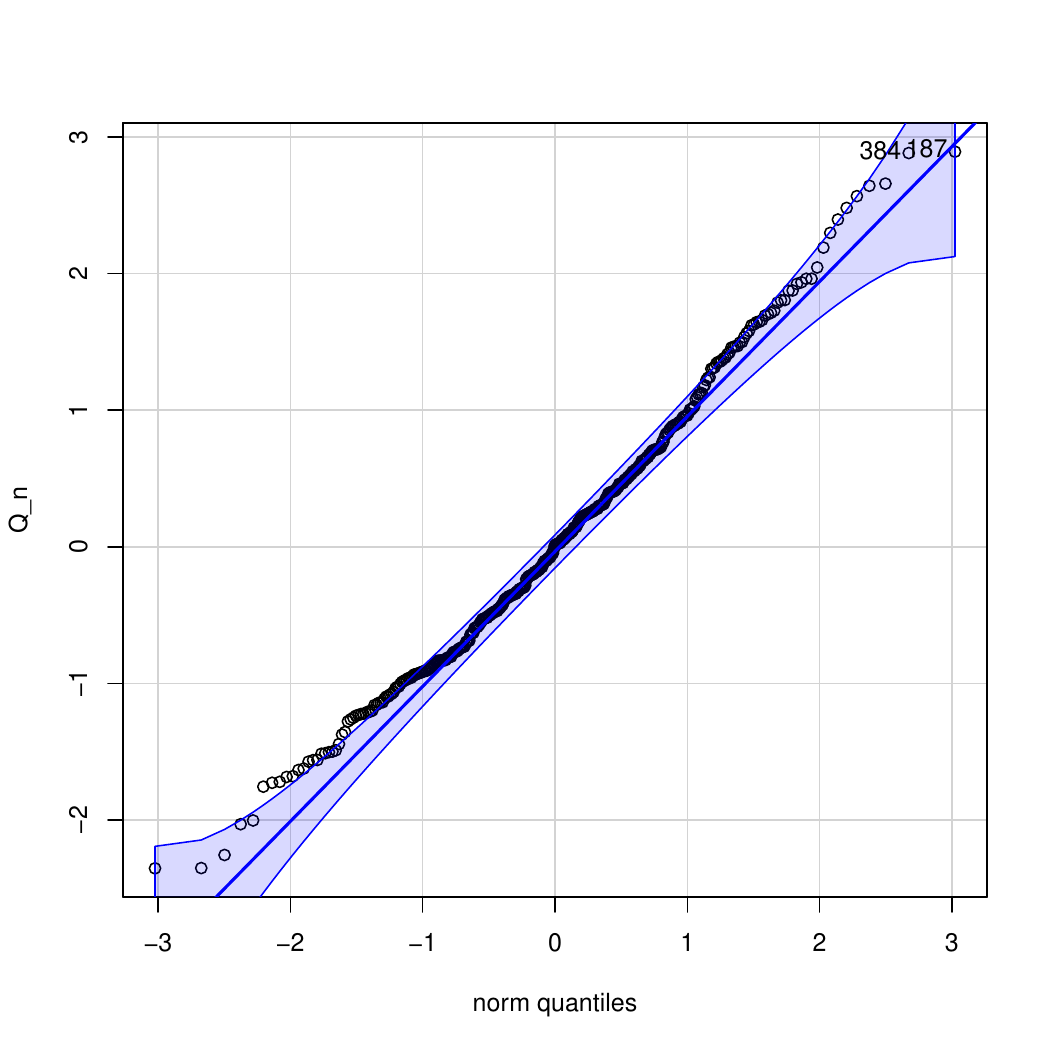}}
	\caption{Q-Q plots of the multi-resolution projection test statistic $\widehat Q_{n}(\widehat\X, \widehat\Y)$ defined in (\ref{Eqn:Q}) when $n=m=25,40$ and $p=20,50,100,200$ in {Case I} for model \eqref{simxynew} in Simulation I, where $n$ and $m$ are the number of random samples for $\X$ and $\Y$ and $p$ is the dimension of $\X$ and $\Y$. }
	\label{qqplot}
\end{figure}

 In the following experiments,  we first choose the projection process  $\gamma$ in \eqref{equvlent-dpi}   to be a stationary Ornstein-Uhlenbeck process, satisfying  $v(s, t) \propto a^{-1} \exp (-a|s-t|)$ with $a=1$.
To  confirm  the asymptotic  distribution of  $\widehat Q_{n}(\widehat\X, \widehat\Y)$ presented in Theorem \ref{Theorem4}, Figure \ref{qqplot} displays the QQ plots in {Case I} for model \eqref{simxynew} under $n=m=25,40$ and $p=20,50,100,200$. From Figure \ref{qqplot}, it can be seen that when the dimension $p$ increases, the asymptotic distribution of $\widehat Q_{n}(\widehat\X, \widehat\Y)$  is close to the standard normal distribution, which implies that the asymptotic distribution of  $\widehat Q_{n}(\widehat\X, \widehat\Y)$ can be well approximated by the normal distribution.

% \begin{figure}[!ht]
% \subfigcapskip=-5pt
% 	\centering
% 	\subfigure[$n=m=25, p=20$]{
% 		\includegraphics[width=0.23\textwidth]{fig2/qq2520n.pdf}}
% 	\subfigure[$n=m=25, p=50$]{
% 		\includegraphics[width=0.23\textwidth]{fig2/qq2550n.pdf}}
%   	\subfigure[$n=m=25, p=100$]{
% 		\includegraphics[width=0.23\textwidth]{fig2/qq25100n.pdf}}
% 	\subfigure[$n=m=25, p=200$]{
% 		\includegraphics[width=0.23\textwidth]{fig2/qq25200n.pdf}}\\
% 	\subfigure[$n=m=40, p=20$]{
% 		\includegraphics[width=0.23\textwidth]{fig2/qq4020n.pdf}}
% 	\subfigure[$n=m=40, p=50$]{
% 		\includegraphics[width=0.23\textwidth]{fig2/qq4050n.pdf}}
%   	\subfigure[$n=m=40, p=100$]{
% 		\includegraphics[width=0.23\textwidth]{fig2/qq40100n.pdf}}
% 	\subfigure[$n=m=40, p=200$]{
% 		\includegraphics[width=0.23\textwidth]{fig2/qq40200n.pdf}}
% 	\caption{Q-Q plots of   $\widehat Q_{n}(\widehat\X, \widehat\Y)$  under $n=m=25,40$ and $p=20,50,100,200$ in {Case I} for model \eqref{simxynew} in Simulation I. }
% 	\label{qqplot}
% \end{figure}

To examine the performance of the size and power of our method, we choose $\bfmu_{1}(t)=(\mu_{1 1}(t), \ldots,\mu_{1 p}(t) )^T$ in the same fashion
as \citet{chenTwosampleTestHighdimensional2010}.
Specifically, the percentages of dimensions such that $\mu_{1 k}(t) = \mu_{2 k}(t)$ for   $k = 1, \ldots,p$ are chosen to be 0\%,   50\%, 75\%, 90\%, 95\% and 98\% and 100\%, respectively. Thus, we can examine the performance of size by experimenting with 100\%,  and the performance of power by experimenting
with the other percentages.
% (Jicai: Please check the accuracy of this expression)

\begin{table}[!htbp]
\begin{center}
\setlength{\tabcolsep}{1.5mm}{
\begin{threeparttable}[b]
\caption{The empirical size and power for the proposed {MRP} test and QCZ test \citep{qiuTwosampleTestsMultivariate2021}  in  {Case I} for model \eqref{simxynew} at the significance level $\alpha=0.05$ in Simulation I for different $(n,m)$ and $p$,
where $n$ and $m$ are the number of random samples for $\X$ and $\Y$, $p$ is the dimension of $\X$ and $\Y$, and the percentages 0\%, 50\%, 75\%, 90\%, 95\%  98\% and 100\% denote the percentages of the dimensions such that $\mu_{1 k}(t) = \mu_{2 k}(t)$. }\label{tab3.1.1}
\footnotesize
\begin{tabular}{cccccccccccccc}
              \hline\hline
&&& \multicolumn{7}{c}{MRP test}&&&\multicolumn{2}{c}{QCZ test}    \\
\cline{4-10}                    \cline{13-14}\\  [-0.15cm]
$(n,m)$& $p$ && $0\%$  &50\%&75\%&90\%&95\% &98\%&100\%(size)&&&$T_{n}$(size)&$T_{n,max}$(size)\\[0.1cm]
					\hline
\multirow{4}{*}{$(25,25)$}
					&$20$ &&1&1&0.985 &0.495  &0.205  &0.050 &0.050 &&&0.730&0.050\\ [0.1cm]
					&$50$ &&1&1&0.995 &0.785  &0.435  &0.135 &0.050 &&&1    &0.560\\ [0.1cm]
					&$100$&&1&1&1     &0.953  &0.488  &0.175 &0.058 &&&1    &0.0\\ [0.1cm]
					&$200$&&1&1&1     &0.993  &0.725  &0.248 &0.056 &&&1    &0.0 \\[0.1cm] 											\hline
\multirow{4}{*}{$(40,40)$}
					&$20$ &&1&1&0.995 &0.785    &0.325  &0.053 &0.053 &&&0.310&0.070\\[0.1cm]
					&$50$ &&1&1&1     &0.950    &0.695  &0.215 &0.048 &&&1    &0.030\\ [0.1cm]
					&$100$&&1&1&1     &1        &0.855  &0.295 &0.054 &&&1    &0.0\\ [0.1cm]
					&$200$&&1&1&1     &1        &0.985  &0.440 &0.048 &&&1    &0.0\\
					\hline\hline
				\end{tabular}
			\end{threeparttable}
		}
	\end{center}
\end{table}
%
% Figure \ref{xyplot} depicts the samples $\X_i$ and $\Y_i$ for different percentages of true null hypotheses $\mu_{1 k}(s)=\mu_{2 k}(s)\equiv0$ for $k = 1,\cdots,p$ when $n=m=40,p=100$.  As can be seen from Figure \ref{xyplot}(a)(b), when $\mu_{1 k}(t) = \mu_{2 k}(t)\equiv0$  for 98\% percentage dimensions, which means that there are only 2 dimensions of the mean functions being different, the samples $\X_i$ and $\Y_i$ are very similar. However, our test can still distinguish the difference, with non-trivial power of 0.295 as reported in Table \ref{tab3.1.1}.
%The difference between $\X_i$ and $\Y_i$ becomes larger and larger as the percentage of true null hypotheses $\mu_{1 k}(s)=\mu_{2 k}(s)\equiv 0$ decreases.
%
%\begin{figure}[!ht]
%	\centering
%	\subfigure[$\Y$(-100\%)]{
%		\includegraphics[width=0.32\textwidth]{fig2/y0_1.pdf}}
%	\subfigure[$\X$-98\%]{
%		\includegraphics[width=0.32\textwidth]{fig2/x98_1.pdf}}
%	\subfigure[$\X$-95\%]{
%		\includegraphics[width=0.32\textwidth]{fig2/x95_1.pdf}}\\
%	\subfigure[$\X$-90\%]{
%		\includegraphics[width=0.32\textwidth]{fig2/x90_1.pdf}}
%	\subfigure[$\X$-50\%]{
%		\includegraphics[width=0.32\textwidth]{fig2/x50_1.pdf}}
%	\subfigure[$\X$-0\%]{
%		\includegraphics[width=0.32\textwidth]{fig2/x0_1.pdf}}
%	\caption{$\X$ and $\Y$ plots for different percentage of true null hypotheses $\mu_{1 k}(s)=\mu_{2 k}(s)=0$ for $k = 1,\cdots,p$ when $n=m=40,p=100$ for \textbf{Case I} of Simulation I}
%	\label{xyplot}
%\end{figure}
%
%
\begin{table}[!htbp]
	\begin{center}
		\setlength{\tabcolsep}{1.3mm}
{
			\begin{threeparttable}[b]
				\caption{The empirical size  and power for the proposed MRP test and QCZ test \citep{qiuTwosampleTestsMultivariate2021} in  {Case II} for model \eqref{simxynew}  at the significance level $\alpha=0.05$ in Simulation I for different $(n,m)$ and $p$,
where $n$ and $m$ are the number of random samples for $\X$ and $\Y$, $p$ is the dimension of $\X$ and $\Y$, and 0\%, 50\%, 75\%, 90\%, 95\%  98\% and 100\% denote the percentages of the dimensions such that $\mu_{1 k}(t) = \mu_{2 k}(t)$. }
				\label{tab3.3.1}\footnotesize
\begin{tabular}{cccccccccccccc}
              \hline\hline
&&& \multicolumn{7}{c}{MRP test}&&&\multicolumn{2}{c}{QCZ test}    \\
\cline{4-10}                    \cline{13-14}\\  [-0.15cm]
$(n,m)$& $p$ && $0\%$  &50\%&75\%&90\%&95\% &98\%&100\%(size)&&&$T_{n}$(size)&$T_{n,max}$(size)\\[0.1cm]
					\hline\multirow{4}{*}{$(25,25)$}
					&$20$ &&1&1&1 &1       &0.998  &0.050 &0.050&&&0.720&0.040\\ [0.1cm]
					&$50$ &&1&1&1 &1       &0.980  &0.225 &0.040&&&1&0.590\\[0.1cm]
					&$100$&&1&1&1 &1       &0.695  &0.195 &0.056&&&1&0\\[0.1cm]
					&$200$&&1&1&1 &0.950   &0.480  &0.110 &0.054&&&1&0 \\ [0.1cm]						
					\hline \multirow{4}{*}{$(40,40)$}
					&$20$ &&1&1&1 &1      &1      &0.055 &0.055&&&0.410&0.09\\[0.1cm]
					&$50$ &&1&1&1 &1      &1      &0.480 &0.045&&&1&0.06\\[0.1cm]
					&$100$&&1&1&1 &1      &1      &0.240 &0.043&&&1&0\\[0.1cm]
					&$200$&&1&1&1 &1      &0.860  &0.143 &0.052&&&1&0\\
					\hline\hline
				\end{tabular}
\end{threeparttable}
	}
\end{center}
\end{table}

Tables  \ref{tab3.1.1}-\ref{tab3.3.1} summarize the empirical size and power for the proposed test based on an Ornstein-Uhlenbeck process at the significance level $\alpha=0.05$ in  Cases I-II  for model \eqref{simxynew} under $n=m=25,40$ and $p=20,50,100,200$, as well as the empirical size of the two QCZ tests. Tables  \ref{tab3.1.1}-\ref{tab3.3.1} show that the empirical sizes of  our test  are
very close to the true significance level, which further confirms our asymptotical properties.
However, the two QCZ test statistics, $T_{n}$ and $T_{n,max}$ don't achieve approximately the true significance level in most settings. Even when the dimension $p$ is smaller than the sample sizes
($p=20,n=m=40$),  the empirical size  is 0.31 in Table   \ref{tab3.1.1} and 0.41 in Table  \ref{tab3.3.1}.  This is perhaps because the QCZ tests are only designed for low-dimensional functional data and need to estimate the inverse of the covariance matrix, which is infeasible when functional data are high-dimensional.

Tables  \ref{tab3.1.1}-\ref{tab3.3.1} also show that the empirical power of our test has a non-trivial performance across all settings.  Especially,   in the setting of   $p=50,n=m=25$ and 98\%, where there is
 only one different component between  $\bfmu_{1}(t)$ and  $\bfmu_{2}(t)$,
the power is 0.135  in Case I and  0.225 in Case II. These imply that our test can effectively distinguish small differences between  $\bfmu_{1}(t)$ and  $\bfmu_{2}(t)$.

From Tables \ref{tab3.1.1}-\ref{tab3.3.1},  an interesting finding is that, for a fixed percentage of dimensions such that $\mu_{1 k}(t) = \mu_{2 k}(t)$, the empirical power of our test increases in Case I but decreases in Case II as  $p$ increases.
This phenomenon can be explained by the theoretical analysis of the power in Theorem \ref{theorem-power} and Corollary \ref{proposition2-2}. Specifically, Theorem \ref{theorem-power} and Corollary \ref{proposition2-2} suggest that the power is asymptotically an increasing function with respective to $\Delta_{nm}(\bfmu_{1}-\bfmu_{2}, \G_{1},\G_{2})$.
In the setting of Case I, where the covariance functions $\G_{1}$ and $\G_{2}$ allow a two-dependence structure,
when $p$ increases, the denominator of $\Delta_{nm}(\bfmu_{1}-\bfmu_{2}, \G_{1},\G_{2})$
is almost invariant but  the numerator, depended on  $\bfmu_{1}-\bfmu_{2}$, becomes larger. Thus, the power becomes higher when $p$  increases. In  Case II, where the covariance functions  have the full dependence structure,  the  denominator of $\Delta_{nm}(\bfmu_{1}-\bfmu_{2}, \G_{1},\G_{2})$  increases faster than  the signal $\bfmu_{1}-\bfmu_{2}$
when $p$  increases, which yields that the power decreases.

\begin{table}[!htbp]
	\begin{center}
		\setlength{\tabcolsep}{3.8mm}
{
			\begin{threeparttable}[b]
				\caption{The empirical size  and power  for the proposed MRP test based on a Wiener process
    in {Case I} for model \eqref{simxynew}  at the significance level $\alpha=0.05$ in Simulation I for different $(n,m)$ and $p$,
where $n$ and $m$ are the number of random samples for $\X$ and $\Y$, $p$ is the dimension of $\X$ and $\Y$, and the percentages 0\%, 50\%, 75\%, 90\%, 95\%  98\% and 100\% denote the percentages of the dimensions such that $\mu_{1 k}(t) = \mu_{2 k}(t)$. }
				\label{tab3.2.1}\footnotesize
				\begin{tabular}{ccccccccc}
					\hline\hline
					$(n,m)$& $p$ & $0\%$ &50\%&75\%&90\%&95\% &98\%&100\%(size)\\ [0.1cm]
					\hline\multirow{4}{*}{$(25,25)$}
					&$20$ &1&1&0.900 &0.330  &0.165  &0.043 &0.043\\[0.1cm]
					&$50$ &1&1&0.985 &0.588  &0.288  &0.113 &0.038\\[0.1cm]
					&$100$&1&1&0.965 &0.890  &0.358  &0.133 &0.042\\[0.1cm]
					&$200$&1&1&1 &0.945      &0.555  &0.203 &0.044 \\ [0.1cm]						
					\hline \multirow{4}{*}{$(40,40)$}
					&$20$ &1&1&0.985 &0.575   &0.245  &0.045 &0.045\\[0.1cm]
					&$50$ &1&1&0.996 &0.795   &0.455  &0.138 &0.039\\[0.1cm]
					&$100$&1&1&1     &0.955   &0.555  &0.175 &0.050\\[0.1cm]
					&$200$&1&1&1     &1       &0.828  &0.285 &0.048\\
					\hline\hline
				\end{tabular}
		\end{threeparttable}
	}
\end{center}
\end{table}

To investigate the effect of the projection process $\gamma$   on the size  and power  of our test,
we further conduct the simulations  in  {Case I} for model \eqref{simxynew} by   choosing $\gamma$ to be a Wiener process, where the covariance function $v(s, t)$  satisfies  $v(s, t) \propto s \wedge t:=\min \{s, t\}$. The results are reported in Table \ref{tab3.2.1}. Table \ref{tab3.2.1} shows that the performance of our test with a Wiener process is basically similar to that with an Ornstein-Uhlenbeck process shown in Table \ref{tab3.1.1}, which implies that the performance of our test is robust to the choice of $\gamma$. Compared Table \ref{tab3.1.1} with  Table \ref{tab3.2.1},  the Ornstein-Uhlenbeck process-based test has slightly higher powers and thus slightly outperforms the Wiener process-based test. Thus, the Ornstein-Uhlenbeck process is suggested as the projection direction in practice.
 \subsection{Simulation II}\label{simulation2}
 In Simulation II, the data is generated from  a more complicated  functional moving average structure than
 model \eqref{simxynew},
investigated by   \citet{kokoszkaDeterminingOrderFunctional2013a}:
\begin{equation}\label{SimulationII}
\begin{aligned}
X_{i k}(t)&=Z_{1i k}(t)+\sum_{l=1}^p\int_{[0,1]} \phi_l(u, t) Z_{1i (k-l)}(u) \mathrm{d} u+\mu_{1 k}(t), t\in[0,1]; \\
Y_{j k}(t)&=Z_{2j k}(t)+\sum_{l=1}^p\int_{[0,1]} \phi_l(u, t) Z_{2j (k-l)}(u) \mathrm{d} u +\mu_{2 k}(t), t\in[0,1],
\end{aligned}
\end{equation}
for $i=1,  \ldots, n$, $j=1,  \ldots, m$ and $k=1,  \ldots, p$,   where $\left\{Z_{1i k}(t)\right\}_{i=1}^n$ and $\left\{Z_{2j k}(t)\right\}_{j=1}^m$ are  independently generated from  a standard Brownian motion defined on $[0,1]$, and $\phi_k(u, v)$s are  defined as
 $\phi_k(u, v)= {c_{k}}\exp \left\{-\left(u^{2}+v^{2}\right)/ 2\right\}/{0.7468}$, where  $c_{k}$s are constants, satisfying $\int_{[0,1]} \int_{[0,1]} \phi_k^{2}(u, v) \mathrm{d} u \mathrm{d} v =c_{k}^2$.
The norm scalar $\{c_{k}\}_{k=1}^p$  is set in the same way as Cases I-II in Simulation I.
Without loss of generality,  we let  $\mu_{2 k}(t)= 0$. The mean functions of $\X$ for  {Case I} and {Case II} are defined  by
\begin{equation*}
\begin{aligned}
	\mu_{1 k,\I}(t)&=t\log(k/p+1)+\sin^2(2\pi t+k/p), \text {  for  } k=1, \ldots,p,\\
	% \label{simmu1}
 \mu_{1 k,\I\I}(t)&=\{t\log(k/p+1)+\sin^2(2\pi t+k/p)\}\log(p^2/2), \text {  for  } k=1, \ldots,p.
 \end{aligned}
\end{equation*}

\begin{figure}[!htbp]
\subfigcapskip=-5pt
	\centering

	\subfigure[$n=m=25, p=20$]{
		\includegraphics[width=0.31\textwidth]{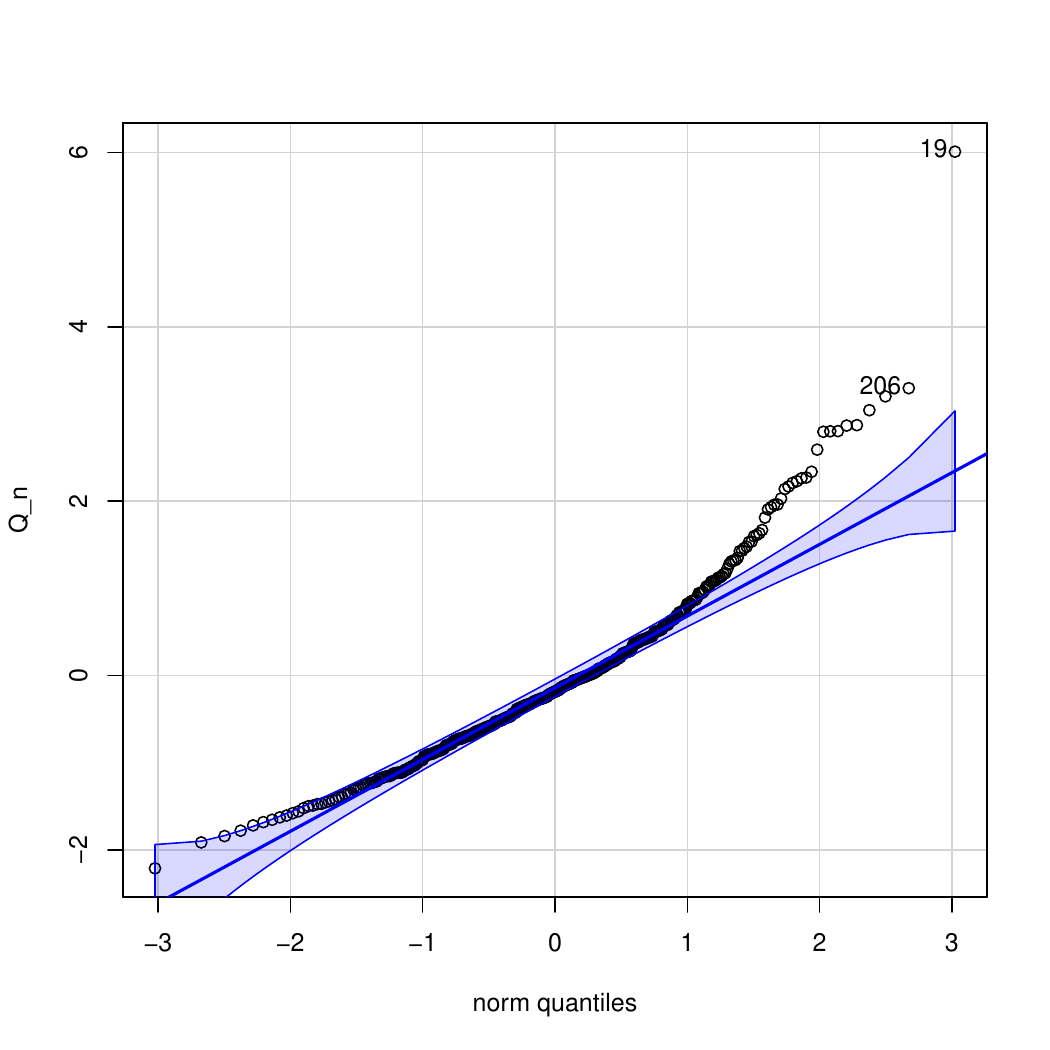}}
    \setcounter{subfigure}{4} 
  \subfigure[$n=m=40, p=20$]{
		\includegraphics[width=0.31\textwidth]{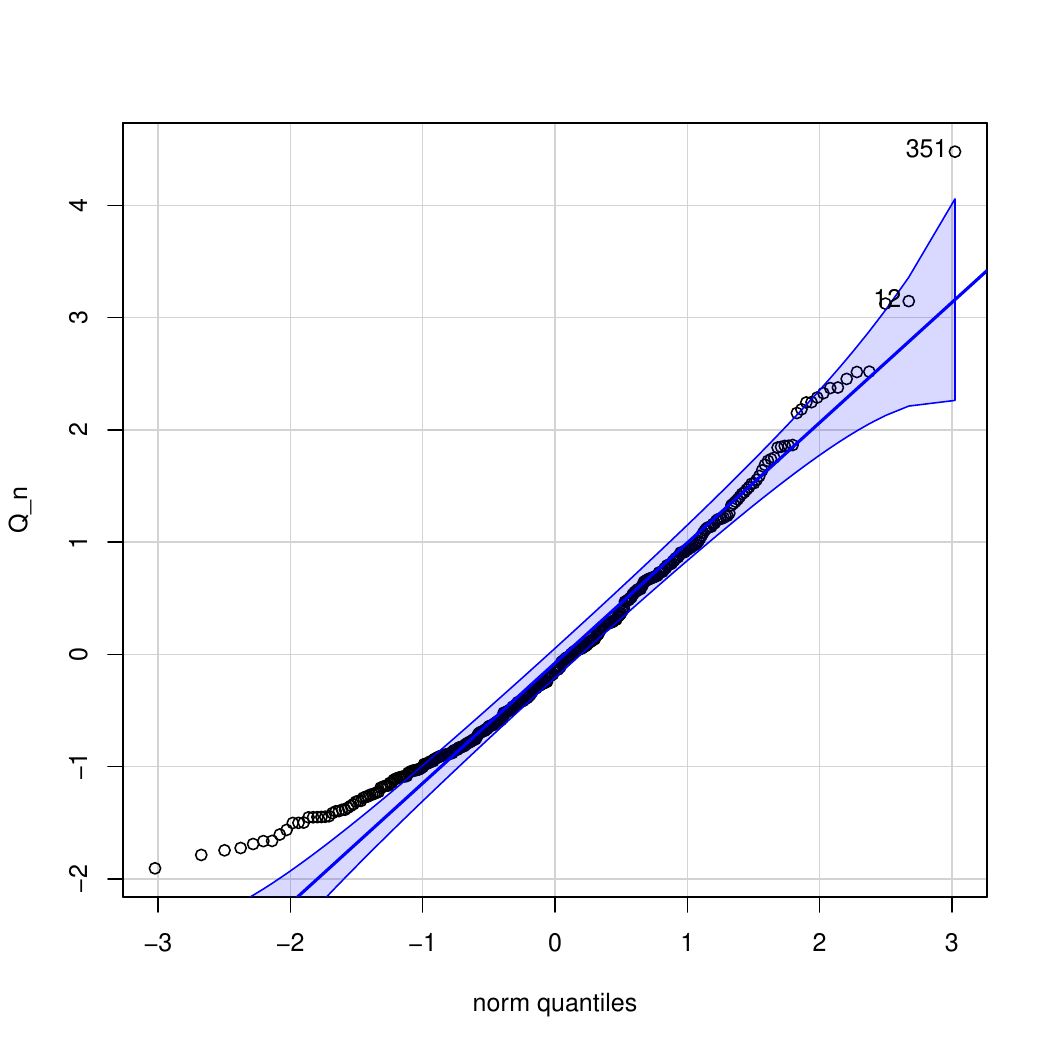}}\\
    \setcounter{subfigure}{1} 
	\subfigure[$n=m=25, p=50$]{
		\includegraphics[width=0.31\textwidth]{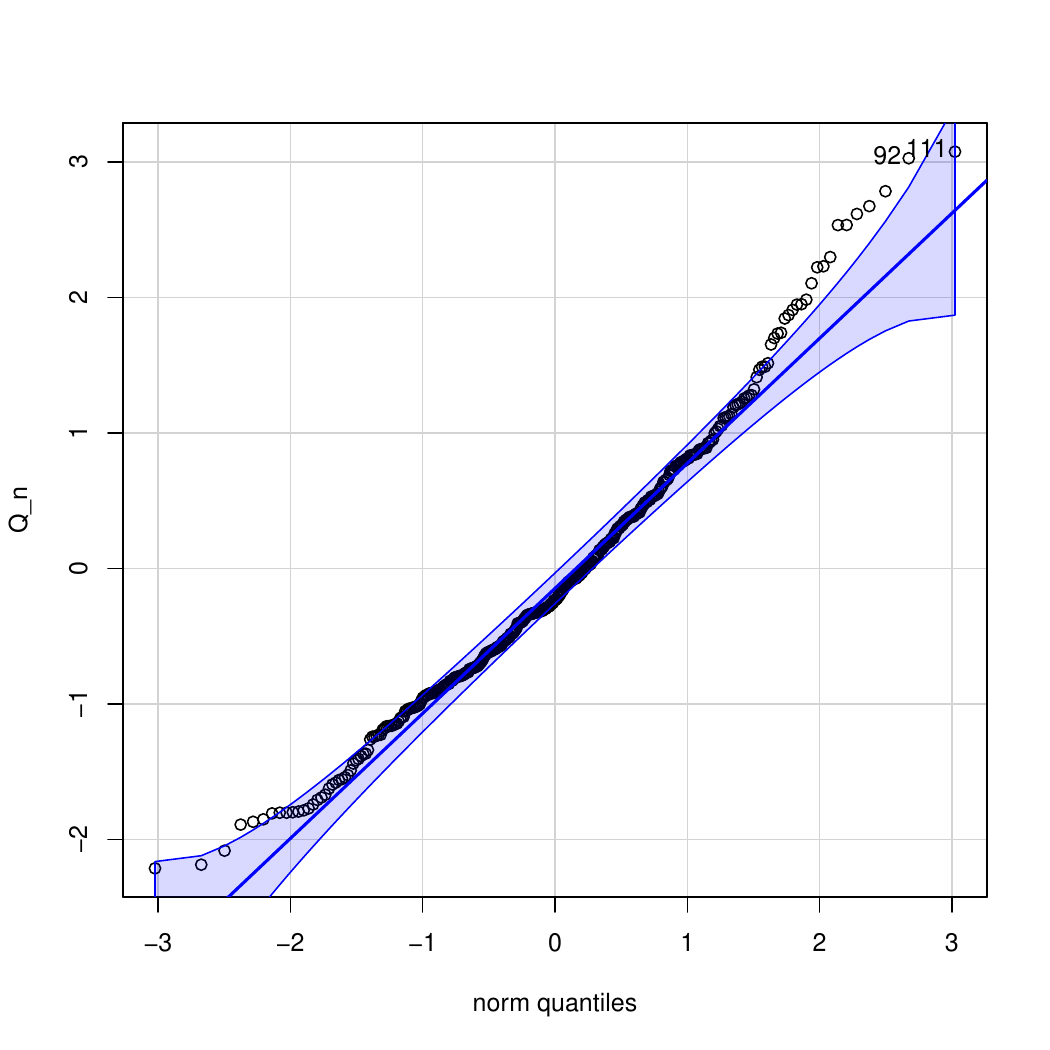}}
    \setcounter{subfigure}{5} 
  	\subfigure[$n=m=40, p=50$]{
		\includegraphics[width=0.31\textwidth]{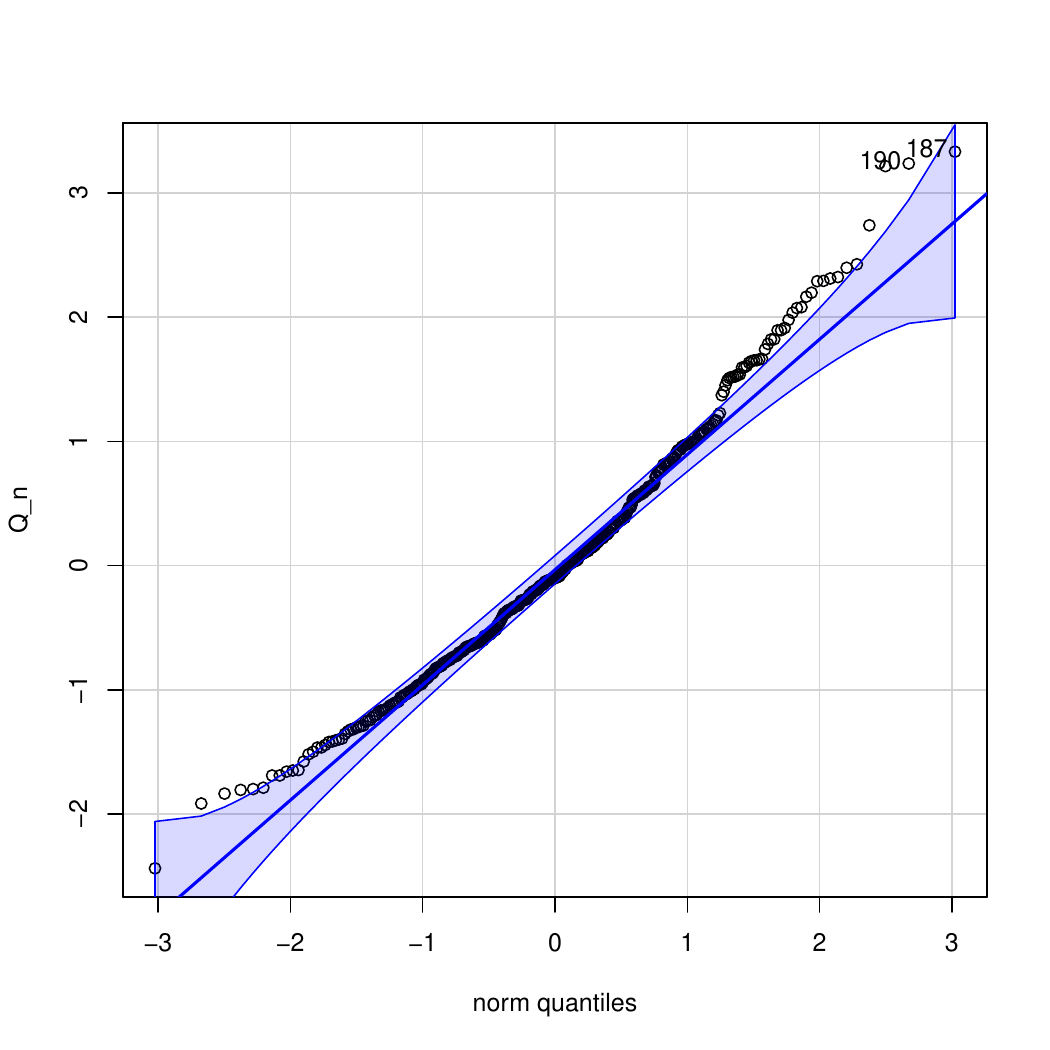}}\\
    \setcounter{subfigure}{2} 
  \subfigure[$n=m=25, p=100$]{
		\includegraphics[width=0.31\textwidth]{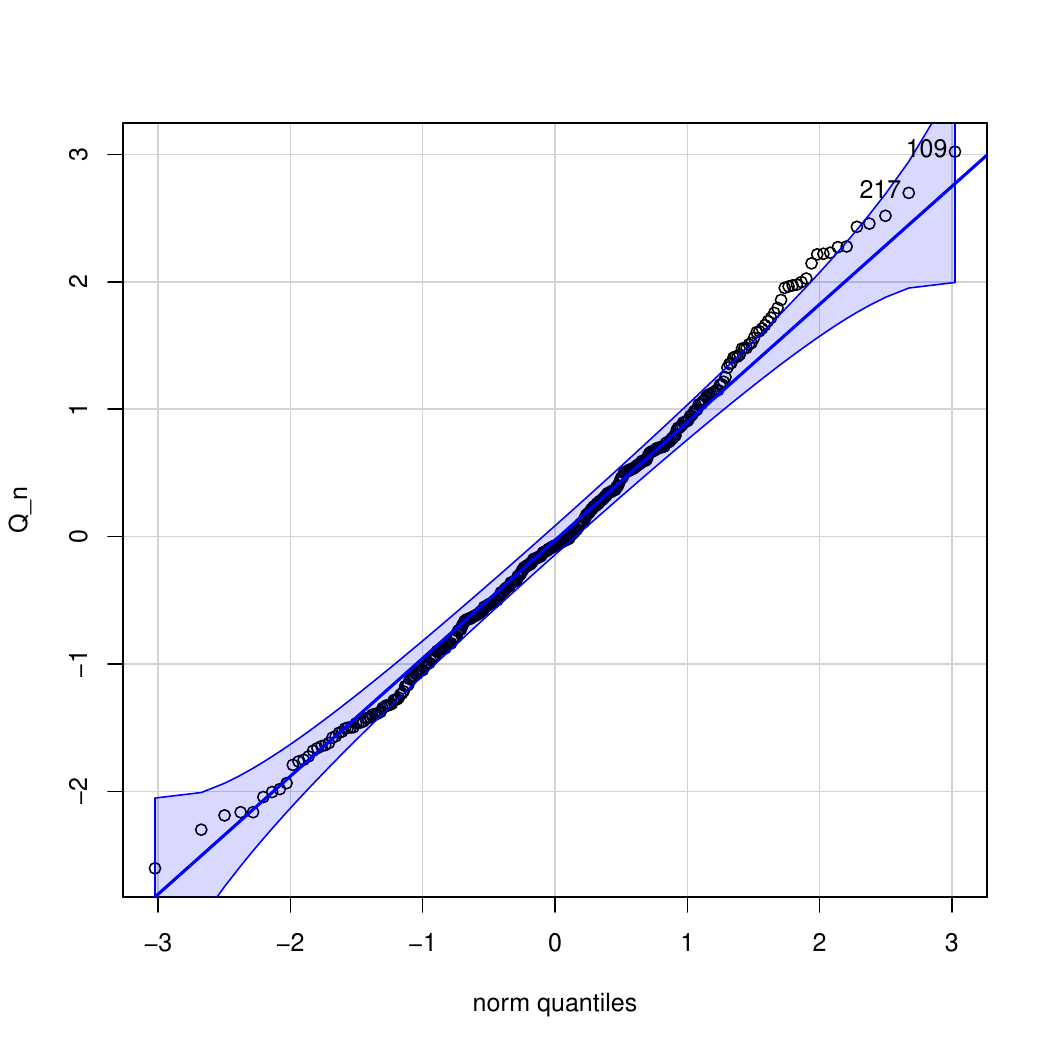}}
    \setcounter{subfigure}{6} 
    	\subfigure[$n=m=40, p=100$]{
		\includegraphics[width=0.31\textwidth]{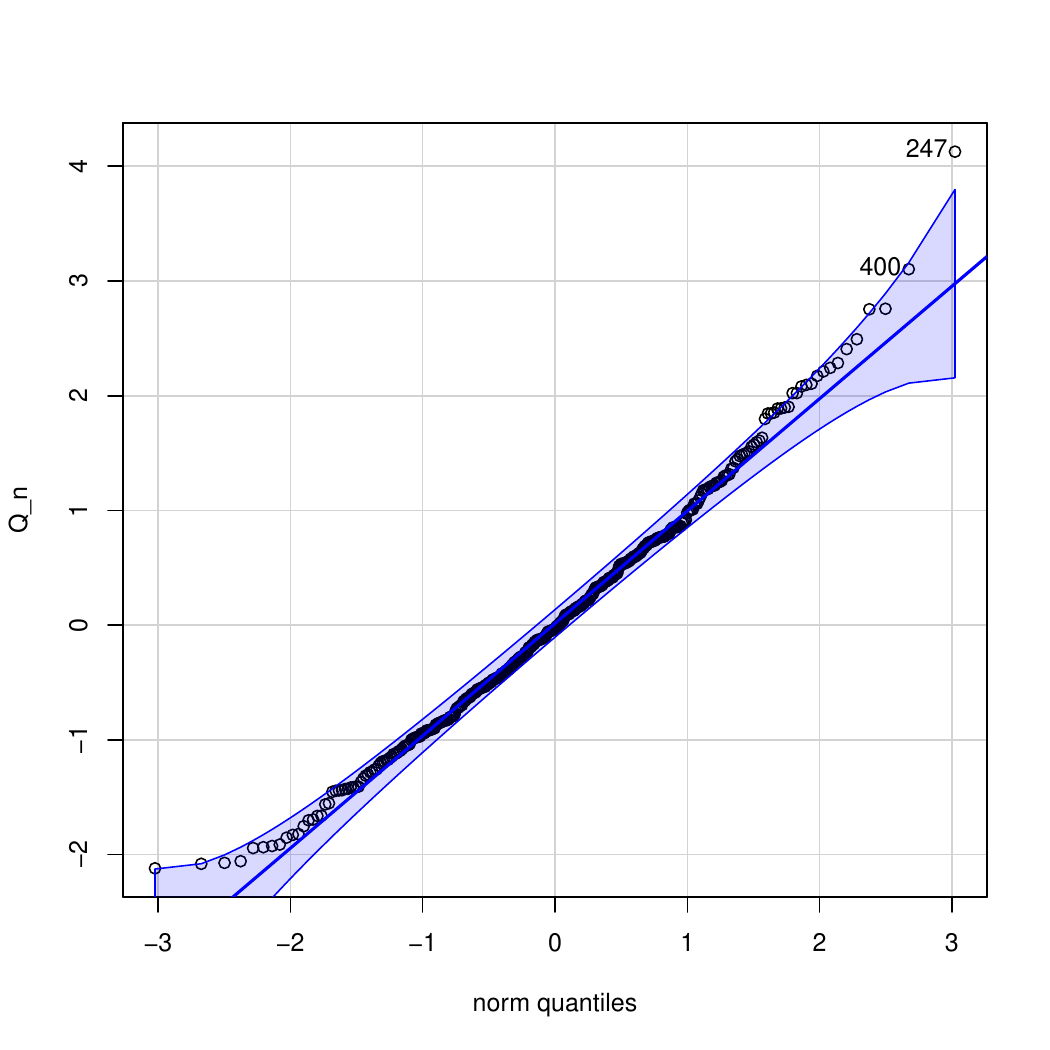}}\\
    \setcounter{subfigure}{3} 
	\subfigure[$n=m=25, p=200$]{
		\includegraphics[width=0.31\textwidth]{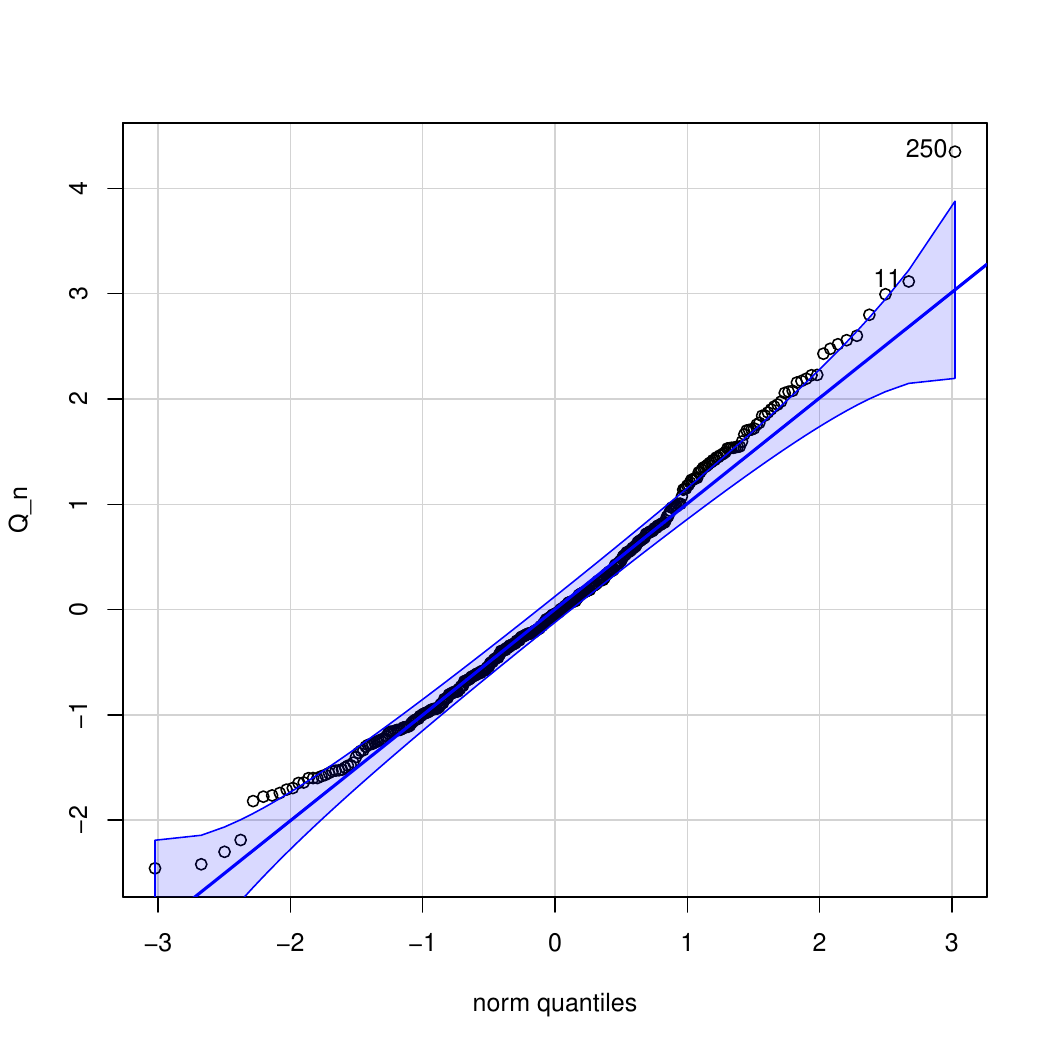}}
    \setcounter{subfigure}{7} 
	\subfigure[$n=m=40, p=200$]{
		\includegraphics[width=0.31\textwidth]{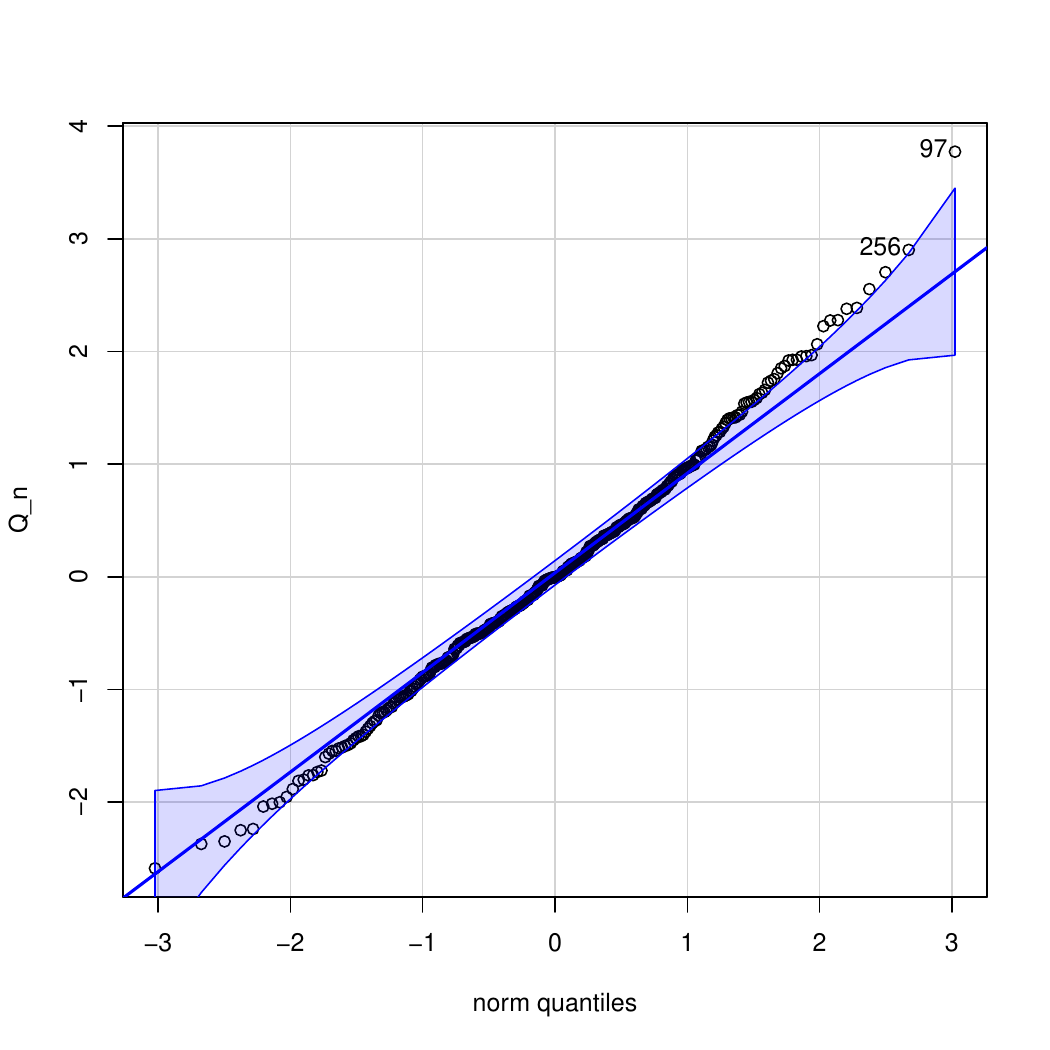}}
\caption{Q-Q plots of   $\widehat Q_{n}(\widehat\X, \widehat\Y)$  under $n=m=25,40$ and $p=20,50,100,200$ for {Case I} for model \eqref{SimulationII} in Simulation II, where $n$ and $m$ are the number of random samples for $\X$ and $\Y$, $p$ is the dimension of $\X$ and $\Y$.}
	\label{qqplot2}
\end{figure}

Figure \ref{qqplot2} shows the QQ plot of  $ {\widehat Q}_{n}(\widehat\X, \widehat\Y)$ with an Ornstein-Uhlenbeck process under $n=m=25,40$ and $p=20,50,100,200$ in {Case I} for model \eqref{SimulationII}.
Similar to Figure \ref{qqplot}, Figure \ref{qqplot2} reveals that the asymptotic distribution of $\widehat Q_{n}(\widehat\X,\widehat\Y)$ is approximated well $N(0, 1)$ when the dimension and the sample sizes increase.
% \begin{figure}[!ht]
% \subfigcapskip=-5pt
% 	\centering
% 	\subfigure[$n=m=25, p=20$]{
% 		\includegraphics[width=0.23\textwidth]{fig/qq2520.pdf}}
% 	\subfigure[$n=m=25, p=50$]{
% 		\includegraphics[width=0.23\textwidth]{fig/qq2550.pdf}}
%   \subfigure[$n=m=25, p=100$]{
% 		\includegraphics[width=0.23\textwidth]{fig/qq25100.pdf}}
% 	\subfigure[$n=m=25, p=200$]{
% 		\includegraphics[width=0.23\textwidth]{fig/qq25200.pdf}}\\
% 	\subfigure[$n=m=40, p=20$]{
% 		\includegraphics[width=0.23\textwidth]{fig/qq4020.pdf}}
% 	\subfigure[$n=m=40, p=50$]{
% 		\includegraphics[width=0.23\textwidth]{fig/qq40502.pdf}}
%   	\subfigure[$n=m=40, p=100$]{
% 		\includegraphics[width=0.23\textwidth]{fig/qq40100.pdf}}
% 	\subfigure[$n=m=40, p=200$]{
% 		\includegraphics[width=0.23\textwidth]{fig/qq40200.pdf}}
% \caption{Q-Q plots of   $\widehat Q_{n}(\widehat\X, \widehat\Y)$  under $n=m=25,40$ and $p=20,50,100,200$ for {Case I} for model \eqref{SimulationII} in Simulation II.}
% 	\label{qqplot2}
% \end{figure}

Tables \ref{tab3.1} and \ref{tab3.3}  report the empirical size and power for the proposed test
with an Ornstein-Uhlenbeck process at the significance level $\alpha=0.05$ in  Cases I-II  for model \eqref{SimulationII}. From these two tables, we can see that their performances are similar to those presented in Tables \ref{tab3.1.1}-\ref{tab3.3.1} for model \eqref{simxynew}. The results indicate that the proposed test still works
well under the more complicated functional moving average model \eqref{SimulationII}, which
generalizes Condition (C1). Thus, the proposed test is suitable for more general high-dimensional functional data with a more generalized Condition (C1).
\begin{table}[!htbp]
\begin{center}
\setlength{\tabcolsep}{1.3mm}{
\begin{threeparttable}[b]
\caption{The empirical size  and power  for the proposed MRP test and QCZ test \citep{qiuTwosampleTestsMultivariate2021}
  in {Case I} for model \eqref{SimulationII}  in Simulation II at the significance level $\alpha=0.05$ for different $(n,m)$ and $p$,
where $n$ and $m$ are the number of random samples for $\X$ and $\Y$, $p$ is the dimension of $\X$ and $\Y$, and 0\%, 50\%, 75\%, 90\%, 95\%  98\% and 100\% denote the percentages of the dimensions such that $\mu_{1 k}(t) = \mu_{2 k}(t)$. }	\label{tab3.1}
\footnotesize
\begin{tabular}{cccccccccccccc}
              \hline\hline
&&& \multicolumn{7}{c}{MRP test}&&&\multicolumn{2}{c}{QCZ test}    \\
\cline{4-10}                    \cline{13-14}\\  [-0.15cm]
$(n,m)$& $p$ && $0\%$  &50\%&75\%&90\%&95\% &98\%&100\%(size)&&&$T_{n}$(size)&$T_{n,max}$(size)\\[0.1cm]
					\hline
					\hline\multirow{4}{*}{$(25,25)$}
					&$20$ &&1&1&0.958 &0.48   &0.225  &0.050 &0.050 &&&0.665&0.042\\ [0.1cm]
					&$50$ &&1&1&1     &0.788  &0.485  &0.125 &0.038 &&&1    &0.633\\ [0.1cm]
					&$100$&&1&1&1     &0.955  &0.580  &0.173 &0.060 &&&1    &0.0\\    [0.1cm]
					&$200$&&1&1&1     &0.998  &0.778  &0.245 &0.055 &&&0.6   &0.0 \\ [0.1cm]						
					\hline \multirow{4}{*}{$(40,40)$}
					&$20$ &&1&1&0.995 &0.875    &0.385  &0.055 &0.055 &&&0.363&0.070\\ [0.1cm]
					&$50$ &&1&1&1     &0.945    &0.740  &0.168 &0.045 &&&1    &0.058\\ [0.1cm]
					&$100$&&1&1&1     &1        &0.865  &0.313 &0.048 &&&1    &0.0\\  [0.1cm]
					&$200$&&1&1&1     &1        &0.985  &0.483 &0.048 &&&1    &0.0\\
					\hline\hline
				\end{tabular}
			\end{threeparttable}
		}
	\end{center}
\end{table}
\begin{table}[!htbp]
\begin{center}
\setlength{\tabcolsep}{1.3mm}{
\begin{threeparttable}[b]
\caption{The empirical size  and power for the proposed MRP test and QCZ test \citep{qiuTwosampleTestsMultivariate2021}
 in {Case II} for model \eqref{SimulationII} in Simulation II at the significance level $\alpha=0.05$ for different $(n,m)$ and $p$,
where $n$ and $m$ are the number of random samples for $\X$ and $\Y$, $p$ is the dimension of $\X$ and $\Y$, and 0\%, 50\%, 75\%, 90\%, 95\%  98\% and 100\% denote the percentages of the dimensions such that $\mu_{1 k}(t) = \mu_{2 k}(t)$. }	\label{tab3.3}
\footnotesize
\begin{tabular}{cccccccccccccc}
              \hline\hline
&&& \multicolumn{7}{c}{MRP test}&&&\multicolumn{2}{c}{QCZ test}    \\
\cline{4-10}                    \cline{13-14}\\  [-0.15cm]
$(n,m)$& $p$ && $0\%$  &50\%&75\%&90\%&95\% &98\%&100\%(size)&&&$T_{n}$(size)&$T_{n,max}$(size)\\[0.1cm]
					\hline
\multirow{4}{*}{$(25,25)$}
					&$20$ &&1&1&1 &1      &0.998  &0.050 &0.050&&&0.660&0.030\\[0.1cm]
					&$50$ &&1&1&1 &1      &0.988  &0.280 &0.055&&&1&0.540\\[0.1cm]
					&$100$&&1&1&1 &1      &0.738  &0.182 &0.052&&&1&0\\ [0.1cm]
					&$200$&&1&1&1 &0.938  &0.445  &0.115 &0.052&&&1&0 \\[0.1cm] 						
					\hline \multirow{4}{*}{$(40,40)$}
					&$20$ &&1&1&1 &1       &1       &0.053 &0.053&&&0.370&0.08\\[0.1cm]
					&$50$ &&1&1&1 &1       &1       &0.578 &0.055&&&1&0.06\\[0.1cm]
					&$100$&&1&1&1 &1       &0.998   &0.320 &0.053&&&1&0\\ [0.1cm]
					&$200$&&1&1&1 &1       &0.824   &0.192 &0.052&&&1&0\\
					\hline\hline
				\end{tabular}
		\end{threeparttable}
	}
\end{center}
\end{table}

Similar to   Table \ref{tab3.2.1},   Table \ref{tab3.2} reports the empirical size  and power  of our test
 in  {Case I} for model \eqref{SimulationII}  at  $\alpha=0.05$
by choosing $\gamma$ to be a Wiener process. The results further indicate that the Ornstein-Uhlenbeck process-based test is slightly superior to the Wiener process-based test.
\begin{table}[!htbp]
	\begin{center}
		\setlength{\tabcolsep}{3.8mm}
{
			\begin{threeparttable}[b]
 \caption{The empirical size  and power for the proposed MRP test based on a  Wiener process
  in  {Case I} for model \eqref{SimulationII}  in Simulation II at the significance level $\alpha=0.05$ for different $(n,m)$ and $p$,
where $n$ and $m$ are the number of random samples for $\X$ and $\Y$, $p$ is the dimension of $\X$ and $\Y$, and the percentages 0\%, 50\%, 75\%, 90\%, 95\%  98\% and 100\% denote the percentages of the dimensions such that $\mu_{1 k}(t) = \mu_{2 k}(t)$. }
				\label{tab3.2}\footnotesize
				\begin{tabular}{ccccccccc}
					\hline\hline
					$(n,m)$& $p$ & $0\%$ &50\%&75\%&90\%&95\% &98\%&100\%(size)\\[0.1cm]
					\hline\multirow{4}{*}{$(25,25)$}
					&$20$ &1&1&0.905 &0.430  &0.185  &0.048 &0.048\\ [0.1cm]
					&$50$ &1&1&0.938 &0.688  &0.418  &0.101 &0.038\\ [0.1cm]
					&$100$&1&1&0.959 &0.845  &0.482  &0.133 &0.059\\ [0.1cm]
					&$200$&1&1&0.987 &0.928  &0.698  &0.205 &0.054 \\ [0.1cm]						
					\hline \multirow{4}{*}{$(40,40)$}
					&$20$ &1&1&0.925 &0.819   &0.315  &0.055 &0.055\\[0.1cm]
					&$50$ &1&1&0.966 &0.825   &0.635  &0.118 &0.044\\[0.1cm]
					&$100$&1&1&0.988 &0.925   &0.785  &0.223 &0.048\\ [0.1cm]
					&$200$&1&1&1     &0.955   &0.895  &0.394 &0.048\\
					\hline\hline
				\end{tabular}
		\end{threeparttable}
	}
\end{center}
\end{table}

\subsection{Simulation III}\label{simulation3}

In the previous two examples, the power is relatively lower when the strengths of the signals $\mu_{2 k}(t)-\mu_{1 k}(t)$ under the alternatives are low relative to the level of noise.
% (Jicai: Please check the accuracy of this expression. In fact, I don't understand this expression)
In this simulation,  we further investigate the performance of our test under the sparse signal model:
\begin{equation}\label{Simulation-III}
	X_{i k}(t)=Z_{1i k}(t)+\mu_{1 k}(t),~~Y_{j k}(t)=Z_{2j k}(t)+\mu_{2 k}(t),
\end{equation}
for $i=1,  \ldots, n$, $j=1, \ldots, m$ and $k=1,  \ldots, p$,   where $\left\{Z_{1i k}(t)\right\}_{i=1}^n$ and $\left\{Z_{2j k}(t)\right\}_{j=1}^m$ are  independently generated from  a standard Brownian motion defined on $[0,1]$.
Let 
% $\mu_{1 k}(t)$ and $\mu_{2 k}(t)$ be
\begin{equation}\label{mu1-sparsity}
\mu_{1 k}(t)=\varepsilon \sqrt{2\log (p)} t  \text{  for  }  1\leq k\leq q= \lfloor p^{c} \rfloor, \text{  and  }  \mu_{2k}(t)=0  \text{  for  } k>q ,
\end{equation}
for some $c \in(0,1)$ and some constant $\varepsilon$, where $c$  controls the sparsity of the alternative hypothesis ($\mu_{1 k}(t)\neq \mu_{2 k}(t)=0$)  and $\varepsilon$ controls the signal strength of each dimension. The alternative hypothesis is more sparse with smaller $c$ and the signal strength of each dimension is stronger with larger $\varepsilon$. Table \ref{tab3.4} summarizes the empirical size and power of our test 
for $p=300$, $n=m=25,40$,  $\varepsilon=0.15,0.25$ and $c=0.35$, $0.45$ and $0.55$.
Table \ref{tab3.4} shows that our test can still maintain high power even in extremely sparse settings ($c=0.35$). We also see that the power increases as $c$ and $\varepsilon$ increase because the strength of the signals becomes stronger. This is consistent with our theoretical power analysis.
\begin{table}[!htbp]
	\begin{center}
		\setlength{\tabcolsep}{6mm}{
			\begin{threeparttable}[b]
  \caption{The empirical size  and power  for the proposed MRP test based on a  Wiener process
     for model \eqref{Simulation-III}  in Simulation III under $n=m=25,40$, $p=300$, where $n$ and $m$ are the number of random samples for $\X$ and $\Y$, $\varepsilon$ controls the signal strength of each dimension, $c$ controls the sparsity of the alternative hypothesis ($\mu_{1 k}(t)\neq \mu_{2 k}(t)$), and $q$ is the number of dimensions such that $\mu_{1 k}(t)\neq\mu_{2 k}(t)$.}
				\label{tab3.4}
				\begin{tabular}{cccccc}
					\hline\hline
				\multirow{2}{*}{$(n,m)$} & \multirow{2}{*}{$\varepsilon$}  & $c=0.35$& $c=0.45$& $c=0.55$&  size \\ 
    & & $(q=7)$& $(q=13)$& $(q=23)$&  $(\bfmu_1(t)=\bfmu_2(t))$\\[0.1cm]
					\hline\multirow{2}{*}{$(25,25)$}
					&$0.15$ &0.179&0.312&0.572&0.032\\[0.1cm]
					&$0.25$ &0.494&0.887&1&0.041\\	[0.1cm]			
					\hline
                 \multirow{2}{*}{$(40,40)$}
					&$0.15$ &0.281&0.543&0.880&0.055\\[0.1cm]
					&$0.25$ &0.820&1&1&0.054\\
					\hline\hline
				\end{tabular}
			\end{threeparttable}
		}
	\end{center}
\end{table}

\section{Real Data Analysis}\label{Realdata}
In this section,  we illustrate the proposed procedure by empirical analysis of two real
datasets.  The first  dataset  is
the  global climate data, provided by NASA Center for Climate
Simulation,  which is available at \url{http://ds.nccs.nasa.gov/thredds/catalog/NEX-GDDP/IND/BCSD/catalog.html}. The second dataset is from an electroencephalogram (EEG) study,  which concerns the relationship between genetic predisposition and the tendency for alcoholism (\url{http://kdd.ics.uci.edu/databases/eeg/eeg.data.html}).

\subsection{Global Climate Data}\label{climate}

In climate change research, the RCPs are designed to provide plausible future scenarios of anthropogenic forcing, such as a low-emission scenario (RCP2.6), two intermediate scenarios (RCP4.5 and RCP6.0), and a high-emission scenario (RCP8.5). These RCP scenarios can be used in climate models for a variety of purposes such as studying the climate dynamic system and predicting future climate.

%\begin{figure}[!ht]
% 	\centering
%  \includegraphics[width=7cm,height=5cm]{fig/CO2.png}
% \includegraphics[width=8cm,height=5cm]{fig/ghs.png}\hspace{-1cm}
%	\caption{Four different scenarios RCP2.6, RCP4.5, RCP6.0 and RCP8.5.  }
%	\label{RCP}
%\end{figure}

%\begin{figure}[!ht]
%	\centering
%	\subfigure{
%		\includegraphics[width=0.4\textwidth]{fig/CO2.png}} \\
%	\subfigure{
%		\includegraphics[width=0.4\textwidth]{fig/ghs.png}}
%	\caption{Four different scenarios RCP2.6, RCP4.5, RCP6.0 and RCP8.5.  }
%	\label{RCP}
%\end{figure}

The global climate data consists of three variables: daily minimum and maximum near-surface air temperature, and daily precipitation across {$1,028,032$} grid points in the whole world from 2006 to 2100 in RCP4.5 and RCP8.5 pathways. Our objective is to test whether each climate variable in RCP4.5  and RCP8.5 is different. We consider the data from 2020 to 2069 in the Arctic area, located by 360 grid points over 66.34N-89.875N and 0.125W-359.875E to analyse the climate change between RCP4.5 and RCP 8.5 pathways. 
There are two reasons for choosing the Arctic area. The Arctic Amplification (AA), which refers to the higher rate of warming in the Arctic than over the rest of the globe (almost twice as large as the global average) \citep{screen2010central}, is a more sensitive indicator of global climate change. Another reason is  to reduce the impact of the different types of climate as different regions of the globe have different types of climate, with the Arctic Circle having a tundra climate.
Each climate variable at 360 grid points can be viewed as a 360-dimensional functional variable.
Taking the daily maximum temperature as an example, let $\X(t)=(X_{1}(t),\ldots,X_{360}(t))^{\T}$ and $\Y(t)=(Y_{1}(t),\ldots,Y_{360}(t))^{\T}$ be the daily maximum temperature over one year in  RCP4.5  and  RCP8.5 at the 360 grid points. Then, the  samples of $\X(t)$ and $\Y(t)$ from 2020 to 2069 can be denoted as    $\{\X_{i}(t)\}_{i=1}^{50}$ and $\{\Y_{j}(t)\}_{j=1}^{50}$,
where  each curve $\X_{ik}(t)$ or  $\Y_{jk}(t)$ is observed on $N=365$ days.
\begin{figure}[!htbp]
\subfigcapskip=-20pt
	\centering
	\subfigure[Sample mean of Tmax for RCP4.5]{
		\includegraphics[width=0.45\textwidth]{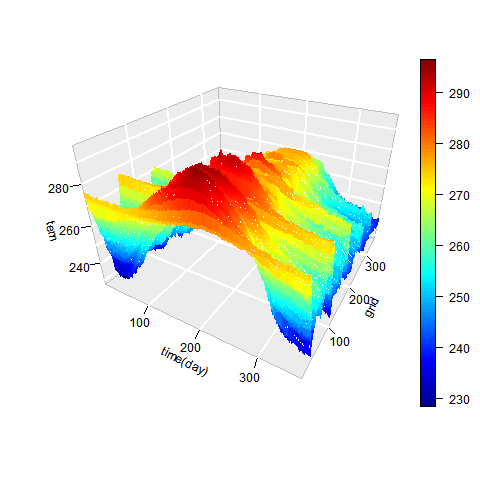}}
	\subfigure[Sample mean of Tmax for RCP8.5]{
		\includegraphics[width=0.45\textwidth]{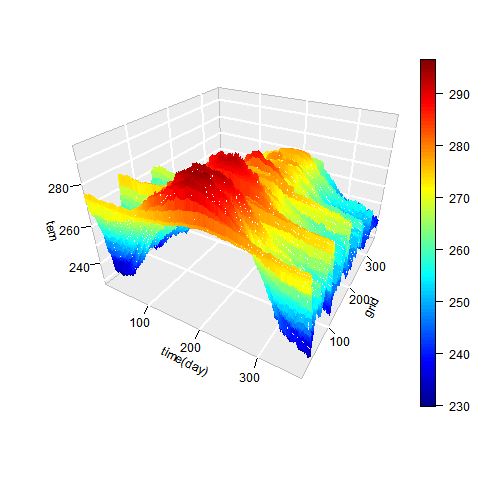}}\\
  \subfigure[Sample mean of Tmin for RCP4.5]{
		\includegraphics[width=0.45\textwidth]{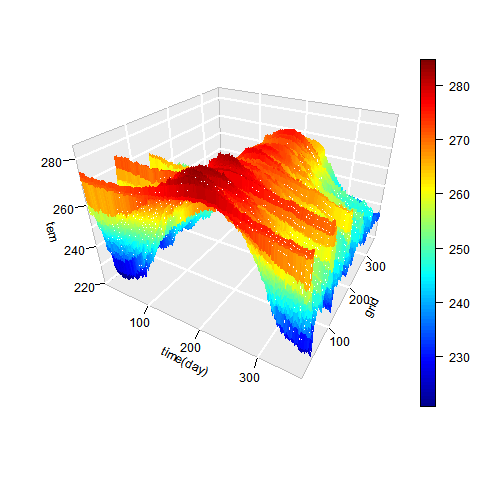}}
	\subfigure[Sample mean of Tmin for RCP8.5]{
		\includegraphics[width=0.45\textwidth]{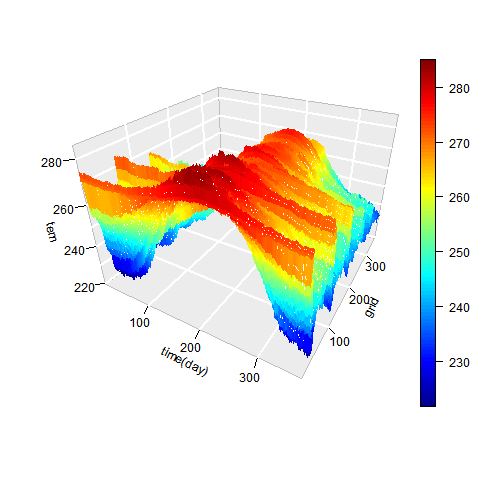}}\\
  	\subfigure[Sample mean of PRCP for RCP4.5]{
		\includegraphics[width=0.45\textwidth]{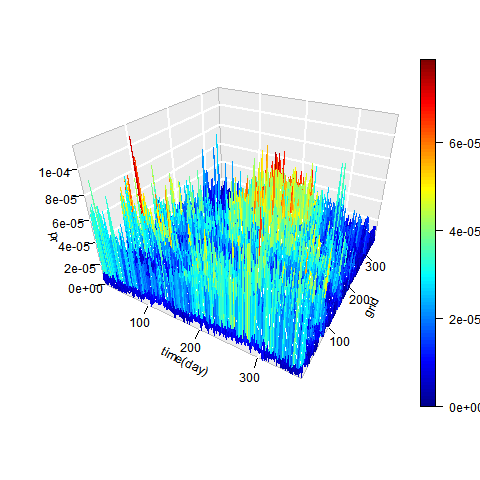}}
	\subfigure[Sample mean of PRCP for RCP8.5]{
	   \includegraphics[width=0.45\textwidth]{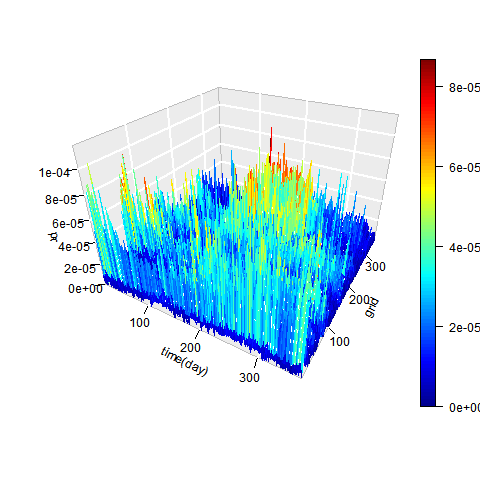}}\\
 %  	\subfigure[Re-sample mean of tasmax for RCP4.5]{
	% 	\includegraphics[width=0.4\textwidth]{fig/sxtemmax15.png}}
	% \subfigure[Re-sample mean of tasmax for RCP8.5]{
	% 	\includegraphics[width=0.4\textwidth]{fig/sytemmax15.png}}
	\caption{Sample mean functions of daily maximum temperature (Tmax), daily minimum temperature (Tmin)  and daily precipitation (PRCP)    from 2020 to 2069 for RCP4.5 and RCP8.5. }
	\label{climateplot1}
\end{figure}
Figure \ref{climateplot1} depicts the plots of the sample mean functions of daily maximum temperature, daily minimum temperature and daily precipitation from 2020 to 2069 for RCP4.5 and RCP8.5, respectively. 
 % As shown in the picture, we almost could not distinguish the mean functions of the daily maximum temperature, daily minimum temperature and daily precipitation between  RCP4.5 and  RCP8.5 by looking at the pictures.
We could barely recognize visually the difference in the mean functions of the daily maximum temperature, daily minimum temperature and daily precipitation between RCP4.5 and RCP8.5.

To further investigate the differences in these climatic features between RCP4.5 and RCP8.5, we apply the proposed MRP test method to analyze this dataset. 
%As presented in Table \ref{tabreal}, we can distinguish the difference between RCP4.5 and RCP8.5 by the proposed test.
Table \ref{tabreal} summarizes  the test statistic and  $p$-values of the proposed  method
for the three climate variables during the three different time ranges, 2020-2069, 2045-2069 and 2020-2044. From Table \ref{tabreal}, we can see that,  for the daily maximum temperature, during the whole studied time range 2020-2069, the $p$-value is almost zero, which implies that the daily maximum temperature in  RCP4.5  and  RCP8.5 pathways are different during the whole time range in the area of  Arctic. We then split the whole time range into two ranges 2020-2044 and 2045-2069 to further explore the main different periods.
For the second 25 years (2045-2069), the difference in the daily maximum temperature between RCP4.5  and  RCP8.5 
is more significant with a larger test statistic (15.44) than that of the whole period (5.88). 
However,  an interesting finding is that the difference is not significant during the first 25 years (2020-2044)  because of the high $p$-value of 0.63. For the daily minimum air temperature and daily precipitation, we can also obtain similar results. Thus, 
%we conclude that,  the climate variables (daily maximum air temperature, daily minimum air temperature  and daily precipitation) between  RCP4.5  and  RCP8.5 are significantly different during  long-time range  although the differenceas may not significant  during a short-time range. 
over the whole studied time range, especially the second 25 years, differences between RCP4.5 and RCP8.5 become more pronounced. These disparities may manifest as higher maximum temperatures, lower minimum  temperatures, and altered patterns of precipitation. However, when examining shorter time frames, such as the first 25 years, the discrepancies between these pathways may not be statistically significant. This suggests that the effects of different greenhouse gas emission scenarios become more apparent as the time frame lengthens. 

These results may be explained by the fact that emissions peak around 2045 under the lower emission pathway RCP4.5, while carbon emissions have been growing under the higher emission pathway RCP8.5. The findings suggest that implementing effective measures to reduce carbon emissions would have a positive impact on climate conditions. Taking action promptly to decrease greenhouse gas emissions is crucial for mitigating climate change and its associated risks. To achieve significant carbon emission reductions, a combination of measures is necessary, including transitioning to renewable energy, improving energy efficiency, adopting sustainable transportation methods, promoting circular economy practices, and implementing nature-based solutions. It's important for governments, organizations, and individuals to work together to implement these measures and ensure a sustainable future for generations to come.

\begin{table}[!ht]
	\begin{center}
		\setlength{\tabcolsep}{2.7mm}{
			\begin{threeparttable}[b]
				\caption{Testing the difference of the mean functions of the projected global climate data based on two Representative Concentration Pathways (
RCP4.5 and RCP8.5) by the proposed MRP test method, where $n$ and $m$ are the number of random samples for $\X$ and $\Y$, and $p$ is the dimension of $\X$ and $\Y$.}
				\label{tabreal}
				\begin{tabular}{ccccr}
					\hline\hline
					Climate Variable& Time Range & $(n,m,p)$  &Test Statistic &$p$-value\\[0.1cm]
					\hline\multirow{3}{*}{Daily Maximum Temperature}
					&2020-2069	&(50,50,360)	&5.88	&$4.01\times10^{-9}$\\[0.1cm]
					&2045-2069	&(25,25,360)	&15.44	&$<10^{-16}$\\[0.1cm]	
					&2020-2044	&(25,25,360)	&-0.48	&0.63\\[0.1cm]					
					\hline\multirow{3}{*}{Daily Minimum Temperature}
					&2020-2069	&(50,50,360)	&5.47 	&$4.54\times10^{-8}$\\[0.1cm]
					&2045-2069	&(25,25,360)	&14.88	&$<10^{-16}$\\[0.1cm]	
					&2020-2044	&(25,25,360)	&0.067	&0.95\\[0.1cm]					
					\hline\multirow{3}{*}{Daily Precipitation}
					&2020-2069	&(50,50,360)	&2.36	&$1.85\times10^{-2}$\\ [0.1cm]					
					&2045-2069	&(25,25,360)	&2.71	&$6.79\times10^{-3}$\\[0.1cm]
					&2020-2044	&(25,25,360)	&1.61	&0.11\\		
					\hline\hline
				\end{tabular}
		\end{threeparttable}
	}
\end{center}
\end{table}

\subsection{EEG Data}\label{EEG}
The EEG  dataset includes two groups of subjects: an alcoholic group of 77 subjects and a control (non-alcoholic) group of 45 subjects. Each subject was exposed to one stimulus or two stimuli while voltage values were measured from 64 channels of electrodes placed on the subject's scalp for 256 time points.  \cite{Li2010on} applied the dataset to study the pattern of voltage over times and channels by  treating   voltage values for each subject  as a $256 \times 64$ matrix.

It is noteworthy that, although the method proposed by \cite{Li2010on} can preserve well the original structure of the data by the way of the matrix-valued objects, it  ignores the  smoothness of  voltage values at the time points. In fact,  it seems to be more natural that  voltage values for each channel should be treated as a continuous function of time. Thus, it may be  more reasonable to view the  voltage values as a random functional curve. From the perspective,  we  represent the data as random functional  data, denoted by   $\{\X_{i}(t)=(X_{i1}(t),\ldots,X_{i64}(t))^T, t\in[0,1]\}_{i=1}^{77}$ for the alcoholic group, and   $\{\Y_{j}(t)=(Y_{j1}(t),\ldots,Y_{j64}(t))^T, t\in[0,1]\}_{j=1}^{45}$ for the control group, where  each  $\X_{ik}(t)$ and $\Y_{jk}(t)$ have $N=256$ observations.

%Moreover,  Figures \ref{xyalcohol} (b) and (d) also present the reconstructed  random curves for the two subjects by the B-spline smoothing method. We can see that, the reconstructed  curves in Figures \ref{xyalcohol} (b) and (d) not only present the same pattern as the original  data in Figures \ref{xyalcohol} (a) and (c) but also maintains the  smoothness of  the data  and  reduces the noise at the time points.

\begin{figure}[!ht]
\subfigcapskip=-20pt
	\centering
% 	\subfigure[\scriptsize Observation for the non-alcoholic Subject 1 ]{
% 		\includegraphics[width=0.45\textwidth]{fig/yalcohol1.png}}
%     \subfigure[\scriptsize Observation for  the alcoholic Subject 1 ]{
% 		\includegraphics[width=0.45\textwidth]{fig/xalcohol1.png}}
% \\
\subfigure[\scriptsize  Sample mean   for the non-alcoholic group]{
		\includegraphics[width=0.45\textwidth]{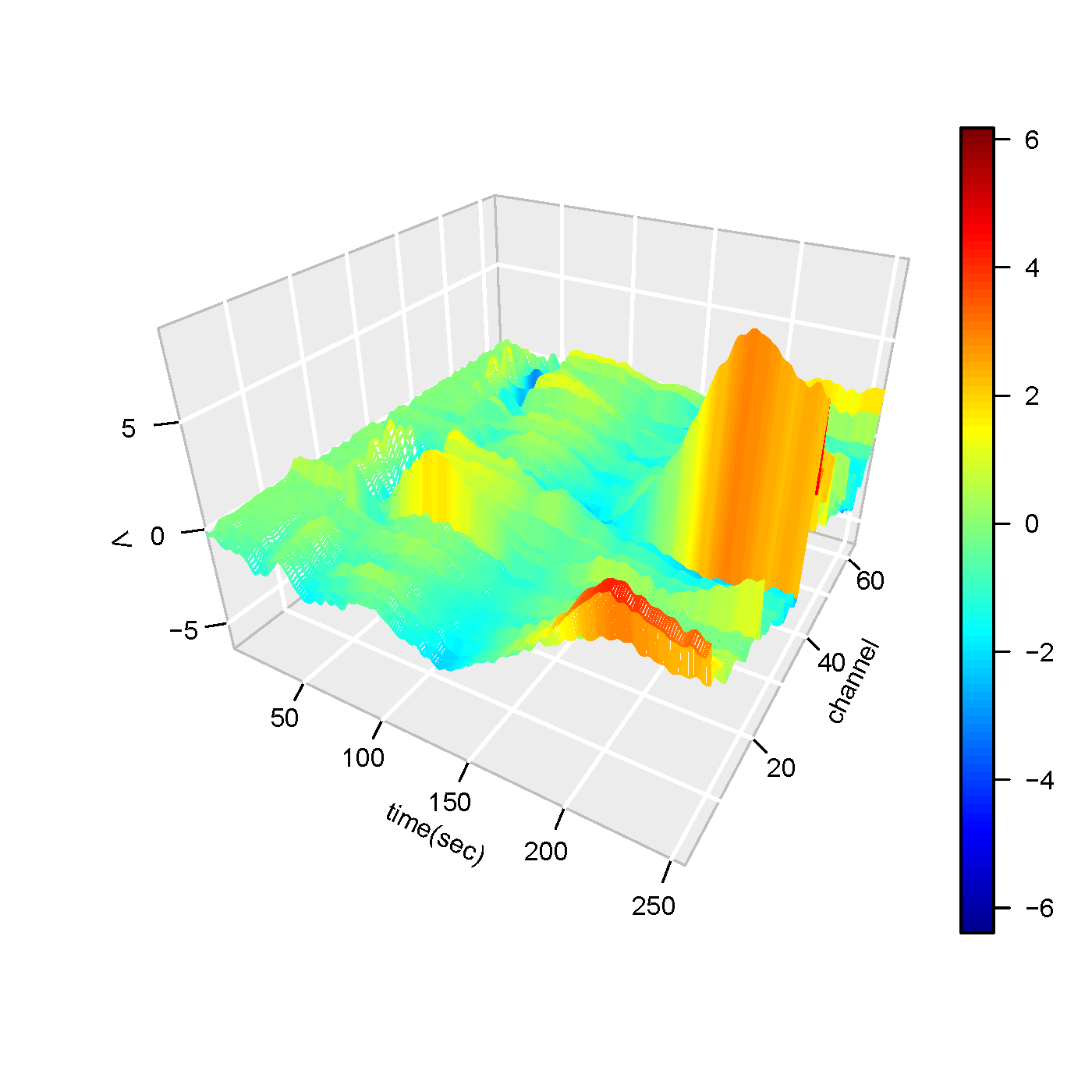}}
    \subfigure[\scriptsize Sample mean for the alcoholic group]{
		\includegraphics[width=0.45\textwidth]{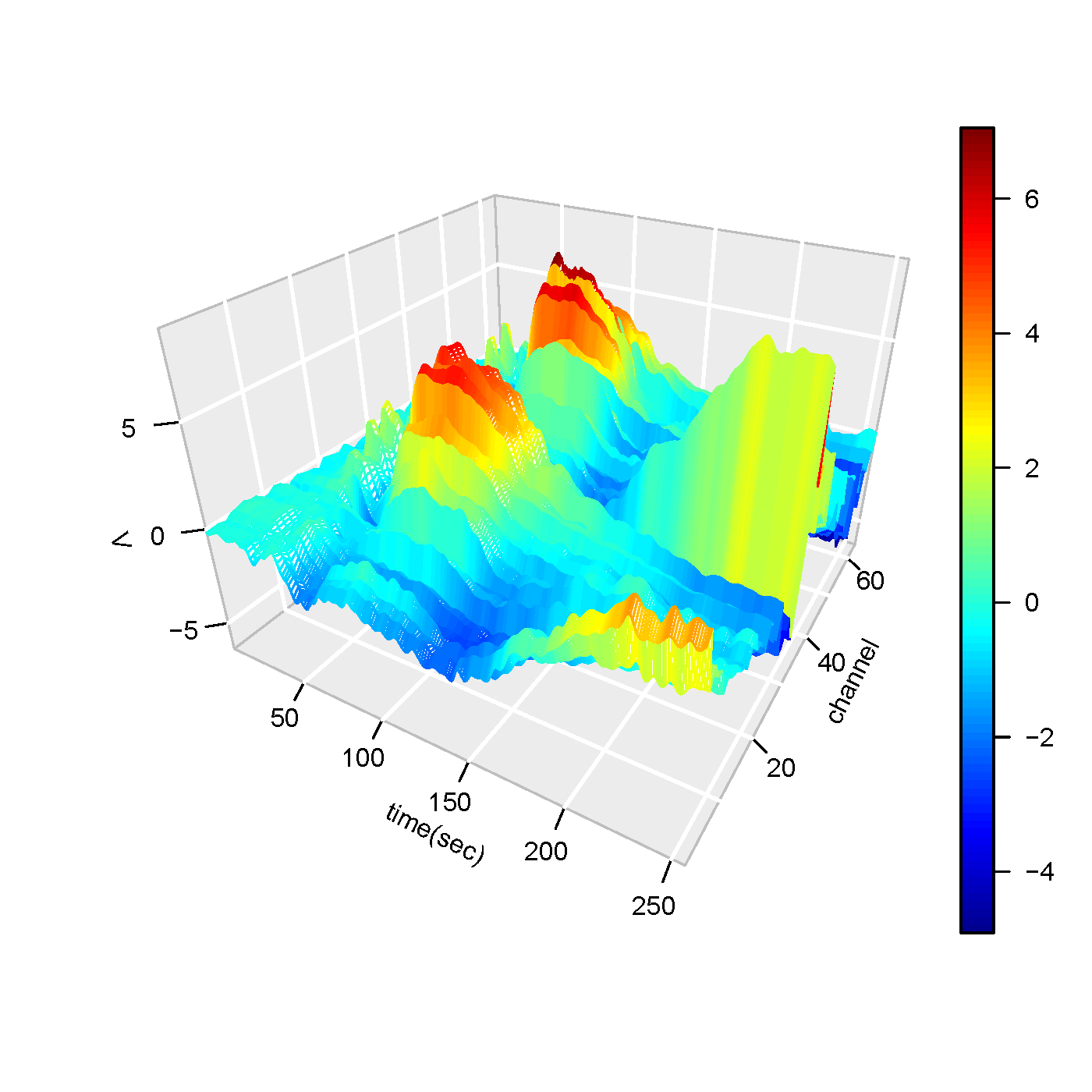}}\\
	\caption{Sample means of the voltage values measured from 64 channels of electrodes placed on the subject's scalp for 256 time points among the non-alcoholic and alcoholic groups.}
	\label{xyalcohol}
\end{figure}

We here are interested in testing whether the EEG patterns
between the control and alcoholic groups are different or not. Using the proposed MRP test, we obtain that the test statistic is 4.680 with the $p$-value approximately equal to  $10^{-8}$. Thus, there is a significant difference in the EEG pattern between the two groups.
The reasonability of the result is supported by their sample mean functions shown in Figure   \ref{xyalcohol} (a) and (b),
% .
% To visualize the data, Figures \ref{xyalcohol} (a) and (b) display  the EEG patterns for the first subjects in the non-alcoholic and alcoholic groups, 
with two horizontal axes representing time and channel and the vertical axis denoting voltage.

\section{Conclusions and Discussion}\label{Discussion}
This paper proposes a new multi-resolution projection approach
to detect the difference between two mean functions of high-dimensional functional data.
We derive the asymptotic properties of the proposed test statistic under the null and alternative hypotheses. The influence of the reconstruction of the functional curves on our test is also explored as observations are discretized in practice which are usually asynchronous. The usefulness of the proposed methods has been demonstrated by simulation studies and
two real datasets.

% Note that  the weight probability measures $G(\bfalpha)$ and $\nu(\gamma)$  in \eqref{equvlent-dpi} are  assumed to be
% Gaussian.  We here remark that the assumption can be relaxed to a more general one.  Specifically, we  can  assume   that the CDF of $\bfalpha$ satisfies   $G(\bfalpha)=\Pi_{k=1}^pG_k(\alpha_k),$ where  $G_k(\alpha_k)$ is a  CDF  with a mean 0 and variance 1, $k=1,\ldots,p$,  and  $\nu(\gamma)$ is any centered process  in  $L_{2}([0, 1])$ with a positive definite  covariance function.  In the general setting, 
% all the  results in the above sections still hold. 

This paper focuses on intensely measured function data. \citet{Zhu2023} considered a two-sample test for equality of distributions for sparsely measured functional data.
They proposed a test of marginal homogeneity by adapting to a sampling
plan and established the corresponding convergence rate. %\sout{An interesting future topic is to extend %our
%current work to sparse high-dimensional functional %data by the sampling plan.}
An interesting future direction is to study the two-sample test for the sparse high-dimensional functional data.
Recently, two-sample tests for equality of distributions have gained more and more attention in the literature of functional data, see  \citet{JMLR:v23:20-1180} and \citet{Zhu2023}. %\sout{Thus,
%another future topic is to extend our multi-resolution %projection-averaging approach to deal with
%this problem.}
Using our multi-resolution projection technique
to deal with this problem is also worthwhile to study.

% Acknowledgements and Disclosure of Funding should go at the end, before appendices and references

\acks{
% All acknowledgements go at the end of the paper before appendices and references.
% Moreover, you are required to declare funding (financial activities supporting the
% submitted work) and competing interests (related financial activities outside the submitted work).
% More information about this disclosure can be found on the JMLR website.
Dr. Shouxia Wang's research was funded by National Natural Science Foundation of China (Nos. 12292980, 12292983) and  China Postdoctoral Science Foundation (No.2023M730090).
Dr. Jiguo Cao's research was supported by the Natural Sciences and Engineering Research Council of Canada (NSERC) Discovery
grant (RGPIN-2023-04057).
Dr. Hua Liu's research was supported by 
National Natural Science Foundation of China (NSFC) (No.12201487), the Project funded by China Postdoctoral Science Foundation (No.2022M722544),
and the Fundamental Research Funds for the Central Universities (SK2022044).
Dr. Jinhong You's research was supported by 
the National Natural Science Foundation of China (NSFC) (No.11971291) and Innovative Research Team of Shanghai University of Finance and Economics. 
}


\newpage

\appendix
%\section{}

\section*{Appendix A: Proofs}\label{Appendix-A}
\setcounter{equation}{0}
\renewcommand{\theequation}{A.\arabic{equation}}
\renewcommand{\thelemma}{A.\arabic{lemma}}

\noindent \textbf{Proof  of  Lemma   \ref{lemma1}.}
	Suppose that $\E\{\X\}=\E\{\Y\}$. For any  $\bfalpha \in\mathbb{R}^{p}$ and  $\gamma \in L_{2}([0, 1])$,  $\langle\bfalpha^{\T}  \mathbf{X}, \gamma\rangle$ and $\langle\bfalpha^{\T}  \mathbf{Y}, \gamma\rangle$ are measurable functions of $\X$ and $\Y$ respectively. Then, we have that
$$\E{\langle\bfalpha^{\T}  \X, \gamma\rangle}
=\bfalpha^{\T}\int_{0}^{1} \E\{\X(t)\}\gamma(t) d t
=\bfalpha^{\T}\int_{0}^{1} \E\{\Y(t)\}\gamma(t) d t=\E{ \langle \bfalpha^{\T}\Y, \gamma\rangle}.$$

If $\E{\langle\bfalpha^{\T}  \X, \gamma\rangle}=\E{ \langle \bfalpha^{\T}\Y, \gamma \rangle}$ holds for any   $\bfalpha \in\mathbb{R}^{p}$ and  $\gamma \in L_{2}([0, 1])$, we have that
$$\bfalpha^{\T}\int_{0}^{1}\E\{[\X(t)-\Y(t)]\}\gamma(t) d t=
\E{\langle\bfalpha^{\T}[\X-\Y], \gamma\rangle}=0.$$
By the arbitrariness  of  $\bfalpha \in\mathbb{R}^{p}$ and  $\gamma \in L_{2}([0, 1])$,   the above result implies that $\E\{\X\}=\E\{\Y\}$.

\hfill\BlackBox

\noindent \textbf{Proof of Theorem  \ref{Theorem11}.} 
 Suppose $\bfalpha=(\alpha_1,\ldots,\alpha_p)^{\T}\sim G(\bfalpha)$, 
 where  $G(\cdot)$  is the CDF of $\bfalpha$,    $G(\bfalpha)=\Pi_{k=1}^pG_k(\alpha_k),$ $G_k(\alpha_k)$ is a  CDF  of $\alpha_k$ with mean 0 and variance 1 for  $k=1,\ldots,p$,  $\nu(\cdot)$ is the CDF of a centered process $\gamma$ in  $L_{2}([0, 1])$ with a positive definite  covariance function   $v(s, t)$,  and $\bfalpha$ and $\gamma$ are independent of  $\X$ and $\Y$. 
 % Suppose $\bfalpha=(\alpha_1,\ldots,\alpha_p)^{\T}\sim N_{p}(0, \mathbf{I}_{p})$   and $\gamma \in L_{2}[0,1]$ is  a centered Gaussian process  with a covariance function $v(s, t)$. 
 Then,  for any fixed functions $\Z_1,\Z_2 \in {L}_{2}^p[0,1]$, we have
	\begin{align}\label{s1-s2-projection}
		&\int_{L_{2} }  \int_{\mathbb{R}^{p } } \langle\bfalpha^{\T} \Z_{1}, \gamma\rangle \langle\bfalpha^{\T} \Z_{2}, \gamma\rangle d G(\bfalpha)\nu(\dif\gamma) \nonumber\\
		%=&\int_{L_{2} }  \int_{\mathbb{R}^{p } } \sum_{k=1}^{p} \langle\alpha_{k} Z_{1k}, \gamma\rangle \sum_{k=1}^{p} \langle\alpha_{k}Z_{2k}, \gamma\rangle dG(\bfalpha)\nu(d\gamma)\nonumber\\
		=&\int_{L_{2} }  \int_{\mathbb{R}^{p } } \Big[\sum_{k=1}^{p} \int_{0}^{1}\alpha_{k} Z_{1k}(t)\gamma(t)\dif t\Big] \Big[\sum_{k=1}^{p} \int_{0}^{1}\alpha_{k} Z_{2k}(t)\gamma(t)\dif t\Big] dG(\alpha_1,\ldots,\alpha_p)\nu(\dif\gamma)\nonumber\\
		=&\int_{L_{2} }  \int_{\mathbb{R}^{p } }\Big[ \sum_{k=1}^{p} \int_{0}^{1}\int_{0}^{1}\alpha_{k}^2 Z_{1k}(t)Z_{2k}(s)\gamma(s)\gamma(t)\dif s \dif t\Big] dG(\alpha_1,\ldots,\alpha_p)\nu(\dif\gamma)\nonumber\\
		=&\int_{L_{2} } \int_{0}^{1}\int_{0}^{1} \Z_{1}(t)^{\T}\Z_{2}(s)\gamma(s)\gamma(t)\dif s \dif t \nu(\dif\gamma)\nonumber\\
		=&\int_{0}^{1}\int_{0}^{1} \Z_{1}(t)^{\T}\Z_{2}(s)v(s,t)\dif s \dif t.
	\end{align}

  Denote $\Z_1=\X_1-\Y_1$ and $\Z_2=\X_2-\Y_2$. Then, by the definition of  $\MRP(\X,\Y)$ and \eqref{s1-s2-projection}, we have
	\begin{eqnarray*}
		 \MRP(\X,\Y) &=&  \int_{L_{2} }  \int_{\mathbb{R}^{p } } [\E\{{\langle\bfalpha^{\T}  (\X-\Y), \gamma\rangle}\}]^2 \dif G(\bfalpha)\nu(\dif\gamma)\\
		&=& \int_{L_{2} }  \int_{\mathbb{R}^{p } } \E\{{\langle\bfalpha^{\T}  (\X_1-\Y_1), \gamma\rangle}\}\E\{{\langle\bfalpha^{\T}  (\X_2-\Y_2), \gamma\rangle}\}\dif G(\bfalpha)\nu(\dif\gamma)\\
		%&=&\int_{L_{2} }  \int_{\mathbb{R}^{p } } \E\left({\langle\bfalpha^{\T}  \Z_1, \gamma\rangle}{\langle\bfalpha^{\T}  \Z_2, \gamma\rangle}\right) d G(\bfalpha)\nu(d\gamma)\\
		&=&\E\Big\{\int_{L_{2} }  \int_{\mathbb{R}^{p } } {\langle\bfalpha^{\T}  \Z_1, \gamma\rangle}{\langle\bfalpha^{\T}  \Z_2, \gamma\rangle} \dif G(\bfalpha)\nu(\dif\gamma)\Big\}\\
		&=&\E\Big\{\int_{0}^{1}\int_{0}^{1} \Z_{1}(t)^{\T}\Z_{2}(s)v(s,t)\dif s \dif t\Big\}\\
		&=& \E \Big\{\int_{0}^{1}\int_{0}^{1}\left(\X_1(t)-\Y_1(t)\right)^{\T}\left(\X_2(s)-\Y_2(s)\right)v(s, t) \dif s \dif t\Big\}\\
%		\\&=&\E\int_{0}^{1}\int_{0}^{1}\left[\X_1(t)^{\T}\X_2(s)+\Y_1(t)^{\T}\Y_2(s)-\X_1(t)^{\T}\Y_2(s)-\Y_1(t)^{\T}\X_2(s)\right]v(s, t) \dif s \dif t
		\\ &=&  \int_{0}^{1}\int_{0}^{1}[\bfmu_{1}(t)-\bfmu_{2}(t)]^{\T}[\bfmu_{1}(s)-\bfmu_{2}(s)]v(s, t) \dif s \dif t.
	\end{eqnarray*}
 
\hfill\BlackBox

\noindent  \textbf{Proof of Theorem  \ref{Theorem2}.}
We first calculate $\E\{\widehat{\MRP}(\X,\Y)\}$ and $\VAR\{\widehat{\MRP}(\X,\Y)\}$. By the definition of $\widehat{\MRP}(\X,\Y)$, it is easy to show that
\begin{eqnarray*}
&&\E\{\widehat{\MRP}(\X,\Y)\}\\
&=&
\int_{0}^{1}\int_{0}^{1}[\E\{\X_1(t)^{\T}\X_2(s)\}+\E\{\Y_1(t)^{\T}\Y_2(s)\}-2\E\{\X_1(t)\}^{\T}\E\{\Y_1(s)\}\} ]v(s, t) \dif s \dif t\\
&=&\int_{0}^{1}\int_{0}^{1}\left(\bfmu_{1}(t)-\bfmu_{2}(t)\right)^{\T}\left(\bfmu_{1}(s)-\bfmu_{2}(s)\right)v(s, t) \dif s \dif t\\
&=&{\MRP}(\X,\Y).
\end{eqnarray*}

  For simplicity,  let
 \begin{eqnarray*}
 	D_{1}&=&  \frac{1}{ n(n-1) }\sum_{i, j=1 \atop i \neq j}^{n}\int_{0}^{1}\int_{0}^{1}\X_i(t)^{\T}\X_j(s)v(s, t) \dif s \dif t,\\
 	D_{2}&=& \frac{1}{ m(m-1) }\sum_{i, j=1 \atop i \neq j}^{m}\int_{0}^{1}\int_{0}^{1}\Y_i(t)^{\T}\Y_j(s)v(s, t) \dif s \dif t,\\
 	D_{3}&=& -\frac{2}{ nm }\sum_{i=1}^{n}\sum_{j=1}^{m}  \int_{0}^{1}\int_{0}^{1}\X_i(t)^{\T}\Y_j(s)v(s, t) \dif s \dif t.
 \end{eqnarray*}
Note that $\VAR\{\widehat{\MRP}(\X,\Y)\}$ can be decomposed into
\begin{align}\label{decompose-var}
\VAR\{\widehat{\MRP}(\X,\Y)\}&=\VAR\{D_{1}\}+\VAR\{D_{2}\}+\VAR\{D_{3}\}\nonumber\\
&\quad\quad+2\operatorname{Cov}\{D_{1}, D_{2}\}-2\operatorname{Cov}\{D_{1}, D_{3}\}-2\operatorname{Cov}\{D_{2}, D_{3}\}.
\end{align}
To obtain $\VAR\{\widehat{\MRP}(\X,\Y)\}$, we next calculate  each term in the above decomposition.

By Lemma \ref{lemmaB1} in Appendix B, we have that
\begin{align}\label{Var-D1}
\operatorname{Var}\{D_{1}\}
&=\frac{1}{n^2 (n-1 )^2} \operatorname{Var}\Big\{\sum_{i, j=1 \atop i \neq j}^{n}\iint\X_i(t)^{\T}\X_j(s)v(s, t) \dif s \dif t\Big\}
	\nonumber\\
&=\frac{2}{n (n-1 )} \iiiint\operatorname{tr}\{\G_{1}(t_1,t)\G_{1}(s_1,s)\} v(s,t)v(s_1,t_1) \dif s \dif t \dif s_1 \dif t_1
	\nonumber\\
&\quad\quad+	\frac{4}{n} \iiiint  \bfmu_{1}(t)^{\T}\G_{1}(s,s_1)\bfmu_{1}(t_1)
	v(s,t)v(s_1,t_1) \dif s \dif t \dif s_1 \dif t_1.
\end{align}
Similarly, it can be shown  that
\begin{align}\label{Var-D2}
\operatorname{Var}\{D_{2}\}&=\frac{2}{m (m-1)} \iiiint\operatorname{tr} \{\G_{2}(t_1,t)\G_{2}(s_1,s) \} v(s,t)v(s_1,t_1) \dif s \dif t \dif s_1 \dif t_1
	\nonumber\\&\quad\quad+	\frac{4}{m} \iiiint   \bfmu_{2}(t)^{\T}\G_{2}(s,s_1)\bfmu_{2}(t_1)
	v(s,t)v(s_1,t_1) \dif s \dif t \dif s_1 \dif t_1,
\end{align}
 \begin{align}\label{Var-D3}
\operatorname{Var}\{D_{3}\}&=\frac{4}{nm} \iiiint\operatorname{tr}\{\G_{1}(s,s_1)\G_{2}(t,t_1) \} v(s,t)v(s_1,t_1) \dif s \dif t \dif s_1 \dif t_1\nonumber\\&\quad\quad+\frac{4}{n} \iiiint  \bfmu_{2}(t)^{\T}\G_{1}(s,s_1)\bfmu_{2}(t_1)
v(s,t)v(s_1,t_1) \dif s \dif t \dif s_1 \dif t_1\nonumber\\
&\quad\quad+\frac{4}{m} \iiiint \bfmu_{1}(t)^{\T}\G_{2}(s,s_1)\bfmu_{1}(t_1)
v(s,t)v(s_1,t_1) \dif s \dif t \dif s_1 \dif t_1,
\end{align}
\begin{align}\label{Var-D1-D3}
\operatorname{Cov}\{D_{1}, D_{3}\}=-\frac{4}{n} \iiiint \bfmu_{1}(t)^{\T}\G_{1}(s,s_1)\bfmu_{2}(t_1)
v(s,t)v(s_1,t_1) \dif s \dif t \dif s_1 \dif t_1,
\end{align}
and
\begin{align}\label{Var-D2-D3}
\operatorname{Cov}\{D_{2}, D_{3}\}=-\frac{4}{m} \iiiint \bfmu_{1}(t)^{\T}\G_{2}(s,s_1)\bfmu_{2}(t_1)
v(s,t)v(s_1,t_1) \dif s \dif t \dif s_1 \dif t_1.
\end{align}
Additionally,  we have that  $\operatorname{Cov}\{D_{1}, D_{2}\}=0$ due to independence between $\{\X_i \}_{i=1}^{n}$ and $\{\Y_i \}_{i=1}^{m}$.
This,  together with \eqref{decompose-var}-\eqref{Var-D1-D3}, yields that
\begin{align*}
	&\VAR\{\widehat{\MRP}(\X,\Y)\}\\
&=\frac{2}{n (n-1)} \iiiint\operatorname{tr}\{\G_{1}(s,s_1)\G_{1}(t,t_1)\} v(s,t)v(s_1,t_1) \dif s \dif t \dif s_1 \dif t_1
	\\&+\frac{2}{m (m-1)} \iiiint\operatorname{tr}\{\G_{2}(s,s_1)\G_{2}(t,t_1)\} v(s,t)v(s_1,t_1) \dif s \dif t \dif s_1 \dif t_1
	\\&+\frac{4}{nm} \iiiint\operatorname{tr}\{\G_{1}(s,s_1)\G_{2}(t,t_1)\} v(s,t)v(s_1,t_1) \dif s \dif t \dif s_1 \dif t_1
	\\&+\frac{4}{n} \iiiint [\bfmu_{1}(t)-\bfmu_{2}(t)]^{\T}\G_{1}(s,s_1)[\bfmu_{1}(t_1)-\bfmu_{2}(t_1)]
	v(s,t)v(s_1,t_1) \dif s \dif t \dif s_1 \dif t_1
	\\&+\frac{4}{m} \iiiint [\bfmu_{1}(t)-\bfmu_{2}(t)]^{\T}\G_{2}(s,s_1)[\bfmu_{1}(t_1)-\bfmu_{2}(t_1)]
	v(s,t)v(s_1,t_1) \dif s \dif t \dif s_1 \dif t_1.
\end{align*}

Let
\begin{align*}
	\sigma_{nm}^{2}(\X,\Y)&=:\frac{2}{n\left(n-1\right)} \iiiint\operatorname{tr}\left\{\G_{1}(s,s_1)\G_{1}(t,t_1)\right\} v(s,t)v(s_1,t_1) \dif s \dif t \dif s_1 \dif t_1
	\\&+\frac{2}{m\left(m-1\right)} \iiiint\operatorname{tr}\left\{\G_{2}(s,s_1)\G_{2}(t,t_1)\right\} v(s,t)v(s_1,t_1) \dif s \dif t \dif s_1 \dif t_1
	\\& +\frac{4}{nm} \iiiint\operatorname{tr}\left\{\G_{1}(s,s_1)\G_{2}(t,t_1)\right\} v(s,t)v(s_1,t_1) \dif s \dif t \dif s_1 \dif t_1.
\end{align*}
Thus, under $H_{0}: \bfmu_{1}= \bfmu_{2}$, we can show that $\VAR\{\widehat{\MRP}(\X,\Y)\}=\sigma_{nm}^{2}(\X,\Y)$.
Under $H_{1}: \bfmu_{1} \neq \bfmu_{2}$, we have that
$\VAR\{\widehat{\MRP}(\X,\Y)\}=\sigma_{nm}^{2}(\X,\Y)\{1+o(1)\}$  by Condition (C3).
Let 
\begin{align*}
	\sigma_{nm_2}^{2}(\X,\Y)
&=\frac{4}{n} \iiiint [\bfmu_{1}(t)-\bfmu_{2}(t)]^{\T}\G_{1}(s,s_1)[\bfmu_{1}(t_1)-\bfmu_{2}(t_1)]
	v(s,t)v(s_1,t_1) \dif s \dif t \dif s_1 \dif t_1
	\\&+\frac{4}{m} \iiiint [\bfmu_{1}(t)-\bfmu_{2}(t)]^{\T}\G_{2}(s,s_1)[\bfmu_{1}(t_1)-\bfmu_{2}(t_1)]
	v(s,t)v(s_1,t_1) \dif s \dif t \dif s_1 \dif t_1.
\end{align*}
Then under Condition (C3'), $\VAR\{\widehat{\MRP}(\X,\Y)\}=\sigma_{nm_2}^{2}(\X,\Y)\{1+o(1)\}$.
% and with condition \eqref{condition2},
%$$
%\VAR\{\widehat{\MRP}(\X,\Y)\}=\sigma_{nm 2}^{2}\{1+o(1)\},
%$$
%where
%\begin{align*}
%\sigma_{nm 2}^{2}&=\frac{4}{n} \iiiint \left(\bfmu_{1}(t)-\bfmu_{2}(t)\right)^{\T}\G_{1}(s,s_1)\left(\bfmu_{1}(t_1)-\bfmu_{2}(t_1)\right)
%	v(s,t)v(s_1,t_1) \dif s \dif t \dif s_1 \dif t_1
%	\\&+\frac{4}{m} \iiiint \left(\bfmu_{1}(t)-\bfmu_{2}(t)\right)^{\T}\G_{2}(s,s_1)\left(\bfmu_{1}(t_1)-\bfmu_{2}(t_1)\right)
%	v(s,t)v(s_1,t_1) \dif s \dif t \dif s_1 \dif t_1
%\end{align*}

We next establish the asymptotic normality of $\widehat\MRP(\X, \Y)$. Note that
$$\widehat\MRP(\X, \Y)=\widehat\MRP_1(\X, \Y)+\widehat\MRP_2(\X, \Y),$$
where
	\begin{eqnarray*}
		\widehat{\MRP}_1 (\X,\Y) &=&   \frac{1}{ n(n-1) }\sum_{i, j=1 \atop i \neq j}^{n}\int_{0}^{1}\int_{0}^{1}\left[\X_i(t)-\bfmu_1(t)\right]^{\T}\left[\X_j(s)-\bfmu_1(s)\right]v(s, t) \dif s \dif t \nonumber\\
		&& +  \frac{1}{ m(m-1) }\sum_{i, j=1 \atop i \neq j}^{m}\int_{0}^{1}\int_{0}^{1}\left[\Y_i(t)-\bfmu_2(t)\right]^{\T}\left[\Y_j(s)-\bfmu_2(s)\right]v(s, t) \dif s \dif t\nonumber\\
		&& -\frac{2}{ nm }\sum_{i=1}^{n}\sum_{j=1}^{m}  \int_{0}^{1}\int_{0}^{1}\left[\X_i(t)-\bfmu_1(t)\right]^{\T}\left[\Y_j(s)-\bfmu_2(s)\right]v(s, t) \dif s \dif t
	\end{eqnarray*}
and
\begin{eqnarray*}
\widehat{\MRP}_2 (\mathbf{X},\mathbf{Y}) &=&   \frac{2}{ n}\sum_{i=1 }^{n}\int_{0}^{1}\int_{0}^{1}\left[\X_i(t)-\bfmu_1(t)\right]^{\T}\left[\bfmu_1(s)-\bfmu_2(s)\right]v(s, t) \dif s \dif t\\
	&&+  \frac{2}{ m }\sum_{i=1}^{m}\int_{0}^{1}\int_{0}^{1}\left[\Y_i(t)-\bfmu_2(t)\right]^{\T}\left[\bfmu_2(s)-\bfmu_1(s)\right]v(s, t) \dif s \dif t\\
	&&+ \int_{0}^{1}\int_{0}^{1}\left[\bfmu_1(t)-\bfmu_2(t)\right]^{\T}\left[\bfmu_1(s)-\bfmu_2(s)\right]v(s, t) \dif s \dif t.
\end{eqnarray*}
It is easy to show that
$$\E\{\widehat{\MRP}_1 (\mathbf{X},\mathbf{Y})\}=0,~~\E\{\widehat{\MRP}_2 (\mathbf{X},\mathbf{Y})\}= {\MRP}(\mathbf{X},\mathbf{Y}),$$
and
\begin{eqnarray*}
	&&\operatorname{Var}\{\widehat{\MRP}_2 (\mathbf{X},\mathbf{Y})\}\\
&=&\frac{4}{n} \iiiint [\bfmu_{1}(t)-\bfmu_{2}(t)]^{\T}\G_{1}(s,s_1)[\bfmu_{1}(t_1)-\bfmu_{2}(t_1)]
	v(s,t)v(s_1,t_1) \dif s \dif t \dif s_1 \dif t_1
	\\&&+\frac{4}{m} \iiiint [\bfmu_{1}(t)-\bfmu_{2}(t)]^{\T}\G_{2}(s,s_1)[\bfmu_{1}(t_1)-\bfmu_{2}(t_1)]
	v(s,t)v(s_1,t_1) \dif s \dif t \dif s_1 \dif t_1.
\end{eqnarray*}

Under Condition (C3),  we obtain that
\begin{align*}
	\operatorname{Var}\left\{\frac{\widehat{\MRP}_2 (\X,\Y)-{\MRP} (\mathbf{X},\mathbf{Y})}{\sigma_{nm}(\X,\Y)}\right\} =o(1).
\end{align*}
Thus, under Condition (C3) we have that
\begin{align}	\frac{\widehat{\MRP}(\X,\Y)-{\MRP}(\X,\Y)}{\sqrt{\VAR\{\widehat{\MRP}(\X,\Y)\}}} &=\frac{\widehat{\MRP}_1(\X,\Y)}{\sigma_{nm}(\X,\Y)\{1+o_{p}(1)\}}
+\frac{\widehat{\MRP}_2(\X,\Y)-{\MRP}(\X,\Y)}{\sigma_{nm}(\X,\Y)\{1+o_{p}(1)\}}\nonumber\\
&=\frac{\widehat{\MRP}_1(\X,\Y)}{\sigma_{nm }(\X,\Y)}+o_{p}(1).
	\label{CLT1}
\end{align}
Under Condition (C3'), we have that
\begin{align}	\frac{\widehat{\MRP}(\X,\Y)-{\MRP}(\X,\Y)}{\sqrt{\VAR\{\widehat{\MRP}(\X,\Y)\}}} &=\frac{\widehat{\MRP}_2(\X,\Y)-{\MRP}(\X,\Y)}{\sigma_{nm_2}(\X,\Y)}
+o_{p}(1).
	\label{CLT1_C3}
\end{align}

To obtain the asymptotic normality of $\widehat{\MRP}_1 (\X,\Y)$, we  assume without loss of generality
that  $\bfmu_{1}(t)=\bfmu_{2}(t)=\0$. For simplicity, define $\W_{i}=\X_{i}$ for $i=1, \ldots, n$ and $\W_{j+n}=\Y_{j}$ for $j=1, \ldots, m$, and for $i \neq j$
$$
\phi_{i j}= \begin{cases}	
\frac{1}{n(n-1)} \iint\W_i(t)^{\T}\W_j(s)v(s, t) \dif s \dif t, & \text {if } i, j \in\left\{1, \ldots, n\right\}, \\
-\frac{1}{nm} \iint\W_i(t)^{\T}\W_j(s)v(s, t) \dif s \dif t, & \text {if } i \in\{1, \ldots, n\},  j \in \{n+1, \ldots, n+m \}, \\
 \frac{1}{m(m-1)} \iint\W_i(t)^{\T}\W_j(s)v(s, t) \dif s \dif t, & \text {if } i, j \in\left\{n+1, \ldots, n+m\right\}.
\end{cases}
$$
Define
 $$V_{j}=\sum_{i=1}^{j-1} \phi_{i j}, \text{ for }j=2,3, \ldots,n+m; ~~~S_{k}=\sum_{j=2}^{k} V_{j};$$
  $$\mathcal{F}_{k}=\sigma\{\W_{1}(t),   \ldots, \W_{k}(t), t\in[0,1]\},$$
 where   $\sigma\{\W_{1}(t),   \ldots, \W_{k}(t), t\in[0,1]\}$ is the $\sigma$-algebra generated by $\{\W_{1}(t),  \ldots, \W_{k}(t)\}$, $t\in[0,1]$. Then, by the definition of $\widehat{\MRP}_1(\X,\Y)$, we have that
$$
\widehat{\MRP}_1 (\X,\Y)=2 \sum_{j=2}^{n+m} \sum_{i=1}^{j-1} \phi_{i j}=2 \sum_{j=2}^{n+m} V_{j}.
$$
Then, by Lemmas \ref{lemma3}-\ref{lemma5} in Appendix B   and Corollary 3.1 of \cite{hallMartingaleLimitTheory1980}, we have that
$$
\widehat \MRP_1(\X,\Y) / \widehat{\sigma}_{nm}(\X,\Y) \stackrel{D}{\longrightarrow} N(0,1), \quad \text { as } p \rightarrow \infty \text { and } \min\{n,m\} \rightarrow \infty.
$$
This, together with  \eqref{CLT1},  completes the proof of Theorem \ref{Theorem2}.
Under condition (C3'), since $\widehat \DPI_2(\X,\Y)$ is the sum of two independent averages, its asymptotic normality can be attained by following the standard means.
 
 \hfill\BlackBox

\noindent  \textbf{Proof of Theorem  \ref{Theorem3}.}
We only show the proof of the ratio consistency of
\begin{align*}
   & \iiiint\widehat{\mathcal{TR}}_{11}(s,s_1,t,t_1) v(s,t)v(s_1,t_1) \dif s \dif t \dif s_1 \dif t_1
    \\&\quad=\iiiint\widehat{\operatorname{tr}\left\{\G_{1}(s,s_1)\G_{1}(t,t_1)\right\} }v(s,t)v(s_1,t_1) \dif s \dif t \dif s_1 \dif t_1
\end{align*}
since the proofs of the other two follow the same procedure.
Consider the following decomposition
\begin{eqnarray}
&&\widehat{\operatorname{tr} \{\G_{1}(s,s_1)\G_{1}(t,t_1)\}}\nonumber\\
&=&\frac{1}{n(n-1)} \operatorname{tr} \Big\{\sum_{j \neq k}^{n} (\X_{j}(s)-\overline{\X}_{(j, k)}(s) ) \X_{ j}^{\T}(s_1) (\X_{k}(t)-\overline{\X}_{(j, k)}(t) ) \X_{k}^{\T}(t_1) \Big\}\nonumber\\
&=&\frac{1}{n(n-1)}\operatorname{tr}\Bigg\{\sum _ { j \neq k }^{ n } \Big\{\left(\X_{j}(s)-\bfmu_1(s)\right)\left(\X_{j}(s_1)-\bfmu_1(s_1)\right)^{\T}\left(\X_{k}(t)-\bfmu_1(t)\right)\left(\X_{k}(t_1)-\bfmu_1(t_1)\right)^{\T}\nonumber\\
	&&-2\left(\overline{\X}_{(j, k)}(s)-\bfmu_1(s)\right)\left(\X_{j}(s_1)-\bfmu_1(s_1)\right)^{\T}\left(\X_{k}(t)-\bfmu_1(t)\right)\left(\X_{k}(t_1)-\bfmu_1(t_1)\right)^{\T}\Big\}\nonumber\\
	&&+\sum_{j \neq k}^{n}\left\{2\left(\X_{j}(s)-\bfmu_1(s)\right) \bfmu_1(s_1)^{\T}\left(\X_{k}(t)-\bfmu_1(t)\right)\left(\X_{k}(t_1)-\bfmu_1(t_1)\right)^{\T}\right.\nonumber\\
	&&\left.-2\left(\overline{\X}_{(j, k)}(s)-\bfmu_1(s)\right) \bfmu_1(s_1)^{\T}\left(\X_{k}(t)-\bfmu_1(t)\right)\left(\X_{k}(t_1)-\bfmu_1(t_1)\right)^{\T}\right\}\nonumber\\
	&&+\sum_{j \neq k}^{n}\left\{\left(\overline{\X}_{(j, k)}(s)-\bfmu_1(s)\right)\left(\X_{j}(s_1)-\bfmu_1(s_1)\right)^{\T}\left(\bar{\X}_{(j, k)}(t)-\bfmu_1(t)\right)\left(\X_{k}(t_1)-\bfmu_1(t_1)\right)^{\T}\right\}\nonumber\\
	&&-\sum_{j \neq k}^{n}\left\{2\left({\X}_{j}(s)-\bfmu_1(s)\right)\bfmu_1(s_1)^{\T}\left(\overline{\X}_{(j, k)}(t)-\bfmu_1(t)\right)\left(\X_{k}(t_1)-\bfmu_1(t_1)\right)^{\T}\right.\nonumber\\
	&&\left.-2\left(\overline{\X}_{(j, k)}(s)-\bfmu_1(s)\right)\left(\X_{j}(s_1)-\bfmu_1(s_1)\right)^{\T}\left(\overline{\X}_{(j, k)}(t)-\bfmu_1(t)\right)\bfmu_1(t_1)^{\T}\right\}\nonumber\\
	&&+\sum_{j \neq k}^{n}\left\{\left({\X}_{j}(s)-\bfmu_1(s)\right)\bfmu_1(s_1)^{\T}\left({\X}_{ k}(t)-\bfmu_1(t)\right)\bfmu_1(t_1)^{\T}\right.\nonumber\\
&&\left.	-2\left(\overline{\X}_{(j, k)}(s)-\bfmu_1(s)\right)\bfmu_1(s_1)^{\T}\left({\X}_{ k}(t)-\bfmu_1(t)\right)\bfmu_1(t_1)^{\T}\right\}\nonumber\\
	&&+\sum_{j \neq k}^{n}\left(\overline{\X}_{(j, k)}(s)-\bfmu_1(s)\right)\bfmu_1(s_1)^{\T}\left(\overline{\X}_{(j, k)}(t)-\bfmu_1(t)\right)\bfmu_1(t_1)^{\T}\Bigg\}\nonumber\\ 
	&\triangleq& \sum_{k=1}^{10} \operatorname{tr}\{\A_{k}\}. \label{trG1G1}
\end{eqnarray}
Simple calculations obtain that  $$\E\{\operatorname{tr}\{\A_{1}\}\}=\operatorname{tr}\{\G_{1}(s,s_1)\G_{1}(t,t_1)\}, ~~~\E\{\operatorname{tr}\{\A_{k}\}\}=0,~~  k=2, \ldots, 9,$$ $$\E \{\operatorname{tr}\{\A_{10} \} \}=\bfmu_1(s_1)^{\T} \G_{1}(s,t)\bfmu_1(t_1) /(n-2).$$
This, together with Condition (C3), yields  that
\begin{align}
 &\E\left\{\iiiint\widehat{\operatorname{tr}\left\{\G_{1}(s,s_1)\G_{1}(t,t_1)\right\} }v(s,t)v(s_1,t_1) \dif s \dif t \dif s_1 \dif t_1\right\}\nonumber
 \\&\quad= \iiiint\E \{\widehat{\operatorname{tr} \{\G_{1}(s,s_1)\G_{1}(t,t_1) \} } \}v(s,t)v(s_1,t_1) \dif s \dif t \dif s_1 \dif t_1\nonumber \\&\quad=\iiiint\operatorname{tr}\left\{\G_{1}(s,s_1)\G_{1}(t,t_1)\right\} v(s,t)v(s_1,t_1) \dif s \dif t \dif s_1 \dif t_1
 \{1+o(1)\}.  \label{etrg1g1}
\end{align}

Note that
\begin{eqnarray}\label{var-sum}
 &&\operatorname{Var}\Big\{ \iiiint\widehat{\operatorname{tr}\{\G_{1}(s,s_1)\G_{1}(t,t_1)\} }v(s,t)v(s_1,t_1) \dif s \dif t d s_1dt_1\Big\} \nonumber\\
 &\leq& 10 \sum_{k=1}^{10}\operatorname{Var}\left\{ \iiiint\operatorname{tr}\{\A_{k}\}v(s,t)v(s_1,t_1) \dif s \dif t d s_1dt_1\right\}.
\end{eqnarray}
We next establish the orders of  each term $\operatorname{Var}\left\{ \iiiint\operatorname{tr}\{\A_{k}\}v(s,t)v(s_1,t_1) \dif s \dif t d s_1dt_1\right\}$
to obtain the   order  of $\operatorname{Var}\Big\{ \iiiint\widehat{\operatorname{tr}\{\G_{1}(s,s_1)\G_{1}(t,t_1)\} }v(s,t)v(s_1,t_1) \dif s \dif t d s_1dt_1\Big\}.$

We first consider the term $\operatorname{Var}\left\{ \iiiint\operatorname{tr}\{\A_{1}\}v(s,t)v(s_1,t_1) \dif s \dif t d s_1dt_1\right\}$.
Since
\begin{align*}
 &\operatorname{E}\left\{\Big[ \iiiint\operatorname{tr}\left\{\A_{1}\right\}v(s,t)v(s_1,t_1) \dif s \dif t d s_1dt_1\Big]^2\right\}\\
 &= \frac{1}{n^{2}(n-1)^{2}} \E\Big\{\iiiint\operatorname{tr} \{\sum_ { j_1 \neq k_1 }^{ n } \left(\X_{j_1}(s)-\bfmu_1(s)\right)\left(\X_{j_1}(s_1)-\bfmu_1(s_1)\right)^{\T}\\ &\quad \quad \times\left(\X_{k_1}(t)-\bfmu_1(t)\right)\left(\X_{k_1}(t_1)-\bfmu_1(t_1)\right)^{\T} \}  v(s,t)v(s_1,t_1) \dif s \dif t \dif s_1 \dif t_1  \\
	&\quad\quad \times \iiiint\operatorname{tr} \{\sum _ { j_2 \neq k_2 }^{ n } \left(\X_{j_2}(s_2)-\bfmu_1(s_2)\right)\left(\X_{j_2}(s_3)-\bfmu_1(s_3)\right)^{\T} \\ & \quad \quad \times\left(\X_{k_2}(t_2)-\bfmu_1(t_2)\right)\left(\X_{k_2}(t_3)-\bfmu_1(t_3)\right)^{\T} \}  v(s_2,t_2)v(s_3,t_3) \dif s_2 \dif t_2 \dif s_3 \dif t_3\Big\},
\end{align*}
we have that
\begin{eqnarray}
 &&\operatorname{Var}\left\{ \iiiint\operatorname{tr}\left\{\A_{1}\right\}v(s,t)v(s_1,t_1) \dif s \dif t d s_1dt_1\right\}\nonumber\\
 &=& \frac{2}{n(n-1)} B_{11}+\frac{4(n-2)}{n(n-1)} B_{12}\nonumber\\
 &&+o\Big\{\Big(\iiiint\operatorname{tr} \{\G_{1}(s,s_1)\G_{1}(t,t_1)\} v(s,t)v(s_1,t_1) \dif s \dif t \dif s_1 \dif t_1\Big)^2\Big\},
\label{vartra1}
\end{eqnarray}
where
\begin{eqnarray*}
B_{11}&=&  \E\Big\{\Big[\iint (\X_{1}(s)-\bfmu_1(s) )^{\T} (\X_{2}(t)-\bfmu_1(t) ) v(s,t) \dif s \dif t \Big]^{4} \Big\},\\
B_{12}&=& \E\Big\{\Big[\iiiint (\X_{1}(s)-\bfmu_1(s))^{\T}\G_{1}(t,t_1) (\X_{1}(s_1)-\bfmu_1(s_1) )  v(s,t)v(s_1,t_1) \dif s \dif t \dif s_1 \dif t_1\Big]^{2}\Big\}.
\end{eqnarray*}

Let $\gamma_{jk}(s,t)$ and $u_{j k}(s,t,t_1,s_1)$ be the $(j, k)$-element of $\bfGamma^{\T}(s) \bfGamma(t)$ and $\bfGamma^{\T}(s) \G(t,t_1) \bfGamma(s_1)$, respectively.
By Condition (C1)  in \eqref{factor}, we have   that
\begin{align*}
	 B_{11} &=\E\Big\{\iint\Z_{1}^{\T}(s)\bfGamma_1(s)^{\T}\bfGamma_1(t)\Z_{2}(t)  \dif s \dif t \Big\}^{4}=\E\Big\{\sum_{j, k=1}^{d} \iint z_{1 j}(s) \gamma_{j k}(s,t) z_{2 k}(t)\dif s \dif t\Big\}^{4} \\ &=\E\Big\{\sum_{j_{1}, j_{2}, j_{3}, j_{4}, k_{1}, k_{2}, k_{3}, k_{4}=1}^{d}\idotsint\gamma_{j_1 k_1}(s,t) \gamma_{j_2 k_2}(s_1,t_1) \gamma_{j_3 k_3}(s_2,t_2)\gamma_{j_4 k_4}(s_3,t_3) \\ &\quad \quad \times z_{1 j_{1}}(s) z_{1 j_{2}}(s_1) z_{1 j_{3}}(s_2) z_{1 j_{4}}(s_3) z_{2 k_{1}}(t) z_{2 k_{2}}(t_1) z_{2 k_{3}}(t_2) z_{2 k_{4}}(t_3)\dif s \dif t \dif s_1 \dif t_1\dif s_2 \dif t_2 \dif s_3 \dif t_3\Big\}
\end{align*}
and
\begin{align*}
B_{12}&=\E\Big\{\iiiint \Z_{1}^{\T}(s)\bfGamma_1(s)^{\T}\G_{1}(t,t_1)\Z_{1}(s_1)\bfGamma_1(s_1)  \dif s \dif t \dif s_1 \dif t_1\Big\}^{2}\\&=\E\Big\{\sum_{j, k=1}^{d}\iiiint    z_{1 j}(s) u_{j k}(s,t,t_1,s_1) z_{1 k}(s_1) \dif s \dif t \dif s_1 \dif t_1\Big\}^{2}\\
&=\E\Big\{\sum_{j_{1}, j_{2}, k_{1}, k_{2}=1}^{d}\idotsint u_{j_{1} k_{1}}(s,t,t_1,s_1) u_{j_{2} k_{2}}(s_2,t_2,t_3,s_3)\\ &\quad \times z_{1 j_{1}}(s) z_{1 j_{2}}(s_1) z_{1 k_{1}}(s_2) z_{1 k_{2}}(s_3)\dif s \dif t \dif s_1 \dif t_1\dif s_2 \dif t_2 \dif s_3 \dif t_3\Big\},
\end{align*}
taking $v(s,t)\equiv 1$ for simplification.
Note that
\begin{align*}
\Big[\iiiint\operatorname{tr}\{\G_{1}(s,s_1)\G_{1}(t,t_1)\} \dif s \dif t \dif s_1 \dif t_1\Big]^2=\Big[\sum_{j, k=1}^{d} \iiiint \gamma_{j k}(s,t)\gamma_{j k}(s_1,t_1)\dif s \dif t \dif s_1 \dif t_1\Big]^{2}
\end{align*}
 and
\begin{align*}&\idotsint\operatorname{tr}\left\{\G_{1}(s,s_1)\G_{1}(t,t_1) \G_{1}(t_2,t_3) \G_{1}(s_2,s_3)\right\}\dif s \dif t\dif s_1 \dif t_1\dif s_2 \dif t_2\dif s_3 \dif t_3\\&=\sum_{k_{1}, k_{2}=1}^{d} \idotsint u_{k_{1} k_{2}}(s,t,t_1,s_1) u_{k_{1} k_{2}}(s_2,t_2,t_3,s_3)\dif s \dif t\dif s_1 \dif t_1\dif s_2 \dif t_2d s_3d t.
\end{align*}
We can show  that
$$B_{11} \leq C \left\{\iiiint\operatorname{tr}\left\{\G_{1}(s,s_1)\G_{1}(t,t_1)\right\} v(s,t)v(s_1,t_1) \dif s \dif t \dif s_1 \dif t_1\right\}^2,$$
for a constant $C>0$. Thus,  we have that
 $$\{n(n-1)\}^{-1} B_{11}=o\left\{\left(\iiiint\operatorname{tr}\left\{\G_{1}(s,s_1)\G_{1}(t,t_1)\right\} v(s,t)v(s_1,t_1) \dif s \dif t \dif s_1 \dif t_1\right)^2\right\}.$$
 It can be shown that
\begin{align*}
	B_{12} &=2\sum_{j, k=1}^{d} \idotsint u_{j k}(s,t,t_1,s_1) u_{jk}(s_2,t_2,t_3,s_3)\dif s \dif t\dif s_1 \dif t_1\dif s_2 \dif t_2\dif s_3 \dif t_3  \\&\quad +\sum_{j, k=1}^{d} \idotsint u_{j j}(s,t,t_1,s_1) u_{kk}(s_2,t_2,t_3,s_3)\dif s \dif t\dif s_1 \dif t_1\dif s_2 \dif t_2\dif s_3 \dif t_3 \\&\quad +\sum_{j=1}^{d} \idotsint u_{j j}(s,t,t_1,s_1) u_{jj}(s_2,t_2,t_3,s_3)\dif s \dif t\dif s_1 \dif t_1\dif s_2 \dif t_2\dif s_3 \dif t_3 \\
	&=2 \idotsint\operatorname{tr}\left\{\G_{1}(s,s_1)\G_{1}(t,t_1) \G_{1}(t_2,t_3) \G_{1}(s_2,s_3)\right\}\dif s \dif t\dif s_1 \dif t_1\dif s_2 \dif t_2\dif s_3 \dif t_3\\&\quad+\left(\iiiint\operatorname{tr}\left\{\G_{1}(s,s_1)\G_{1}(t,t_1)\right\}  \dif s \dif t \dif s_1 \dif t_1\right)^2\\&\quad+\Delta \sum_{j=1}^{d} \idotsint u_{j j}(s,t,t_1,s_1) u_{jj}(s_2,t_2,t_3,s_3)\dif s \dif td s_1d _1d s_2d t3d t\\
	& \leq(2+\Delta) \idotsint\operatorname{tr}\left\{\G_{1}(s,s_1)\G_{1}(t,t_1) \G_{1}(t_2,t_3) \G_{1}(s_2,s_3)\right\}\dif s \dif t\dif s_1 \dif t_1\dif s_2 \dif t_2\dif s_3 \dif t_3\\&\quad+\Big(\iiiint\operatorname{tr}\left\{\G_{1}(s,s_1)\G_{1}(t,t_1)\right\}  \dif s \dif t \dif s_1 \dif t_1\Big)^2.
\end{align*}
By \eqref{vartra1}, we have that
\begin{align*}
&\operatorname{Var}\Big\{ \iiiint \operatorname{tr}\{\A_{1} \} v(s,t)v(s_1,t_1) \dif s \dif t d s_1dt_1\Big\}\\& \leq \frac{2}{n(n-1)} C \left(\iiiint\operatorname{tr}\{\G_{1}(s,s_1)\G_{1}(t,t_1)\} v(s,t)v(s_1,t_1) \dif s \dif t \dif s_1 \dif t_1\right)^2\\&\quad \quad +\frac{4(n-2)}{n(n-1)}\Big\{(2+\Delta) \idotsint\operatorname{tr}\{\G_{1}(s,s_1)\G_{1}(t,t_1) \G_{1}(t_2,t_3) \G_{1}(s_2,s_3)\}\\&\quad  \times v(s, t)v(s_1, t_1)v(s_2, t_2)v(s_3, t_3)\dif s \dif t\dif s_1 \dif t_1\dif s_2 \dif t_2\dif s_3 \dif t_3\\&\quad\quad +\Big(\iiiint\operatorname{tr}\{\G_{1}(s,s_1)\G_{1}(t,t_1)\} v(s,t)v(s_1,t_1) \dif s \dif t \dif s_1 \dif t_1\Big)^2\Big\}
\\&= o\Big\{\Big(\iiiint\operatorname{tr}\{\G_{1}(s,s_1)\G_{1}(t,t_1)\} v(s,t)v(s_1,t_1) \dif s \dif t \dif s_1 \dif t_1\Big)^2\Big\}.
\end{align*}
By similar procedures, we    obtain  that
\begin{align*}
	&\operatorname{Var}\left\{ \iiiint {\operatorname{tr}\{\A_{k}\} }v(s,t)v(s_1,t_1) \dif s \dif t d s_1dt_1\right\}\\&= o\Big\{\Big(\iiiint\operatorname{tr} \{\G_{1}(s,s_1)\G_{1}(t,t_1) \} v(s,t)v(s_1,t_1) \dif s \dif t \dif s_1 \dif t_1\Big)^2\Big\}, ~~ k=2,\ldots,9.
\end{align*}
These, together with \eqref{var-sum}, yield  that
\begin{eqnarray*}
 &&\operatorname{Var}\Big\{ \iiiint\widehat{\operatorname{tr}\{\G_{1}(s,s_1)\G_{1}(t,t_1)\} }v(s,t)v(s_1,t_1) \dif s \dif t d s_1dt_1\Big\} \nonumber\\
 &=&  o\Big\{\Big(\iiiint\operatorname{tr} \{\G_{1}(s,s_1)\G_{1}(t,t_1) \} v(s,t)v(s_1,t_1) \dif s \dif t \dif s_1 \dif t_1\Big)^2\Big\}.
\end{eqnarray*}
 By \eqref{etrg1g1}, we obtain that
 \begin{align*}
 & \iiiint\widehat{\operatorname{tr}\left\{\G_{1}(s,s_1)\G_{1}(t,t_1)\right\} }v(s,t)v(s_1,t_1) \dif s \dif t \dif s_1 \dif t_1   \\&\quad=\iiiint\operatorname{tr}\left\{\G_{1}(s,s_1)\G_{1}(t,t_1)\right\} v(s,t)v(s_1,t_1) \dif s \dif t \dif s_1 \dif t_1
 \{1+o(1)\}.
\end{align*}
 This completes the proof.
 
 \hfill\BlackBox

\noindent  \textbf{Proof  of  Theorem \ref{theorem-power}.}
By  Theorem   \ref{Theorem3} and Condition (C3), we have that
 \begin{eqnarray*}
&&\widehat{\sigma}_{nm}^{2}(\X, \Y)\\
&=&\frac{2(n+m)^2}{n^2m^2}\Big[\frac{m^2}{(n+m)^2} \iiiint {\operatorname{tr}\{\G_{1}(s,s_1)\G_{1}(t,t_1) \}} v(s,t)v(s_1,t_1) \dif s \dif t \dif s_1 \dif t_1
\nonumber
\\&& +\frac{n^2}{(n+m)^2} \iiiint
 {\operatorname{tr}\{\G_{2}(s,\!s_1)\G_{2}(t,t_1) \}} v(s,t)v(s_1,t_1) \dif s \dif t \dif s_1 \dif t_1
\nonumber
\\&& +\frac{nm}{(n+m)^2}\iiiint {\operatorname{tr}\{\G_{1}(s,s_1)\G_{2}(t,t_1)\}} v(s,t)v(s_1,t_1) \dif s \dif t \dif s_1 \dif t_1\Big]\{1+o_p(1)\}\\
&=&\frac{2(n+m)^2}{n^2m^2}\iiiint\operatorname{tr}\{\G_\tau(s,s_1)\G_\tau(t,t_1)\} v(s,t)v(s_1,t_1) \dif s \dif t \dif s_1 \dif t_1  \{1+o_p(1)\},
\end{eqnarray*}
where  $\G_\tau(s,t)=(1-\tau) \G_{1}(s,t)+\tau\G_{2}(s,t)$.
Thus, by  Corollary \ref{Cor-2},   under $H_1: \boldsymbol{\mu}_1 \neq \boldsymbol{\mu}_2$, we have
 $$
\begin{aligned}
& \lim_{p, n, m  \rightarrow \infty}\operatorname{P}\{ \widehat Q_{n}(\X, \Y)> z_{\alpha}\} \\
= & \lim _{m, n, p \rightarrow \infty} P\Big\{\frac{\widehat\MRP(\X, \Y)-\MRP(\X,\Y)}{\widehat{\sigma}_{nm}(\X, \Y)}  > z_{\alpha}-\frac{\MRP(\X,\Y)}{\widehat{\sigma}_{nm}(\X, \Y)}\Big\} \\
= & \lim _{m, n, p \rightarrow \infty} \Phi\Big\{-z_{\alpha}+\frac{(n+m) \tau(1-\tau)\MRP(\X,\Y)}{\sqrt{2 \iiiint\operatorname{tr}\{\G_\tau(s,s_1)\G_\tau(t,t_1)\} v(s,t)v(s_1,t_1) \dif s \dif t \dif s_1 \dif t_1}}\Big\}.
\end{aligned}
$$
 The power under Condition (C3')  is
 $$
 \begin{aligned}
& \lim_{p, n, m  \rightarrow \infty}\operatorname{P}\{ \widehat Q_{n}(\X, \Y)> z_{\alpha}\} \\
= & \lim _{m, n, p \rightarrow \infty} P\Big\{\frac{\widehat\MRP(\X, \Y)-\MRP(\X,\Y)}{\widehat{\sigma}_{nm_2}(\X, \Y)}  > \frac{\widehat{\sigma}_{nm}(\X, \Y)}{\widehat{\sigma}_{nm_2}(\X, \Y)}z_{\alpha}-\frac{\MRP(\X,\Y)}{\widehat{\sigma}_{nm_2}(\X, \Y)}\Big\} \\
= & \lim _{m, n, p \rightarrow \infty} \Phi\Big\{-\frac{{\sigma}_{nm}(\X, \Y)}{{\sigma}_{nm_2}(\X, \Y)}z_{\alpha}+\frac{\MRP(\X,\Y)}{\sigma_{nm_2}(\X, \Y)}\Big\}.
\end{aligned}
$$

As $\sigma_{nm} / \sigma_{nm_2} \rightarrow 0$ under condition Condition (C3'),
then 
\begin{align*}
 \lim_{p, n, m  \rightarrow \infty}\operatorname{P}\{ \widehat Q_{n}(\X, \Y)> z_{\alpha}\}= \lim _{m, n, p \rightarrow \infty} \Phi\Big\{\frac{\MRP(\X,\Y)}{\sigma_{nm_2}(\X, \Y)}\Big\}.
\end{align*}
Substitute the expression for $\sigma_{nm_2}$, and we have that 
 the power under Condition (C3')  is
$$
\begin{aligned}
& \lim_{p, n, m  \rightarrow \infty}\operatorname{P}\{ \widehat Q_{n}(\X, \Y)> z_{\alpha}\} 
\\&\quad=  \lim _{m, n, p \rightarrow \infty} \Phi\bigg\{\frac{\sqrt{(n+m) \tau(1-\tau)}\MRP(\X,\Y)}{2\sqrt{\iiiint \left(\bfmu_{1}(t)-\bfmu_{2}(t)\right)^{\T}\G_{\tau}(s,s_1)\left(\bfmu_{1}(t_1)-\bfmu_{2}(t_1)\right)
		v(s,t)v(s_1,t_1) \dif s \dif t \dif s_1 \dif t_1}}\bigg\}.
\end{aligned}
$$
 \hfill\BlackBox

\noindent  \textbf{Proof  of  Theorem \ref{Theorem4}.}
Note that
\begin{align*}	
		&|\widehat{\MRP}(\mathbf{X},\mathbf{Y})-\widehat\MRP(\widehat\X, \widehat\Y )| \\=&   \Bigg|\frac{1}{ n(n-1) }\sum_{i, j=1 \atop i \neq j}^{n}\int_{0}^{1}\int_{0}^{1}\{[\X_i(t)-\widehat\X_i(t)]^{\T}\X_j(s)+\widehat\X_i(t)^{\T}[\X_j(s)-\widehat\X_j(s)]\}v(s, t) \dif s \dif t\\
		&+  \frac{1}{ m(m-1) }\sum_{i, j=1 \atop i \neq j}^{m}\int_{0}^{1}\int_{0}^{1}\{[\Y_i(t)-\widehat\Y_i(t)]^{\T}\Y_j(s)+\widehat\Y_i(t)^{\T}[\Y_j(s)-\widehat\X_j(s)]\}v(s, t) \dif s \dif t\\
		&-\frac{2}{ nm }\sum_{i=1}^{n}\sum_{j=1}^{m}  \int_{0}^{1}\int_{0}^{1}\{[\X_i(t)-\widehat\X_i(t)]^{\T}\Y_j(s)+\widehat\X_i(t)^{\T}[\Y_j(s)-\widehat\Y_j(s)]\}v(s, t) \dif s \dif t\Bigg|
		\\ &\leq \frac{4 C_1}{n} \sum_{i=1}^{n}\sum_{k=1}^{p}\int_{0}^{1} |\widehat X_{ik}(t)-X_{ik}(t) |\dif t+\frac{4 C_2}{m} \sum_{i=1}^{m}\sum_{k=1}^{p}\int_{0}^{1} |\widehat Y_{ik}(t)-Y_{ik}(t) |\dif t,
	\end{align*}
where  $C_1$ and $C_2$ are two   constants.
From a similar calculation and the proof of Theorem  \ref{Theorem3}, we could get 
\begin{align*}	
&\widehat\sigma_{nm}^2(\widehat\X, \widehat\Y)/\sigma_{nm}^2(\X, \Y)\\&=1+C\iiiint\operatorname{tr}\{\E(\widehat \X_{j}(t)-\X_{j}(t))(\widehat \X_{j}(t_1)-\X_{j}(t_1))^{\T} \G_{2}(s,s_1)\}\dif s \dif t \dif s_1 \dif t_1/\iiiint\operatorname{tr}\{\G_{1}(s,s_1)\G_{1}(t,t_1)\}\dif s \dif t \dif s_1 \dif t_1
\\&=1+o_p(1).
% &\leq
% \frac{C_3}{n(n-1)} \sum_{i\neq j}^{n}(\sum_{k=1}^{p}\int_{0}^{1} |\widehat X_{ik}(t)-X_{ik}(t) |\dif t)(\sum_{k=1}^{p}\int_{0}^{1} |\widehat X_{jk}(t)-X_{jk}(t) |\dif t)\\&+\frac{C_4}{m(m-1)} \sum_{i\neq j}^{m}(\sum_{k=1}^{p}\int_{0}^{1} |\widehat Y_{ik}(t)-Y_{ik}(t) |\dif t)(\sum_{k=1}^{p}\int_{0}^{1} |\widehat Y_{jk}(t)-Y_{jk}(t) |\dif t),
	\end{align*}
Thus, when
\begin{align*}
	&\sum_{k=1}^{p}\int_{0}^{1}\E\left|\widehat X_{k}(t)-X_{k}(t)\right|\dif t+\sum_{k=1}^{p}\int_{0}^{1}\E\left|\widehat Y_{k}(t)-Y_{k}(t)\right|\dif t=o(\sigma_{nm}(\X,\Y)),
\end{align*}
% we have that  $ \widehat\sigma_{nm}(\widehat\X, \widehat\Y)= \sigma_{nm}(\X, \Y)\{1+o_p(1)\}.$
% % $\widehat\MRP(\widehat\X, \widehat\Y)-\widehat\MRP(\X, \Y)=o_p(1)$ by Markov's inequality.
% % Similarly, we can obtain 
% These,
together with  Theorem \ref{Theorem2},  yield that
\begin{align*}
&\frac{\widehat\MRP(\widehat\X, \widehat\Y)-\MRP(\X, \Y)}{\widehat\sigma_{nm}(\widehat\X, \widehat\Y)}\\
&=\frac{\widehat\MRP(\widehat\X, \widehat\Y)-\widehat\MRP(\X, \Y)}{\widehat\sigma_{nm}(\widehat\X, \widehat\Y)}+
\frac{\widehat\MRP(\X,\Y)-\MRP(\X, \Y)}{\widehat\sigma_{nm}(\widehat\X, \widehat\Y)}\\
&=\frac{\widehat\MRP(\widehat\X, \widehat\Y)-\widehat\MRP(\X, \Y)}{\sigma_{nm}(\X, \Y)\{1+o_p(1)\}}+
\frac{\widehat\MRP(\X,\Y)-\MRP(\X, \Y)}{\sigma_{nm}(\X, \Y)\{1+o_p(1)\}}\\
&\stackrel{ {D}}{\longrightarrow} N(0,1),
\end{align*}
 as $p, n, m  \rightarrow \infty$.

\hfill\BlackBox

\section*{Appendix B:  Lemmas \ref{lemmaB1}-\ref{lemma5} }\label{Appendix-B}
\setcounter{equation}{0}
\renewcommand{\theequation}{B.\arabic{equation}}
\renewcommand{\thelemma}{B.\arabic{lemma}}

The following lemmas are needed to prove Theorem \ref{Theorem2}.
\begin{lemma}\label{lemmaB1} We have that
\begin{align*}
  &\operatorname{Var}\Big\{\sum_{i, j=1 \atop i \neq j}^{n}\iint\X_i(t)^{\T}\X_j(s)v(s, t) \dif s \dif t\Big\}
	\\&=2n(n-1) \iiiint\operatorname{tr}\left\{\G_{1}(t_1,t)\G_{1}(s_1,s)\right\} v(s,t)v(s_1,t_1) \dif s \dif t \dif s_1 \dif t_1
	\\&\quad\quad+	4n(n-1)^2 \iiiint  \bfmu_{1}(t)^{\T}\G_{1}(s,s_1)\bfmu_{1}(t_1)
	v(s,t)v(s_1,t_1) \dif s \dif t \dif s_1 \dif t_1.
\end{align*}
\end{lemma}
\noindent  \textbf{Proof  of  Lemma   \ref{lemmaB1}.}
\begin{align}\label{var-1}
 &\operatorname{Var}\Big\{\sum_{i, j=1 \atop i \neq j}^{n}\iint\X_i(t)^{\T}\X_j(s)v(s, t) \dif s \dif t\Big\}\nonumber \\
&=\operatorname{E}\Big\{\Big[\sum_{i, j=1 \atop i \neq j}^{n}\iint\X_i(t)^{\T}\X_j(s)v(s, t) \dif s \dif t\Big]^2\Big\}-\Big[\operatorname{E}\Big\{\sum_{i, j=1 \atop i \neq j}^{n}\iint\X_i(t)^{\T}\X_j(s)v(s, t) \dif s \dif t\Big\} \Big]^2\nonumber\\
&=\operatorname{E}\Big\{ \sum_{i \neq j}^{n}\sum_{i' \neq j'}^{n}\iint\X_i(t)^{\T}\X_j(s)v(s, t) \dif s \dif t \iint\X_{i'}(t)^{\T}\X_{j'}(s)v(s, t) \dif s \dif t\Big\}\nonumber\\
&\quad\quad\quad-\Big[ n(n-1) \iint\bfmu_1(t)^{\T}\bfmu_1(s)v(s, t) \dif s \dif t  \Big]^2.
\end{align}
We  can divide the first term on the right hand side of  \eqref{var-1}    into   the  following cases:

\textbf{Case (1)}:  For $\{i\neq j\neq i'\neq j'\}$,    we have that
 \begin{align}\label{var-1-1}
 & \operatorname{E}\Big\{ \sum_{i\neq j\neq i'\neq j'}^{n} \iint\X_i(t)^{\T}\X_j(s)v(s, t) \dif s \dif t \iint\X_{i'}(t)^{\T}\X_{j'}(s)v(s, t) \dif s \dif t\Big\}\nonumber\\
 &=  n(n-1)(n-2)(n-3)\Big[ \iint\bfmu_1(t)^{\T}\bfmu_1(s)v(s, t) \dif s \dif t  \Big]^2.
\end{align}

\textbf{Case (2)}: For $\{i=i',j\neq j'\}\cup\{i=j',j\neq i'\}\cup\{j=i',i\neq j'\}\cup\{j=j',i\neq i' \}$, we have that
\begin{equation}\label{var-1-2}
 \begin{split}
 & \operatorname{E}\Big\{ \sum_{i \neq j}^{n}\sum_{i' \neq j'}^{n} \iint\X_i(t)^{\T}\X_j(s)v(s, t) \dif s \dif t \iint\X_{i'}(t)^{\T}\X_{j'}(s)v(s, t) \dif s \dif t\Big\}\nonumber\\
 &=  4\operatorname{E}\Big\{ \sum_{i=i',j\neq j'}^{n}  \iint\X_i(t)^{\T}\X_j(s)v(s, t) \dif s \dif t \iint\X_{i'}(t)^{\T}\X_{j'}(s)v(s, t) \dif s \dif t\Big\}\nonumber\\
  &=  4n(n-1)(n-2)  \iiiint\operatorname{tr}\{ \operatorname{E} \{\X_i(t)^{\T}\X_j(s)   \X_{i}(t_1)^{\T}\X_{j'}(s_1) \}\} v(s, t)v(s_1,t_1) \dif s \dif t \dif s_1 \dif t_1
  \nonumber\\
  &= 4n(n-1)(n-2)  \iiiint\operatorname{tr}\{ \operatorname{E}\{ \X_{i}(t)\X_i(t_1)^{\T}\}\operatorname{E}\{\X_{j'}(s_1)\X_j(s)^{\T} \}\} v(s, t)v(s_1,t_1) \dif s \dif t \dif s_1 \dif t_1 \nonumber\\
  &= 4n(n-1)(n-2)  \iiiint\operatorname{tr}\{[\G_{1}(t,t_1)+\bfmu_{1}(t)\bfmu_{1}(t_1)^{\T}]\bfmu_{1}(s_1)\bfmu_{1}(s)^{\T} \} v(s, t)v(s_1,t_1) \dif s \dif t \dif s_1 \dif t_1\nonumber\\
  &=  4n(n-1)(n-2)  \iiiint \bfmu_{1}(s)^{\T}\G_{1}(t,t_1)\bfmu_{1}(s_1)  v(s, t)v(s_1,t_1) \dif s \dif t \dif s_1 \dif t_1\nonumber\\
  & \quad\quad +4n(n-1)(n-2) \Big[ \iint\bfmu_1(t)^{\T}\bfmu_1(s)v(s, t) \dif s \dif t  \Big]^2.
\end{split}
\end{equation}
\textbf{Case (3)}:  For  $\{i=i',j= j'\}\cup\{i=j',j= i'\}$, we have that
  \begin{align}\label{var-1-3}
 &  \operatorname{E}\Big\{ \sum_{i \neq j}^{n}\sum_{i' \neq j'}^{n} \iint\X_i(t)^{\T}\X_j(s)v(s, t) \dif s \dif t \iint\X_{i'}(t)^{\T}\X_{j'}(s)v(s, t) \dif s \dif t\Big\}\nonumber\\
 &= 2\operatorname{E}\Big\{ \sum_{i=i',j= j'}^{n}  \iint\X_i(t)^{\T}\X_j(s)v(s, t) \dif s \dif t \iint\X_{i}(t)^{\T}\X_{j}(s)v(s, t) \dif s \dif t\Big\}\nonumber\\
  &=  2n(n-1)  \iiiint\operatorname{tr}\Big\{ \operatorname{E}\{ \X_{i}(t)\X_i(t_1)^{\T}\}\operatorname{E}\{\X_{j}(s_1)\X_j(s)^{\T} \}\Big\} v(s, t)v(s_1,t_1) \dif s \dif t \dif s_1 \dif t_1 \nonumber\\
  &=  2n(n-1)  \iiiint\operatorname{tr}\Big\{[\G_{1}(t,t_1)+\bfmu_{1}(t)\bfmu_{1}(t_1)^{\T}]
  [\G_{1}(s,s_1)+\bfmu_{1}(s)\bfmu_{1}(s_1)^{\T}]\Big\}\nonumber\\
  &\quad\quad\times v(s, t)v(s_1,t_1) \dif s \dif t \dif s_1 \dif t_1\nonumber\\
  &=  2n(n-1)\iiiint \operatorname{tr} \{\G_{1}(t,t_1)\G_{1}(s,s_1) \}  v(s, t)v(s_1,t_1) \dif s \dif t \dif s_1 \dif t_1\nonumber\\
  & \quad\quad +4n(n-1)\iiiint \bfmu_{1}(s)^{\T}\G_{1}(t,t_1)\bfmu_{1}(s_1)  v(s, t)v(s_1,t_1) \dif s \dif t \dif s_1 \dif t_1\nonumber\\
  &\quad\quad  +2n(n-1) \Big[ \iint\bfmu_1(t)^{\T}\bfmu_1(s)v(s, t) \dif s \dif t  \Big]^2.
\end{align}

By \eqref{var-1}-\eqref{var-1-3}, we have that
\begin{eqnarray*}
 &&\operatorname{Var}\Big\{\sum_{i, j=1 \atop i \neq j}^{n}\iint\X_i(t)^{\T}\X_j(s)v(s, t) \dif s \dif t\Big\}\nonumber \\
&=&n(n-1)(n-2)(n-3)\Big[ \iint\bfmu_1(t)^{\T}\bfmu_1(s)v(s, t) \dif s \dif t  \Big]^2\nonumber\\
&& +4n(n-1)(n-2)  \iint \bfmu_{1}(s)^{\T}\G_{1}(t,t_1)\bfmu_{1}(s_1)  v(s, t)v(s_1,t_1) \dif s \dif t \dif s_1 \dif t_1\nonumber\\
&& +4n(n-1)(n-2) \Big[ \iint\bfmu_1(t)^{\T}\bfmu_1(s)v(s, t) \dif s \dif t  \Big]^2\nonumber\\
&& +2n(n-1)\iiiint \operatorname{tr} \{\G_{1}(t,t_1)\G_{1}(s,s_1) \}  v(s, t)v(s_1,t_1) \dif s \dif t \dif s_1 \dif t_1\nonumber\\
&& +4n(n-1)\iint \bfmu_{1}(s)^{\T}\G_{1}(t,t_1)\bfmu_{1}(s_1)  v(s, t)v(s_1,t_1) \dif s \dif t \dif s_1 \dif t_1\nonumber\\
  && +2n(n-1) \Big[ \iint\bfmu_1(t)^{\T}\bfmu_1(s)v(s, t) \dif s \dif t  \Big]^2\nonumber\\
&&-\Big[ n(n-1) \iint\bfmu_1(t)^{\T}\bfmu_1(s)v(s, t) \dif s \dif t  \Big]^2\nonumber\\
&=&2n(n-1)\iiiint \operatorname{tr} \{\G_{1}(t,t_1)\G_{1}(s,s_1) \}  v(s, t)v(s_1,t_1) \dif s \dif t \dif s_1 \dif t_1\nonumber\\
&&+4n(n-1)^2\iint \bfmu_{1}(s)^{\T}\G_{1}(t,t_1)\bfmu_{1}(s_1)  v(s, t)v(s_1,t_1) \dif s \dif t \dif s_1 \dif t_1.
\end{eqnarray*}

\hfill\BlackBox

\begin{lemma}\label{lemma3}
 For any $N$, $\{S_{j}, \mathcal{F}_{j}\}_{j=1}^{N}$ is the sequence of zero mean and a square integrable martingale.
\end{lemma}
\noindent  \textbf{Proof  of  Lemma   \ref{lemma3}.}
By the definitions of $S_{j}$ and $\mathcal{F}_{j}$,
it is easy to show that $\mathcal{F}_{j-1} \subseteq \mathcal{F}_{j}$ and $S_{j}$ is of zero mean and square integrable, for any $1 \leq j \leq n+m$.

We next  show that $\{S_{j}\}_{j=1}^{N}$ is a martingale with respective to the filter  $\{\mathcal{F}_{j}\}_{j=1}^{N}$.
We consider the following two cases:
 \begin{itemize}
 \item[Case (1):]  When $j\leq k$, we have that
  $$\E\{V_{j} \mid \mathcal{F}_{k}\}=\sum_{i=1}^{j-1} \E\{\phi_{i j} \mid \mathcal{F}_{k}\}=\sum_{i=1}^{j-1} \phi_{i j}=V_{j}.$$

  \item[Case (2):]  When $j> k$, we have that
 \begin{align*}
 \E\{V_{j} \mid \mathcal{F}_{k}\}
  &=
  \sum_{i=1}^{k} \E\{\phi_{i j} \mid \mathcal{F}_{k}\}+\sum_{i=k+1}^{j-1} \E\{\phi_{i j} \mid \mathcal{F}_{k}\}\\
  &=C_1\sum_{i=1}^{k}\E\{\iint\W_i(t)^{\T}\W_j(s)v(s, t) \dif s \dif t \mid \mathcal{F}_{k}\}\\
  &+
  C_2\sum_{i=k+1}^{j-1}\E\{\iint\W_i(t)^{\T}\W_j(s)v(s, t) \dif s \dif t \mid \mathcal{F}_{k}\}
  \\
  &=C_1\sum_{i=1}^{k}\iint\W_i(t)^{\T}\E\{\W_j(s)\mid \mathcal{F}_{k}\}v(s, t) \dif s \dif t \\
  &+
  C_2\sum_{i=k+1}^{j-1}\iint\E\{\W_i(t)\}^{\T}\E\{\W_j(s)\}v(s, t) \dif s \dif t
=0,
  \end{align*}
where $C_1$ $C_2\in\{{\frac{1}{n(n-1)},-\frac{1}{nm},\frac{1}{m(m-1)}}\}.$
  \end{itemize}
 Then, for any $q \geq k$, we have that
  $$\E\{S_{q} \mid \mathcal{F}_{k}\}=\sum_{j=2}^{q} \E\{V_{j} \mid \mathcal{F}_{k}\}=
  \sum_{j=2}^{k}  V_{j}=S_{k}.$$
Thus, $\{S_{j}\}_{j=1}^{N}$ is a martingale with respective to the filter  $\{\mathcal{F}_{j}\}_{j=1}^{N}$ for any $N$.
\hfill\BlackBox

\begin{lemma}\label{lemma4}
	Under Condition (C3), we have that
	 $$
	  {\sum_{j=2}^{n+m} \E\{V_{j}^{2} \mid \mathcal{F}_{j-1}\}}/{\sigma_{nm}^{2}(\X,\Y)} \stackrel{ {P}}{\longrightarrow}  \frac{1}{4}.
	$$ 	
\end{lemma}

\noindent  \textbf{Proof  of  Lemma   \ref{lemma4}.}
By the definition of $V_{j}$ and $\phi_{i j},$ we have that
\begin{eqnarray*}
 \E\{V_{j}^{2} \mid \mathcal{F}_{j-1}\}
 =\E\{[\sum_{i=1}^{j-1}\phi_{i j}]^{2} \mid \mathcal{F}_{j-1}\}
 =\sum_{i_{1}, i_{2}=1}^{j-1}\E \{\phi_{i_{1} j}\phi_{i_{2} j}\mid \mathcal{F}_{j-1}\},
 \end{eqnarray*}
which  can be  divided  into  the following three cases:
 \begin{itemize}
 \item[Case (1):]  When $j\leq n$, we have that
 \begin{align*}
 &\E\{V_{j}^{2} \mid \mathcal{F}_{j-1}\}\\
 &=  \frac{1}{n^2(n-1)^2}\E\Big\{ \sum_{i_{1}, i_{2}=1}^{j-1}
 \iiiint\W_{i_1}(t)^{\T}\W_j(s)\W_j(s_1)^{\T}\W_{i_2}(t_1)  v(s, t)\\
 &\quad \quad  \times v(s_1, t_1) \dif s \dif t\dif s_1 \dif t_1   \mid \mathcal{F}_{j-1}\Big\}\\
  &=
 \frac{1}{n^2(n-1)^2}\sum_{i_{1}, i_{2}=1}^{j-1} \iiiint \W_{i_1}(t)^{\T}{\G}_{1}(s,s_1) \W_{i_2}(t_1)  v(s, t)v(s_1, t_1) \dif s \dif t\dif s_1 \dif t_1.
 \end{align*}

  \item[Case (2):] When $j=n+1$, we have that
 \begin{align*}
 &\E\{V_{j}^{2} \mid \mathcal{F}_{j-1}\}\\
 &=  \frac{1}{n^2m^2}\E\Big\{ \sum_{i_{1}, i_{2}=1}^{n}
 \iiiint\W_{i_1}(t)^{\T}\W_n(s)\W_n(s_1)^{\T}\W_{i_2}(t_1)  v(s, t)\\
 &\quad \quad  \times v(s_1, t_1) \dif s \dif t\dif s_1 \dif t_1   \mid \mathcal{F}_{j-1}\Big\}\\
  &=
 \frac{1}{n^2m^2}\sum_{i_{1}, i_{2}=1}^{n} \iiiint \W_{i_1}(t)^{\T}{\G}_{2}(s,s_1) \W_{i_2}(t_1)  v(s, t)v(s_1, t_1) \dif s \dif t\dif s_1 \dif t_1.
 \end{align*}

 \item[Case (3):] When $j \geq n+2$, we have that
    \begin{align*}
 &\E\{V_{j}^{2} \mid \mathcal{F}_{j-1}\}\\
 &=
 \frac{1}{n^2m^2}\sum_{i_{1}, i_{2}=1}^{n} \iiiint \W_{i_1}(t)^{\T}{\G}_{2}(s,s_1) \W_{i_2}(t_1)  v(s, t)v(s_1, t_1) \dif s \dif t\dif s_1 \dif t_1 \\
 &+\frac{1}{m^2(m-1)^2}\sum_{i_{1}, i_{2}=n+1}^{j-1} \iiiint \W_{i_1}(t)^{\T}{\G}_{2}(s,s_1) \W_{i_2}(t_1)  v(s, t)v(s_1, t_1) \dif s \dif t\dif s_1 \dif t_1.
 \end{align*}
 \end{itemize}

 Denote  $$\eta_{n+m}=\sum_{j=2}^{n+m} E\{V_{j}^{2} \mid \mathcal{F}_{j-1}\}.$$
  Based on the above Cases (1)-(3), we obtain that
\begin{eqnarray}\label{Eeta}
&&\E\{\eta_{n+m}\}\nonumber\\
 &=&\frac{1}{n^2(n-1)^2}\sum_{j=2}^{n}\sum_{i_{1}, i_{2}=1}^{j-1} \iiiint \E\{ \W_{i_1}(t)^{\T}{\G}_{1}(s,s_1) \W_{i_2}(t_1) \} v(s, t)v(s_1, t_1) \dif s \dif t\dif s_1 \dif t_1\nonumber\\
&&+\frac{1}{n^2m^2}\sum_{i_{1}, i_{2}=1}^{n} \iiiint \E\{ \W_{i_1}(t)^{\T}{\G}_{2}(s,s_1) \W_{i_2}(t_1)\}  v(s, t)v(s_1, t_1) \dif s \dif t\dif s_1 \dif t_1
	\nonumber\\&& +\sum_{j=n+2}^{n+m}\Big[
 \frac{1}{n^2m^2}\sum_{i_{1}, i_{2}=1}^{n} \iiiint \E\{ \W_{i_1}(t)^{\T}{\G}_{2}(s,s_1) \W_{i_2}(t_1)\}  v(s, t)v(s_1, t_1) \dif s \dif t\dif s_1 \dif t_1 \nonumber\\
 &&+\frac{1}{m^2(m-1)^2}\sum_{i_{1}, i_{2}=n+1}^{j-1} \iiiint\E\{  \W_{i_1}(t)^{\T}{\G}_{2}(s,s_1) \W_{i_2}(t_1) \} v(s, t)v(s_1, t_1) \dif s \dif t\dif s_1 \dif t_1\Big]\nonumber\\
 &=&\frac{1}{n^2(n-1)^2}\frac{n (n-1 )}{2}
   \iiiint\operatorname{tr}\left\{\G_{1}(s,s_1)\G_{1}(t,t_1)\right\} v(s,t)v(s_1,t_1) \dif s \dif t \dif s_1 \dif t_1
	\nonumber\\&&+\frac{1}{m^2(m-1)^2}\frac{m(m-1 )}{2} \iiiint\operatorname{tr}\left\{\G_{2}(s,s_1)\G_{2}(t,t_1)\right\} v(s,t)v(s_1,t_1) \dif s \dif t \dif s_1 \dif t_1
	\nonumber\\&& +\frac{1}{n^2m^2}mn \iiiint\operatorname{tr}\left\{\G_{1}(s,s_1)\G_{2}(t,t_1)\right\} v(s,t)v(s_1,t_1) \dif s \dif t \dif s_1 \dif t_1
	\nonumber\\&=&\frac{1}{4} \sigma_{nm}^{2}(\X,\Y).
\end{eqnarray}
Moreover, we have that
\begin{eqnarray*}\label{Eeta--1}
&&\E\{\eta^{2}_{n+m}\}\nonumber\\
 &=&\E\Big\{ \frac{1}{n^2(n-1)^2}\sum_{j=2}^{n}\sum_{i_{1}, i_{2}=1}^{j-1} \iiiint \W_{i_1}(t)^{\T}{\G}_{1}(s,s_1) \W_{i_2}(t_1)   v(s, t)v(s_1, t_1) \dif s \dif t\dif s_1 \dif t_1\nonumber\\
&&+\frac{m}{n^2m^2}\sum_{i_{1}, i_{2}=1}^{n} \iiiint   \W_{i_1}(t)^{\T}{\G}_{2}(s,s_1) \W_{i_2}(t_1)   v(s, t)v(s_1, t_1) \dif s \dif t\dif s_1 \dif t_1
	\nonumber\\&& +\frac{1}{m^2(m-1)^2}\sum_{j=n+2}^{n+m}\sum_{i_{1}, i_{2}=n+1}^{j-1} \iiiint   \W_{i_1}(t)^{\T}{\G}_{2}(s,s_1) \W_{i_2}(t_1) v(s, t)v(s_1, t_1) \dif s \dif t\dif s_1 \dif t_1\Big\}^2\nonumber\\
 &=& \sigma_{nm}^{4}(\X,\Y)\{1+o(1)\}.
\end{eqnarray*}
Thus, we have
\begin{equation}\label{Veta}
\operatorname{Var}\{\eta_{n+m}\}=\E\{\eta_{n+m}^2\}-(\E\{\eta_{n+m}\})^2=o(\sigma_{nm}^{4}(\X,\Y)).
\end{equation}

By \eqref{Eeta} and \eqref{Veta}, and we have
$$
 \E\Big\{\sum_{j=1}^{n+m} \E\{V_{j}^{2} \mid \mathcal{F}_{j-1}\}\Big\}/ \sigma^2_{nm}(\X,\Y)=
\E\{\eta_{n+m}\}/\sigma^2_{nm}(\X,\Y)=\frac{1}{4},
$$
and
$$
 \operatorname{Var}\Big\{\sum_{j=1}^{n+m} \E\{(V_{j}^{2} \mid \mathcal{F}_{ j-1}\}\Big\}/\sigma^4_{nm}(\X,\Y)= \operatorname{Var}\{\eta_{n+m}\}/\sigma^4_{nm}(\X,\Y)=o(1).
$$
This completes the proof of Lemma \ref{lemma4}.
\hfill\BlackBox

\begin{lemma}\label{lemma5}
	Under Conditions (C1) and  (C4), for any $\epsilon>0$, we have  that
	$$
	\sum_{j=2}^{n+m}   \E\{V_{j}^{2} I(|V_{j}|>\epsilon \sigma_{nm}(\X,\Y)) \mid \mathcal{F}_{j-1}\}/{\sigma_{nm}^{2}(\X,\Y)} \stackrel{ {P}}{\longrightarrow} 0 .
	$$
\end{lemma}

\noindent  \textbf{Proof  of  Lemma  \ref{lemma5}.}
Note that
$$
\sum_{j=2}^{n+m} \sigma_{nm}^{-2}(\X,\Y)    \E\{V_{j}^{2} I(|V_{j}|>\epsilon \sigma_{nm}(\X,\Y)) \mid \mathcal{F}_{j-1}\}   \leq \sigma_{nm1}^{-4}(\X,\Y) \epsilon^{-2} \sum_{j=2}^{n+m}\E\{V_{j}^{4} \mid \mathcal{F}_{j-1}\}
$$
and
\begin{eqnarray*}
 \E\{V_{j}^{4} \mid \mathcal{F}_{j-1}\}
 =\E\Big\{\Big[\sum_{i=1}^{j-1}\phi_{i j}\Big]^{4} \mid \mathcal{F}_{j-1}\Big\}
 =\sum_{i_{1}, i_{2},i_{3}, i_{4}=1}^{j-1}\E \{\phi_{i_{1} j}\phi_{i_{2} j}\phi_{i_{3} j}\phi_{i_{4} j}\mid \mathcal{F}_{j-1}\}.
 \end{eqnarray*}
Therefore, we have
\begin{align}\label{first-order-V}
\E\Big\{\sum_{j=2}^{n+m} \E\{V_{j}^{4} \mid \mathcal{F}_{j-1}\}\Big\}&=
\sum_{j=2}^{n+m} \sum_{i_{1}, i_{2},i_{3}, i_{4}=1}^{j-1}\E \{\phi_{i_{1} j}\phi_{i_{2} j}\phi_{i_{3} j}\phi_{i_{4} j}\}\nonumber\\
&= O\left((n+m)^{-8}\right) \sum_{j=2}^{n+m} \sum_{i_{1}, i_{2},i_{3}, i_{4}=1}^{j-1}\E \{\phi_{i_{1} j}\phi_{i_{2} j}\phi_{i_{3} j}\phi_{i_{4} j}\}\nonumber\\
&= O\left((n+m)^{-8}\right)(3 Q+P),
\end{align}
where
\begin{align*}
Q&= \sum_{j=2}^{n+m} \sum_{i_1 \neq i_2}^{j-1} \E\Big\{\idotsint\W_{j}(t)^{\T}\W_{i_1}(s)\W_{i_1}(s_1)^{\T}\W_{j}(t_1)\W_{j}(t_2)^{\T}\W_{i_2}(s_2)
\\&\quad\quad \times \W_{i_2}(s_3)^{\T}\W_{j}(t_3) v(s, t)v(s_1, t_1)v(s_2, t_2)v(s_3, t_3)\dif s \dif t\dif s_1 \dif t_1
\dif s_2 \dif t_2 \dif s_3 \dif t_3  \Big\}
\end{align*}
and
$$P= \sum_{j=2}^{n+m} \sum_{i=1}^{j-1} \E\Big\{\iint\W_{i}(s)^{\T}\W_{j}(t)v(s, t)\dif s \dif t\Big\}^{4}.$$

Note that
\begin{align*}
	Q &=\sum_{j=2}^{n+m} \sum_{i_1 \neq i_2}^{j-1} \idotsint\E\Big\{\operatorname{tr}\{\W_{j}(t_3)\W_{j}(t)^{\T}\W_{i_1}(s)\W_{i_1}(s_1)^{\T}\W_{j}(t_1)
\W_{j}(t_2)^{\T}\\&\quad \times
	\W_{i_2}(s_2)\W_{i_2}(s_3)^{\T}\}\Big\}v(s, t)v(s_1, t_1)v(s_2, t_2)v(s_3, t_3)\dif s \dif t\dif s_1 \dif t_1
	\dif s_2 \dif t_2 \dif s_3 \dif t_3
\\&=\sum_{j=2}^{n} \sum_{i_1 \neq i_2}^{j-1} \idotsint\E\Big\{\W_{j}(t)^{\T}\G_1(s,s_1)\W_{j}(t_1)\W_{j}(t_2)^{\T}\G_1(s_2,s_3)
	\W_{j}(t_3)\Big\} \\&\quad \times v(s, t)v(s_1, t_1)v(s_2, t_2)v(s_3, t_3)\dif s \dif t\dif s_1 \dif t_1
	\dif s_2 \dif t_2 \dif s_3 \dif t_3  \\
 \end{align*}
 \begin{align*}
 &+  \sum_{i_1 \neq i_2}^{n} \idotsint\E\Big\{\W_{n+1}(t)^{\T}\G_1(s,s_1)\W_{n+1}(t_1)\W_{n+1}(t_2)^{\T}\G_1(s_2,s_3)
	\W_{n+1}(t_3)\Big\} \\&\quad \times v(s, t)v(s_1, t_1)v(s_2, t_2)v(s_3, t_3)\dif s \dif t\dif s_1 \dif t_1
	\dif s_2 \dif t_2 \dif s_3 \dif t_3  \\
 &+  \sum_{j=n+2}^{n+m} \sum_{i_1 \neq i_2}^{j-1} \idotsint\E\Big\{\W_{j}(t)^{\T}\G_{2}(s,s_1)\W_{j}(t_1)\W_{j}(t_2)^{\T}\G_{2}(s_2,s_3)
	\W_{j}(t_3)\Big\}\\&\quad \times  v(s, t)v(s_1, t_1)v(s_2, t_2)v(s_3, t_3)\dif s \dif t\dif s_1 \dif t_1
	\dif s_2 \dif t_2 \dif s_3 \dif t_3  \\
	&=o\left((n+m)^{8}\sigma_{nm}^{4}(\X,\Y)\right)
\end{align*}
and
\begin{align*}
P&= \sum_{j=2}^{n+m} \sum_{i=1}^{j-1} \E\Big\{\iint\W_{i}(s)^{\T}\W_{j}(t)v(s, t)\dif s \dif t\Big\}^{4} \\
&= \sum_{j=2}^{n} \sum_{i=1}^{j-1} \E\Big\{\iint\W_{i}(s)^{\T}\W_{j}(t)v(s, t)\dif s \dif t\Big\}^{4} +  \sum_{j=n+1}^{n+m} \sum_{i=1}^{n} \E\Big\{\iint\W_{i}(s)^{\T}\W_{j}(t)v(s, t)\dif s \dif t\Big\}^{4} \\
 &+  \sum_{j=n+1}^{n+m} \sum_{i=n+1}^{j-1} \E\Big\{\iint\W_{i}(s)^{\T}\W_{j}(t)v(s, t)\dif s \dif t\Big\}^{4}\\
&= \sum_{j=2}^{n} \sum_{i=1}^{j-1} \E\Big\{\iint\X_{i}(s)^{\T}\X_{j}(t)v(s, t)\dif s \dif t\Big\}^{4}+
 \sum_{j=n+1}^{n+m} \sum_{i=1}^{n} \E\Big\{\iint\X_{i}(s)^{\T}\Y_{j-n}(t)v(s, t)\dif s \dif t\Big\}^{4}\\
 &+\sum_{j=n+1}^{n+m} \sum_{i=n+1}^{j-1} \E\Big\{\iint\Y_{i-n}(s)^{\T}\Y_{j-n}(t)v(s, t)\dif s \dif t\Big\}^{4}\\
&=  P_{1}+P_{2}+P_{3},
\end{align*}
where
\begin{align*}
	&P_1=  \sum_{j=2}^{n} \sum_{i=1}^{j-1} \E\Big\{\iint\X_{i}(s)^{\T}\X_{j}(t)v(s, t)\dif s \dif t\Big\}^{4},\\
	&P_2=  \sum_{j=n+1}^{n+m} \sum_{i=1}^{n} \E\Big\{\iint\X_{i}(s)^{\T}\Y_{j-n}(t)v(s, t)\dif s \dif t\Big\}^{4},
	\\&	P_3= \sum_{j=n+1}^{n+m} \sum_{i=n+1}^{j-1} \E\Big\{\iint\Y_{i-n}(s)^{\T}\Y_{j-n}(t)v(s, t)\dif s \dif t\Big\}^{4}.
\end{align*}

We first consider the term $P_2$. Let $\bfGamma_{1}^{\T}(s) \bfGamma_{2}(t)=\left(a_{i j}(s,t)\right)_{d \times d}$. By Condition (C1),  we have  the following inequalities:
\begin{align*}
&\sum_{i=1}^{d}\sum_{j=1}^{d}\Big[\iint a_{ij}(s,t)v(s, t)\dif s \dif t\Big]^{4}\\
 & \leq\Big(\sum_{i=1}^{d}\sum_{j=1}^{d}\Big[\iint a_{ij}(s,t)v(s, t)\dif s \dif t\Big]^{2}\Big)^{2}\\
 & =\Big(\sum_{i=1}^{d}\sum_{j=1}^{d} \iiiint a_{ij}(s,t)a_{ij}(s_1, t_1)v(s, t)v(s_1, t_1)\dif s \dif t\dif s_1 \dif t_1 \Big)^{2}\\
 &=\Big(\iiiint\operatorname{tr}\{\bfGamma_{1}(s)^{\T} \bfGamma_{2}(t) \Gamma_{2}^{\T}(t_1) \bfGamma_{1}(s_1)\}v(s, t)v(s_1, t_1)\dif s \dif t\dif s_1 \dif t_1\Big)^{2} \\
&=\Big(\iiiint\operatorname{tr}\{ \G_{2}(t,t_1) \G_{1}(s_1,s)\}v(s, t)v(s_1, t_1)\dif s \dif t\dif s_1 \dif t_1\Big)^{2},
\end{align*}
and
\begin{align*}
	& \sum_{i=1}^{d} \sum_{j_{1} \neq j_{2}}^{d}\Big[\iint a_{ij_1}(s,t)v(s, t)\dif s \dif t\Big]^{2}\Big[\iint a_{ij_2}(s_1,t_1)v(s_1, t_1)\dif s_1 \dif t_1\Big]^{2}  \\
   & \leq\Big(\sum_{i=1}^{d} \sum_{j_{1} \neq j_{2}}^{d} \Big[\iint a_{ij}(s,t)v(s, t)\dif s \dif t\Big]^{2}\Big)^{2}\\
	&\leq\Big(\iiiint\operatorname{tr}\{ \G_{2}(t,t_1) \G_{1}(s_1,s)\}v(s, t)v(s_1, t_1)\dif s \dif t\dif s_1 \dif t_1\Big)^{2}
\end{align*}
and
\begin{align*}
& \sum_{i_{1} \neq i_{2}}^{d} \sum_{j_{1} \neq j_{2}}^{d} \iint a_{i_1j_1}(s,t)v(s, t)\dif s \dif t\iint a_{i_1j_2}(s_1,t_1)v(s_1, t_1)\dif s_1 \dif t_1\\
&\quad \quad \quad\times\iint a_{i_2j_1}(s_2,t_2)v(s_2, t_2)\dif s_2 \dif t_2\iint a_{i_2j_2}(s_3,t_3)v(s_3, t_3)\dif s_3 \dif t_3  \\& \leq\idotsint\operatorname{tr} \{\bfGamma_{1}(s)^{\T}\G_{2}(t,t_1) \bfGamma_{1}(s_1) \Gamma_{1}^{\T}(t_2) \G_{2}(s_2,s_3)\bfGamma_{1}(t_3) \}
	\\&\quad \quad \quad  \times  v(s, t)v(s_1, t_1)v(s_2, t_2)v(s_3, t_3)\dif s \dif t\dif s_1 \dif t_1\dif s_2 \dif t_2\dif s_3 \dif t_3 \\& =\idotsint\operatorname{tr} \{\G_{1}(s,s_1)\G_{2}(t,t_1) \G_{1}(t_2,t_3) \G_{2}(s_2,s_3) \}\\&\quad \quad \quad  \times v(s, t)v(s_1, t_1)v(s_2, t_2)v(s_3, t_3)\dif s \dif t\dif s_1 \dif t_1\dif s_2 \dif t_2\dif s_3 \dif t_3.
\end{align*}
These, together with Condition (C1), yield that
\begin{align*}
&\E\Big\{\iint\X_{i}(s)^{\T}\Y_{j-n}(t)v(s, t)\dif s \dif t\Big\}^{4}\\
&=\E\Big\{\iint\Z_{1}(s)^{\T}\bfGamma_{1}^{\T}(s) \bfGamma_{2}(t)\Z_{2}(t)v(s, t)\dif s \dif t\Big\}^{4}\\
&= \sum_{i=1}^{d} \sum_{j=1}^{d}(3+\Delta_1)(3+\Delta_2)\Big[\iint a_{ij}(s,t)v(s, t)\dif s \dif t\Big]^{4} \\&+\sum_{i=1}^{d}(3+\Delta_2) \sum_{j_{1} \neq j_{2}}^{d}  \Big[\iint a_{ij_1}(s,t)v(s, t)\dif s \dif t\Big]^{2}\Big[\iint a_{ij_2}(s_1,t_1)v(s_1, t_1)\dif s_1 \dif t_1\Big]^{2}  \\
	&+\sum_{j=1}^{d}(3+\Delta_1) \sum_{i_{1} \neq i_{2}}^{d} \Big[\iint a_{i_1j}(s,t)v(s, t)\dif s \dif t\Big]^{2}\Big[\iint a_{i_2j}(s_1,t_1)v(s_1, t_1)\dif s_1 \dif t_1\Big]^{2}  \\
 &+9 \sum_{i_{1} \neq i_{2}}^{d} \sum_{j_{1} \neq j_{2}}^{d}  \iint a_{i_1j_1}(s,t)v(s, t)\dif s \dif t\iint a_{i_1j_2}(s_1,t_1)v(s_1, t_1)\dif s_1 \dif t_1\\&\quad \quad\quad\quad \times\iint a_{i_2j_1}(s_2,t_2)v(s_2, t_2)\dif s_2 \dif t_2\iint a_{i_2j_2}(s_3,t_3)v(s_3, t_3)\dif s_3 \dif t_3  \\
 \end{align*}
 \begin{align*}
&= O\Big\{\Big(\iiiint\operatorname{tr}\{ \G_{2}(t,t_1) \G_{1}(s_1,s)\}v(s, t)v(s_1, t_1)\dif s \dif t\dif s_1 \dif t_1\Big)^{2}\Big\}\\&+O\Big\{\idotsint\operatorname{tr}\{\G_{1}(s,s_1)\G_{2}(t,t_1) \G_{1}(t_2,t_3) \G_{2}(s_2,s_3)\}\\&\quad \quad\quad\quad  \times v(s, t)v(s_1, t_1)v(s_2, t_2)v(s_3, t_3)\dif s \dif t\dif s_1 \dif t_1\dif s_2 \dif t_2\dif s_3 \dif t_3\Big\}.
\end{align*}
Thus, by Condition (C4),  we obtain that
\begin{align*}
&O((n+m)^{-8})P_{2}\\
 %&=\sum_{j=n+1}^{n+m} \sum_{i=1}^{n}\Bigg[ O\left\{\left(\iiiint\operatorname{tr}\left\{ \G_{2}(t,t_1) \G_{1}(s_1,s)\right\}v(s, t)v(s_1, t_1)\dif s \dif t\dif s_1 \dif t_1\right)^{2}\right\}\\&+O\Bigg\{\idotsint\operatorname{tr}\left\{\G_{1}(s,s_1)\G_{2}(t,t_1) \G_{1}(t_2,t_3) \G_{2}(s_2,s_3)\right\}\\&\quad  \times v(s, t)v(s_1, t_1)v(s_2, t_2)v(s_3, t_3)\dif s \dif t\dif s_1 \dif t_1\dif s_2 \dif t_2\dif s_3 \dif t_3\Bigg\}
%\Bigg] \\
	&=O((n+m)^{-6})\Big[O\Big\{\Big(\iiiint\operatorname{tr}\{ \G_{2}(t,t_1) \G_{1}(s_1,s)\}v(s, t)v(s_1, t_1)\dif s \dif t\dif s_1 \dif t_1\Big)^{2}\Big\}\\&\quad\quad+O\Big\{\idotsint\operatorname{tr}\{\G_{1}(s,s_1)\G_{2}(t,t_1) \G_{1}(t_2,t_3) \G_{2}(s_2,s_3)\}\\&\quad \quad\quad\quad  \times v(s, t)v(s_1, t_1)v(s_2, t_2)v(s_3, t_3)\dif s \dif t\dif s_1 \dif t_1\dif s_2 \dif t_2\dif s_3 \dif t_3\Big\}
	\Big]\\
	&=o(\sigma_{nm}^{4}(\X,\Y)).
\end{align*}
Similarly, we have
 $$O((n+m)^{-8})P_{1}=o(\sigma_{nm}^{4}(\X,\Y)),  ~~O((n+m)^{-8})P_{3}=o(\sigma_{nm}^{4}(\X,\Y)).$$
By Markov's inequality and \eqref{first-order-V}, we have
 \begin{align*}
\sum_{j=2}^{n+m}    \E\{V_{j}^{4} \mid \mathcal{F}_{j-1}\}=o(\sigma_{nm}^{4}(\X,\Y)).
\end{align*}
 Therefore, we have
 $$
	\sum_{j=2}^{n+m}   \E\{V_{j}^{2} I(|V_{j}|>\epsilon \sigma_{nm}(\X,\Y)) \mid \mathcal{F}_{j-1}\}/{\sigma_{nm}^{2}(\X,\Y)} \stackrel{ {P}}{\longrightarrow} 0 .
	$$
 \hfill\BlackBox

%\hfill\BlackBox

\vskip 0.2in
\bibliography{reference}

\end{document}